%% file: JFM-template.tex
\documentclass[nolineno]{jfm}

\usepackage{graphicx}
\usepackage{comment}
\usepackage{newtxtext}
\usepackage{newtxmath}
\usepackage{natbib}
\usepackage{amsmath,bm}
\usepackage{amsfonts}
\usepackage{upgreek}
\usepackage{float}
\usepackage{hyperref}
\usepackage{amssymb}
\usepackage{caption} 
\usepackage{subcaption}
\usepackage{tikz}
\usepackage[mode=buildnew]{standalone}
\usepackage{pgfplots}
\pgfplotsset{compat=1.18}
\usetikzlibrary{positioning} 
\usepackage{xcolor}
\hypersetup{
    colorlinks = true,
    urlcolor   = blue,
    citecolor  = black,
}

\newcommand{\RomanNumeralCaps}[1]
\linenumbers

\definecolor{babypink}{HTML}{fa9fb5}
\definecolor{pink}{HTML}{dd3497}
\definecolor{purple}{HTML}{49006a}
\definecolor{lightgreen}{HTML}{addd8e}
\definecolor{green}{HTML}{78c679}
\definecolor{darkgreen}{HTML}{238443}
\definecolor{lightblue}{HTML}{9ebcda}
\definecolor{lavender}{HTML}{8c6bb1}
\definecolor{darkpurple}{HTML}{4d004b}

\title{Turbulence transport in moderately dense gas--particle compressible flows}
\author{Archana Sridhar\aff{1}
  \corresp{\email{arsridha@umich.edu}},
  Rodney O. Fox\aff{2}
 \and Jesse Capecelatro\aff{3,}\aff{1}}

\affiliation{\aff{1} Department of Aerospace Engineering, University of Michigan, Ann Arbor, MI 48109, USA
\aff{2} Department of Chemical and Biological Engineering, Iowa State University, Ames, IA 50011, USA
\aff{3} Department of Mechanical Engineering, University of Michigan, Ann Arbor, MI 48109, USA}

\def \u{\bm{u}}

\definecolor{pd}{HTML}{d73027}
\definecolor{pdv}{HTML}{fee090}
\definecolor{transport}{HTML}{fdae61}
\definecolor{press-dil}{HTML}{7fbc41}
\definecolor{viscdiss}{HTML}{3288bd}
\definecolor{ps}{HTML}{8073ac}
\DeclareRobustCommand\full{\tikz[baseline=-0.6ex]\draw[thick] (0,0)--(0.5,0);}
\DeclareRobustCommand\mar{\tikz[baseline=-0.6ex]{\draw[ultra thick] (0,0)--(0.3,0);\fill (0.15,0) circle (0.04cm);}}
\DeclareRobustCommand\dotted{\tikz[baseline=-0.6ex]\draw[thick,dotted] (0,0)--(0.54,0);}
\DeclareRobustCommand\dashed{\tikz[baseline=-0.6ex]\draw[thick,dashed] (0,0)--(0.54,0);}

\makeatletter
\newcommand{\textsize}{\f@size pt}
\makeatother

\begin{document}
\maketitle

\begin{abstract}
This study employs three-dimensional particle-resolved simulations of planar shocks passing through a suspension of stationary solid particles to study wake-induced gas-phase velocity fluctuations, termed pseudo-turbulence. Strong coupling through interphase momentum and energy exchange generates unsteady wakes and shocklets in the interstitial space between particles. A Helmholtz decomposition of the velocity field shows that the majority of pseudo-turbulence is contained in the solenoidal component from particle wakes, whereas the dilatational component corresponds to the downstream edge of the particle curtain where the flow chokes. One-dimensional phase-averaged statistics of pseudo-turbulent kinetic energy (PTKE) are quantified at various stages of flow development. Reduction in PTKE is observed with increasing shock Mach number due to decreased production, consistent with single-phase compressible turbulence. The anisotropy in Reynolds stresses is found to be relatively constant through the curtain and consistent over all the conditions simulated. Analysis of the budget of PTKE shows that the majority of turbulence is produced through drag and balanced by viscous dissipation. The energy spectra of the streamwise gas-phase velocity fluctuations reveal an inertial subrange that begins at the mean interparticle spacing and decays with a power law of $-5/3$ and steepens to $-3$ at scales much smaller than the particle diameter. A two-equation model is proposed for PTKE and its dissipation. The model is implemented within a hyperbolic Eulerian-based two-fluid model and shows excellent agreement with the particle-resolved simulations.
\end{abstract}



\section{Introduction}
\label{sec:headings}
High-speed flows through particulate media occur in diverse applications, such as detonation blasts~\citep{zhang2001explosive}, volcanic eruptions~\citep{chojnicki2006shock,lube2020multiphase}, coal-dust explosions~\citep{sapko2000experimental,zheng2009statistical}, pulsed-detonation engines~\citep{roy2004pulse,chang2003shock}, and plume-surface interactions during interplanetary landings~\citep{Morris2011,Plemmons2009,capecelatro2022modeling}. In these examples, turbulence plays a crucial role in governing processes like reactant mixing and particle dispersion. However, the nature of this turbulence is distinct from both single-phase compressible turbulence and low-speed multiphase turbulence, posing a challenge to the accuracy of existing models.


Compressibility effects in turbulent flows are often characterized using the turbulent Mach number~\citep{sagaut2008homogeneous,jagannathan_reynolds_2016}. For values of $M_t\leq 0.3$, large scale separation exists between acoustics and turbulence. This results in a nearly incompressible flow called the quasi-isentropic regime. For higher values of $M_t$ (i.e. $0.3< M_t\leq 0.6$), dilatational effects are significant, leading to a nonlinear subsonic regime. The flows considered in the present study predominantly fall within this regime.

Since the 1970s, numerous studies have investigated the role of compressibility in the development of turbulent mixing layers and the generation of turbulent kinetic energy~\citep{bradshaw1977compressible,Brown_Roshko_1974,sarkar1991analysis}. Early work by \cite{sarkar1991analysis} and \cite{zeman90} examined the effects of dilatational dissipation, $\epsilon_d$, finding that its increase with $M_t$ leads to a reduction in turbulent kinetic energy, thereby decreasing turbulent mixing. They suggested that the suppression of growth rate is linked to increased $\epsilon_d$ due to shocklets. They developed a mathematical model to incorporate this effect into Reynolds stress closure models. However, \citet{sarkar_stabilizing_1995} later showed, using direct numerical simulations of turbulent homogeneous shear flow, that the reduction of turbulent kinetic energy is primarily due to decreased turbulence production, rather than directly caused by dilatational dissipation. Subsequent studies by \citet{vreman1996compressible} and \citet{pantano2002study} corroborated this finding, showing that dilatational dissipation is negligible.  Instead, the reduced growth rate of turbulence is linked to diminished pressure fluctuations and, consequently, lower turbulence production resulting from a reduction in the pressure-strain term.

\cite{kida_energy_1990} were among the first to analyse the kinetic energy spectrum in forced compressible turbulence, observing that its scaling is largely independent of Mach number. \cite{donzis_fluctuations_2013} also found that the turbulent kinetic energy spectrum in compressible isotropic turbulence follows a $-5/3$ power law in the inertial range for $0.1 \le M_t \le 0.6$, consistent with the classical Kolmogorov scaling for incompressible flows~\citep{kolmogorov1941b}. Further insights into compressibility scaling emerge from a Helmholtz decomposition of the velocity field $\u$ into its solenoidal component $\u_s$ and dilatational component $\u_d$~\citep{kida_energy_1990, donzis_fluctuations_2013, wang2011effect, wang2012effect, san_stratified_2018}. Compressibility effects are typically attributed to $\boldsymbol{u}_d$, and both \citet{donzis_fluctuations_2013} and \citet{wang2011effect} observed that the majority of turbulent kinetic energy resides in the solenoidal component, with $\boldsymbol{u}_d$ increasing with $M_t$. However, all of these studies have focused on single-phase compressible turbulent flows in the absence of particles.

In multiphase flows, interphase coupling introduces additional complexity that significantly influences energy transfer and turbulence characteristics. Fluid velocity fluctuations induced by particle wakes are referred to as pseudo-turbulence~\citep{Lance_Bataille_1991, mehrabadi2015pseudo}, a term also applied to bubble-induced turbulence (BIT) in liquid flows~\citep{risso_agitation_2018}. \citet{Lance_Bataille_1991} first demonstrated that a homogeneous swarm of bubbles generates pseudo-turbulence with a spectral subrange exhibiting a $-3$ power law. They showed that at statistically steady state, this spectral scaling results from a balance between viscous dissipation and energy production due to drag forces from rising bubbles. Similar scaling has since been observed in other bubbly flows~\citep{risso_agitation_2018, mercado2010bubble, mezui2022buoyancy, mezui2023experimental}. Subsequent experimental studies coupling BIT with shear-induced turbulence have found that the spectra of liquid velocity fluctuations follow a $-3$ scaling at small wave numbers, transitioning to a $-5/3$ scaling at higher wave numbers, suggesting a single-phase signature is preserved at the smallest scales~\citep{risso_agitation_2018}. Numerical simulations of gas--particle turbulent channel flow reveal that two-way coupling between the phases results in reduction in fluid-phase turbulent kinetic energy at the scale of individual particles, while a broadband reduction over all scales is observed at moderate to high mass loading~\citep{Capecelatro_Desjardins_Fox_2018}.

Over the past few decades, turbulence models have evolved to incorporate the effects of particles~\citep{troshko2001two,fox2014multiphase,ma_direct_2017}. A production term must be included to account for generation of turbulence through drag. A dissipation time scale is often employed based on the relative velocity between the phases ($u_r$) and particle diameter ($d_p$), given by $\tau = d_p / u_r$. The use of two-equation transport models for gas--solid flows dates back to the work of \citet{elghobashi}, who derived a rigorous set of equations for dilute concentrations of particles in incompressible flow using a two-fluid approach. Since then, models have been proposed for denser regimes in shear turbulence~\citep{ahmadi_ma}. \citet{crowe1996} provided a review of numerical models for turbulent kinetic energy in two-phase flows. However, these models are limited to \textit{intrinsic turbulence} whereby the carrier-phase turbulence would exist even in the absence of particles, as opposed to pseudo-turbulence that is entirely generated by the particle phase. \citet{mehrabadi2015pseudo} recently developed an algebraic model for pseudo-turbulent kinetic energy (PTKE) based on particle-resolved simulation data that depends on the slip Reynolds number and particle volume fraction. A limitation of algebraic models is that PTKE can only be predicted in regions of finite volume fraction. In cases where turbulence is generated within a suspension of particles and advects downstream into the surrounding gas, transport equations for PTKE are more appropriate \citep{shallcross2020volume}.

Particle-laden compressible flows challenge numerical models due to the strong coupling between shock waves, particles, and turbulence over a wide range of scales. Using particle-resolved simulations of compressible homogeneous flows past random arrays of particles, \citet{khalloufi2023drag} found that both $M_t$ and PTKE increase with particle volume fraction for a fixed free-stream Mach number. Experimental and numerical studies of particle-laden underexpanded jets have demonstrated significant modification of shock structures due to the two-way coupling between the gas and particles even at low volume fractions where one-way coupling would be deemed appropriate for single-phase flow \citep{sommerfeld1994structure,patel2024experimental}. \cite{regele_unsteady_2014,hosseinzadeh-nik_investigation_2018} conducted two-dimensional particle-resolved simulations of a shock interacting with a moderately dense particle curtain and found PTKE to be of the same order of magnitude as the resolved kinetic energy. \cite{mehta2020pseudo} conducted three-dimensional inviscid simulations and observed fluctuations of the order of 50\% of the kinetic energy based on the mean velocity. They also observed that the strength of the fluctuations increases with the shock Mach number, $M_s$, and volume fraction $\Phi_p$. \cite{osnes2019computational} performed particle-resolved simulations of shock--particle interactions and proposed an algebraic model for PTKE based on the mean flow speed and particle volume fraction.  \citet{shallcross2020volume} presented a volume-filtered framework for the multiphase compressible Navier--Stokes equations. They proposed a one-equation model for PTKE containing a production term due to drag and an algebraic closure for dissipation. The dissipation model used a time-scale similar to~\cite{ma_direct_2017} with a blending function to account for regions devoid of particles informed by two-dimensional simulations.

This study focuses on turbulence transport at moderate volume fractions and Mach numbers. The paper is organized as follows. In \S~\ref{sec:setup}, the simulation configuration and governing equations are presented. Simulation results are provided in \S~\ref{sec:results}, starting with a qualitative assessment of the flow, followed by one-dimensional phase-averaged statistics of the gas-phase velocity. The budget of PTKE is presented next, revealing key production and dissipation mechanisms. The energy spectra within the particle curtain is then presented and separate contributions from solenoidal and dilatational components highlight the sources of PTKE. In \S~\ref{sec:model}, a two-equation turbulence model for PTKE and its dissipation is proposed and implemented within a hyperbolic two-fluid model. Results from \S~\ref{sec:results} are used to guide closure. An a-posteriori analysis is performed and first- and second-order statistics are compared. Key findings and results are summarized in \S~\ref{sec:conclusions}.

\section{Simulation setup and methods}\label{sec:setup}
\subsection{Flow configuration}
\begin{figure}
    \centering     \includegraphics[width=0.75\linewidth]{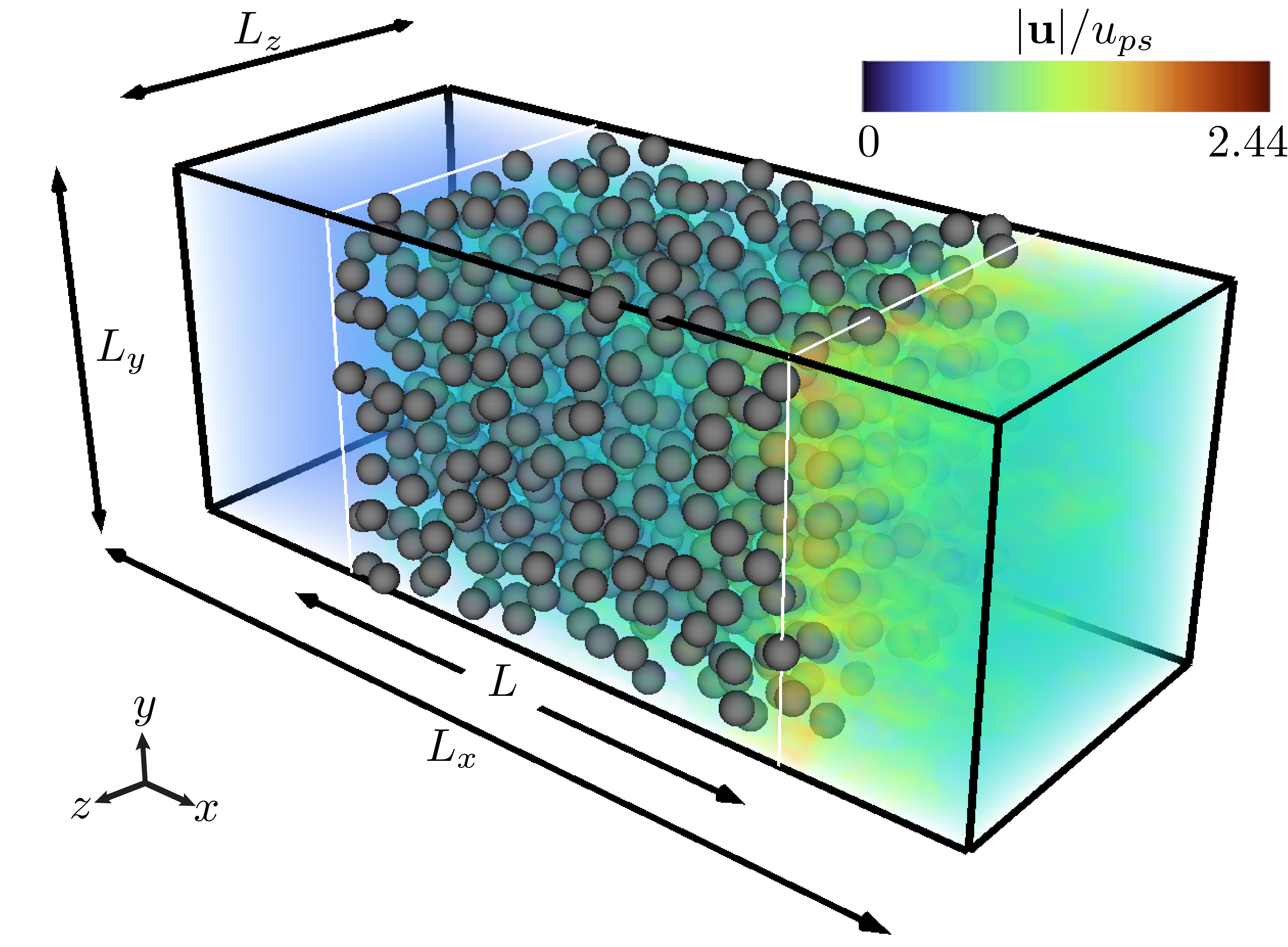}
     \caption{The simulation domain showing particle position and a volume rendering of the gas-phase velocity magnitude after the shock has passed the curtain ($t/\tau_L = 2$) with $\Phi_p=0.2$ and $M_s=1.66$.}
     \label{fig:schem}
\end{figure}

To isolate shock--particle--turbulence interactions, three-dimensional particle-resolved simulations of a planar shock passing through a suspension of particles are conducted. These simulations are designed to emulate the multiphase shock-tube experiments of~\cite{wagner2012multiphase}. Figure~\ref{fig:schem} shows a volume rendering of the gas-phase velocity magnitude within the simulation domain at a moment when the shock has advanced significantly beyond the curtain and exited the domain.  The velocity increases across the particle curtain with maximum values at the downstream curtain edge where the flow chokes due to the sudden change in volume fraction.

Particles with diameter $D=115$~$\upmu$m and density $2520 \ \rm{kg}/\rm{m}^3$ are randomly distributed within a curtain of thickness $L=2$~mm ($L=17.4D$). A minimum of two grid points is maintained between particle surfaces. A planar shock is initially placed at a non-dimensional length of $x=5.5D$ with the flow direction parallel to the $x$-axis. The upstream edge of the curtain is placed at $x=7D$. Periodic boundary conditions are imposed in the two spanwise ($y$ and $z$) directions. The domain size for all but one case is $[L_x\times L_y\times  L_z]=[30\times 12\times 12] D$. $L_y$ and $L_z$ were chosen based on a domain size independence study summarized in Appendix~\ref{appA}. The domain is discretized with uniform grid spacing $\Delta x=D/40$, corresponding to $[1201\times480\times 480]$ grid points.

 The pre-shock gas-phase density is $\rho_\infty=0.987 \ \rm{kg}/\rm{m}^3$, pressure $P_\infty=82.7\ {\rm kPa}$, sound speed $c_\infty=343 \ \rm{m}/\rm{s}$ and velocity $u_\infty=0 \ \rm{m}/\rm{s}$. Post-shock properties, denoted by the subscript \textit{ps}, are obtained via the Rankine--Hugoniot conditions. The shock Mach number is defined as ${M_s} = u_s/c_\infty$, where $u_s$ is the shock speed. A reference time-scale based on the distance (in terms of particle curtain length) that the shock travels is defined as $\tau_L=L/u_s$. The particle Reynolds number based on post-shock properties is defined as $\Rey_{ps} = \rho_{ps}u_{ps}D / \mu_{ps}$, where $\mu_{PS}$ is the gas-phase viscosity at temperature $T_{ps}$. The number of particles $N_p$ within the curtain is determined from the average volume fraction, $\Phi_p$. A summary of the cases considered in this study is given in Table~\ref{tab:kd}. Cases $1-9$ represent different combinations of $M_s$ and $\Phi_p$. Case $10$ exhibits a longer domain length to study turbulence transport downstream of the particle curtain.
  
  \begin{table}
    \begin{center}
  \def~{\hphantom{0}}
    \begin{tabular}{ccccccc}
        Case No.  & ~$M_s$~   &   ~$\Phi_p$~  & ~$\Rey_{ps}$~ & ~$N_p$~ & ~$L_x/D$~ & ~$L_y/D$~\\[3pt]
         1   & 1.2 & 0.1 & 813 & 467 & 30 & 12\\ 
         2   & 1.2 & 0.2 & 813 & 935 & 30 & 12\\
         3  & 1.2 & 0.3 & 813 & 1402 & 30 & 12\\
         4   & 1.66 & 0.1 & 3251 & 467 & 30 & 12\\ 
         5 & 1.66 & 0.2 & 3251 & 935 & 30 & 12\\ 
         6 & 1.66 & 0.3 & 3251 & 1402 & 30 & 12\\
         7 & 2.1 & 0.1 & 5591 & 467 & 30 & 12\\ 
         8 & 2.1 & 0.2 & 5591 & 935 & 30 & 12\\
         9 & 2.1 & 0.3 & 5591 & 1402 & 30 & 12\\
         10 & 1.66 & 0.3 & 3251 & 1402 & 58 & 12 \\
    \end{tabular}
    \caption{Parameters for the various runs used in this study.}
    \label{tab:kd}
    \end{center}
  \end{table}

\subsection{Governing equations}\label{sec:eqs}
The gas-phase is governed by the viscous compressible Navier--Stokes, given by
\begin{equation}
\frac{\partial \rho}{\partial t}  + \nabla \cdot (\rho \u) = 0,
\end{equation}
\begin{equation}
\frac{\partial \rho \u}{\partial t} + \nabla \cdot (\rho \u \otimes \u + p \mathbb{I} - \boldsymbol{\sigma})=0
\end{equation}
and
\begin{equation}
\frac{\partial \rho E}{\partial t} + \nabla \cdot (\{ \rho E + p \} \u + \boldsymbol{q} - \u \cdot \boldsymbol{\sigma}) = 0,
\end{equation}
where $\rho$ is the gas-phase density, $\u=(u,v,w)$ is the velocity and $E$ is the total energy. The viscous stress tensor is
\begin{equation}
    \boldsymbol{\sigma} = \mu (\nabla \u + \nabla \u^T) + \lambda \nabla \cdot \u
\end{equation}
and the heat flux is
\begin{equation}
    \boldsymbol{q} = - k \nabla T
\end{equation}
where $k$ is the thermal conductivity. The dynamic viscosity is modelled as a power law, $\mu= \mu_0 [(\gamma-1)T/T_0]^n$, where $\gamma=1.4$ is the ratio of specific heats and $n=0.666$. The second coefficient of viscosity is $\lambda = \mu_B - 2/3 \mu$ where the bulk viscosity $\mu_B=0.6 \mu$ is chosen as a model for air. The thermal conductivity is varied with a similar power law as viscosity to maintain a constant Prandtl number of $0.7$. Thermodynamic relations for temperature and pressure are given by
\begin{equation}
    T = \frac{\gamma p}{(\gamma -1)\rho} \quad\text{and}\quad p= (\gamma-1)(\rho E - \frac{1}{2} \rho \u \cdot \u).
\end{equation}
No-slip, adiabatic boundary conditions are enforced at the particle surfaces. Details on the numerical implementation are provided in the following section.

\subsection{Numerics}
The simulations are performed using the compressible multiphase flow solver \texttt{jCODE}~\citep{jcode}. Spatial derivatives are approximated using narrow-stencil finite-difference operators that satisfy the summation by parts (SBP) property~\citep{Strand1994,Svard}. A sixth-order centered finite-difference scheme is used for the interior points, and a fourth-order, one-sided finite difference is applied at the boundaries.  Kinetic energy preservation is achieved using a skew-symmetric-type splitting of the inviscid fluxes~\citep{pirozzoli2011stabilized}, providing nonlinear stability at low Mach number. To ensure proven temporal stability, the SBP scheme is combined with the simultaneous approximation-term (SAT) treatment that weakly enforces characteristic boundary conditions at the inflow and outflow~\citep{Svard}. High-order SBP dissipation operators~\citep{Mattsson2004} are employed to dampen spurious high-wavenumber modes. Localized artificial diffusivity is used as a means of shock capturing by following the `LAD-D2-0' formulation in~\citet{kawai2010assessment}. To limit the artificial diffusivity to regions of high compression (shocks), we employ the sensor originally proposed by~\citet{ducros1999large} and later improved by~\citet{hendrickson2018improved}. More details can be found in \citet{khalloufi2023drag}. The equations are advanced in time using a standard fourth-order Runge--Kutta scheme, with a constant Courant--Friedrichs--Lewy (CFL) number of 0.8.

A ghost-point immersed boundary method originally proposed by~\citet{mohd1997combined} and later extended to compressible flows by~\citet{chaudhuri2011use} is employed to enforce boundary conditions at the surface of the particles. Values of the conserved variables at ghost points residing within the solid are assigned after each Runge-–Kutta sub-iteration to enforce no-slip, adiabatic boundary conditions. The framework was validated in our previous study~\citep{khalloufi2023drag}, demonstrating that 40 grid points across the particle diameter is sufficient to capture drag and PTKE. An assessment of the domain size and sensitivity to random particle placement is reported in Appendix~\ref{appA}.


\section{Results}\label{sec:results}
\subsection{Flow visualisation}\label{sec:qualitative}
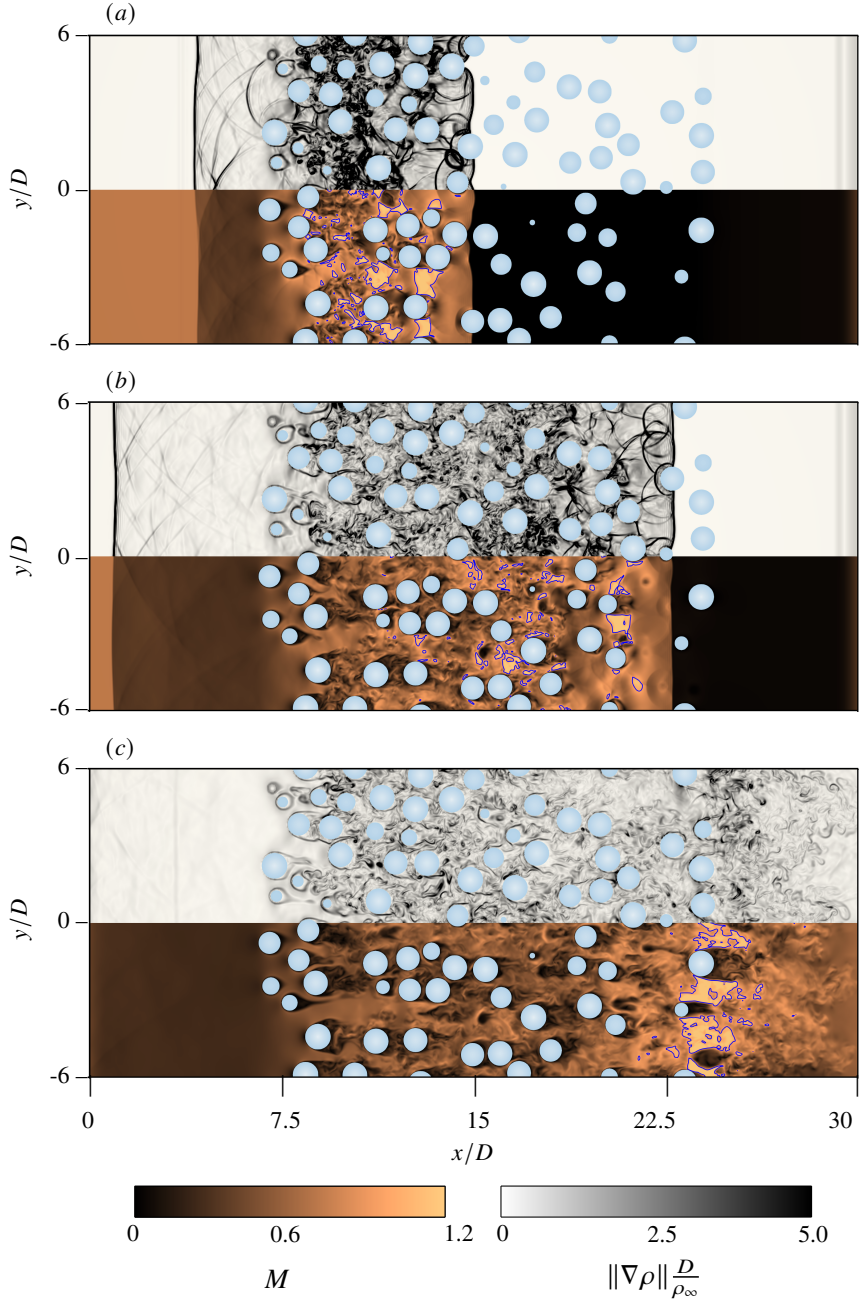
\begin{figure}
  \centering
  \hspace{-24pt}\resizebox{0.85\textwidth}{!}{\input{0Caseshockphysics.tex}}
  \caption{Two-dimensional planes showing Schlieren (top half of each plot) and local Mach number $M=\lVert\u\rVert/c$ (bottom half) at (a) $t/\tau_L=0.5$, (b) $t/\tau_L=1$ and (c) $t/\tau_L=2$ for ${M}_s=1.66$ and $\Phi_p=0.2$. Contour lines of $M=1$ shown in purple. Blue circles depict particle cross sections.}
  \label{fig:qualitative}
\end{figure}
Instantaneous snapshots of the flow field corresponding to Case 2 (${M}_s=1.66, \Phi_p=0.2$) at $t/\tau_L=0.5$, $1$, and $2$ are presented in figure~\ref{fig:qualitative}. A two-dimensional slice in the $x-y$ plane shows the local gas-phase Mach number and numerical Schlieren in the vicinity of the particles. The incident shock travels in the positive $x$ direction and impinges the particle curtain located at $x_0$ at time $t=0$. Upon impact, the shock splits into a weaker transmitted shock that penetrates the curtain, as shown in figure~\ref{fig:qualitative}$(a)$. At the upstream edge of the curtain, the arrival of the shock generates multiple shocklets at the surface of each particle, which coalesce into a reflected shock wave. Shock--particle interactions are seen to generate significant fluctuations in the gas-phase velocity. Contour lines of $M=1$ (shown in purple) demarcate local supersonic regions. In figure~\ref{fig:qualitative}$(b)$, the shock has nearly reached the downstream curtain edge, and the local supersonic regions move downstream with the flow. Figure~\ref{fig:qualitative}$(c)$ shows that the flow has stabilized with both the transmitted and reflected shocks having exited the domain boundaries. The particles restrict the area of the transmitted shock, causing the gas phase to choke near the downstream edge of the curtain due to the abrupt change in volume fraction, followed by a supersonic expansion. Velocity fluctuations induced by the particles advect downstream from the curtain and dissipate, akin to grid-generated turbulence.

\subsection{Averaging operations}\label{sec:avg}
The flows under consideration are unsteady and statistically homogeneous in the two spanwise directions. Averaged quantities depend solely on one spatial dimension ($x$) and time. Due to the presence of particles and gas-phase density variations, special attention must be given to the averaging process. To facilitate statistical phase-averaging, an indicator function is defined as
\begin{equation}\label{ind}
    \mathcal{I}(\bm{x}) = \begin{cases}
        1 & \text{if } \bm{x} \in \text{gas phase},\\
        0 & \text{if } \bm{x} \in \text{particle}.
    \end{cases}
\end{equation}
Spatial averages are taken as integrals over $y-z$ slices. The integration of the indicator function yields a volume fraction $\alpha$ (or area fraction in this case) that depends solely on $x$ (time is omitted since the particles being stationary), given by
\begin{equation}
    \alpha_g (x) =\langle\mathcal{I}\rangle\equiv\frac{1}{L_y L_z}\int_{L_y}\int_{L_z}\mathcal{I} \ {\rm d}y\,{\rm d}z,
\end{equation}
where angled brackets denote a spatial average. Two other important averaging operations that will be used throughout this study are phase averages and density-weighted (Favre) averages. If $\psi(\bm{x},t)$ represents a random field variable, these averages are defined as
\begin{equation}
\begin{aligned}
    &\text{Spatial-average: } \langle\psi\rangle (x,t) \equiv \frac{1}{L_y L_z}\int_{L_y}\int_{L_z}  \psi \ {\rm d}y\,{\rm d}z, \\
    &\text{Phase-average: } \overline{\psi}(x,t) \equiv \frac{\langle \mathcal{I}\psi\rangle}{\langle \mathcal{I}\rangle}\equiv \frac{\langle \mathcal{I}\psi\rangle}{\alpha_g}, \\
    &\text{Favre-average: } \widetilde{\psi}(x,t) \equiv \frac{\langle\mathcal{I} \rho \psi \rangle}{\langle \mathcal{I}\rho \rangle} \equiv \frac{\overline {\rho \psi}}{\overline{\rho}}.
\end{aligned}
\end{equation}
Spatial averages and phase averages are related via $\langle{\mathcal{I}\psi} \rangle = \alpha_g \overline{\psi}$ and similarly $\overline{\rho\psi}=\overline{\rho}\widetilde{\psi}$.  A field variable can be decomposed into its phase-average and a fluctuating quantity as $\psi = \overline{\psi} + \psi'$. Similarly, the Favre decomposition is $\psi = \widetilde{\psi} + \psi''$. 

\subsection{Mean velocity, fluctuations and anisotropy}\label{sec:1dstats}

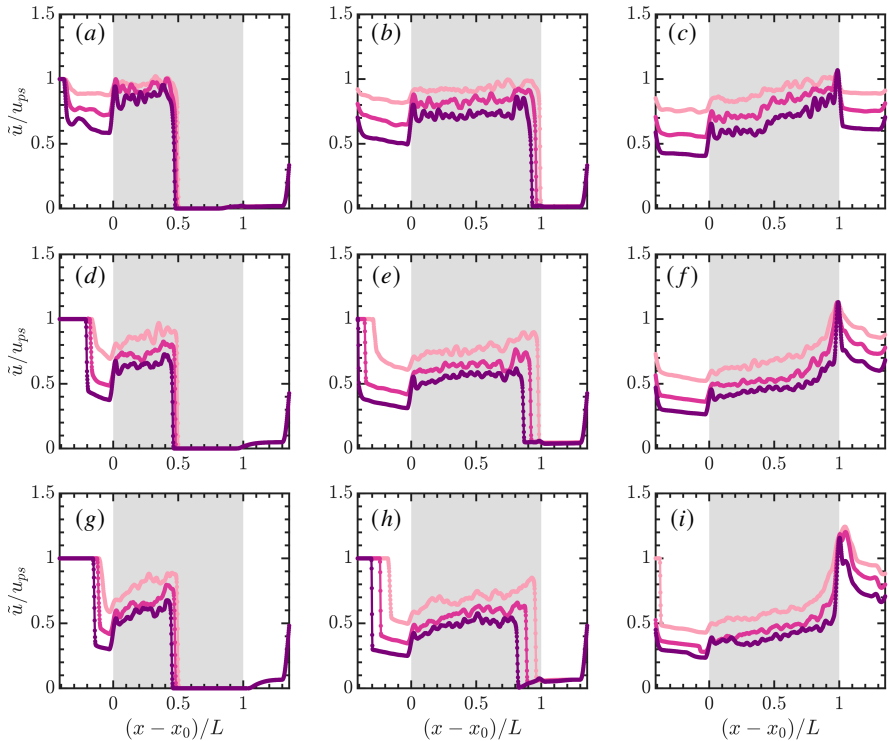
\begin{figure}
  \centering
  \begin{tabular}{ccc}
      \subfloat{
          \includegraphics[width=0.28\textwidth]{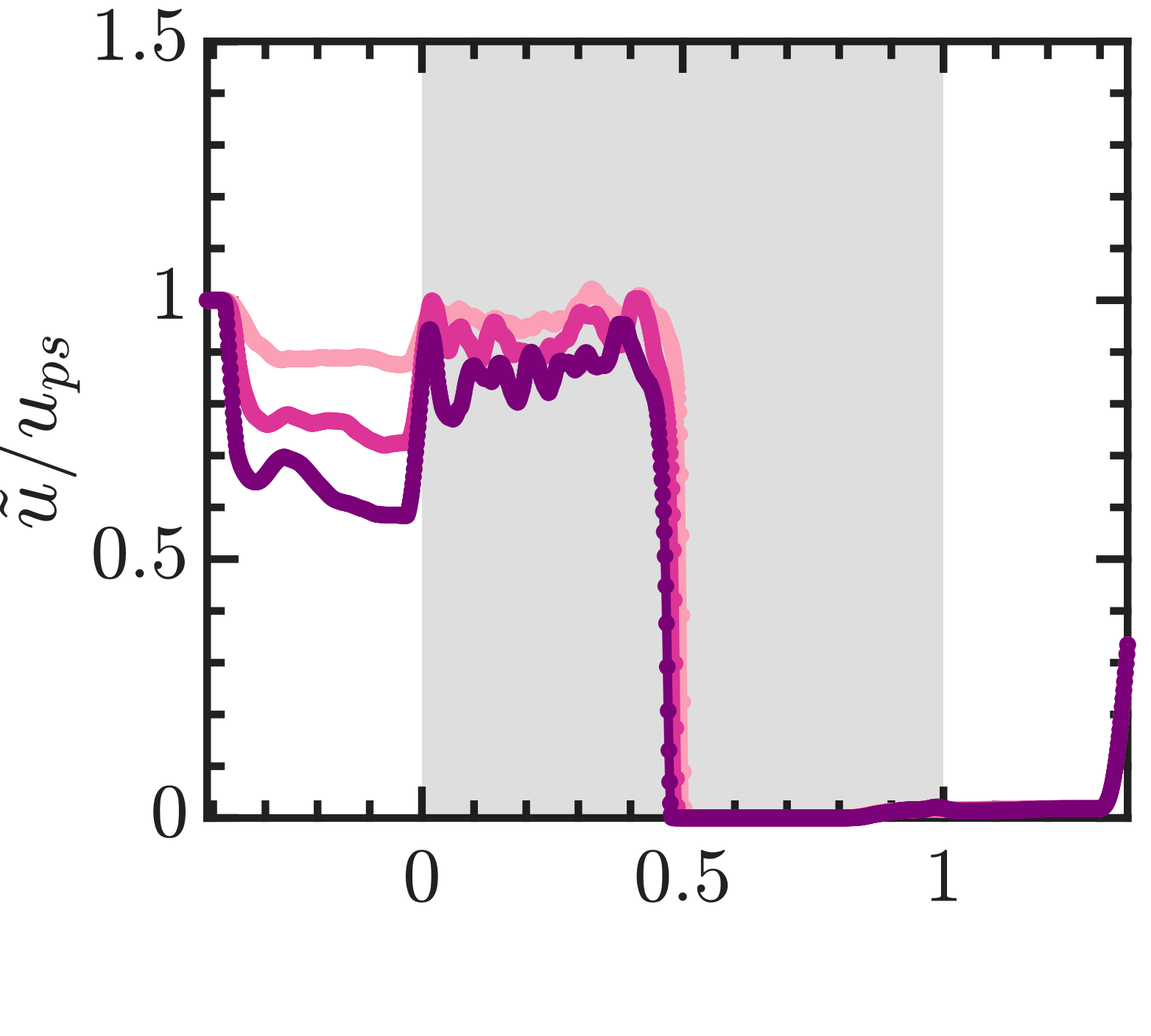}
      } 
      \subfloat{
          \includegraphics[width=0.28\textwidth]{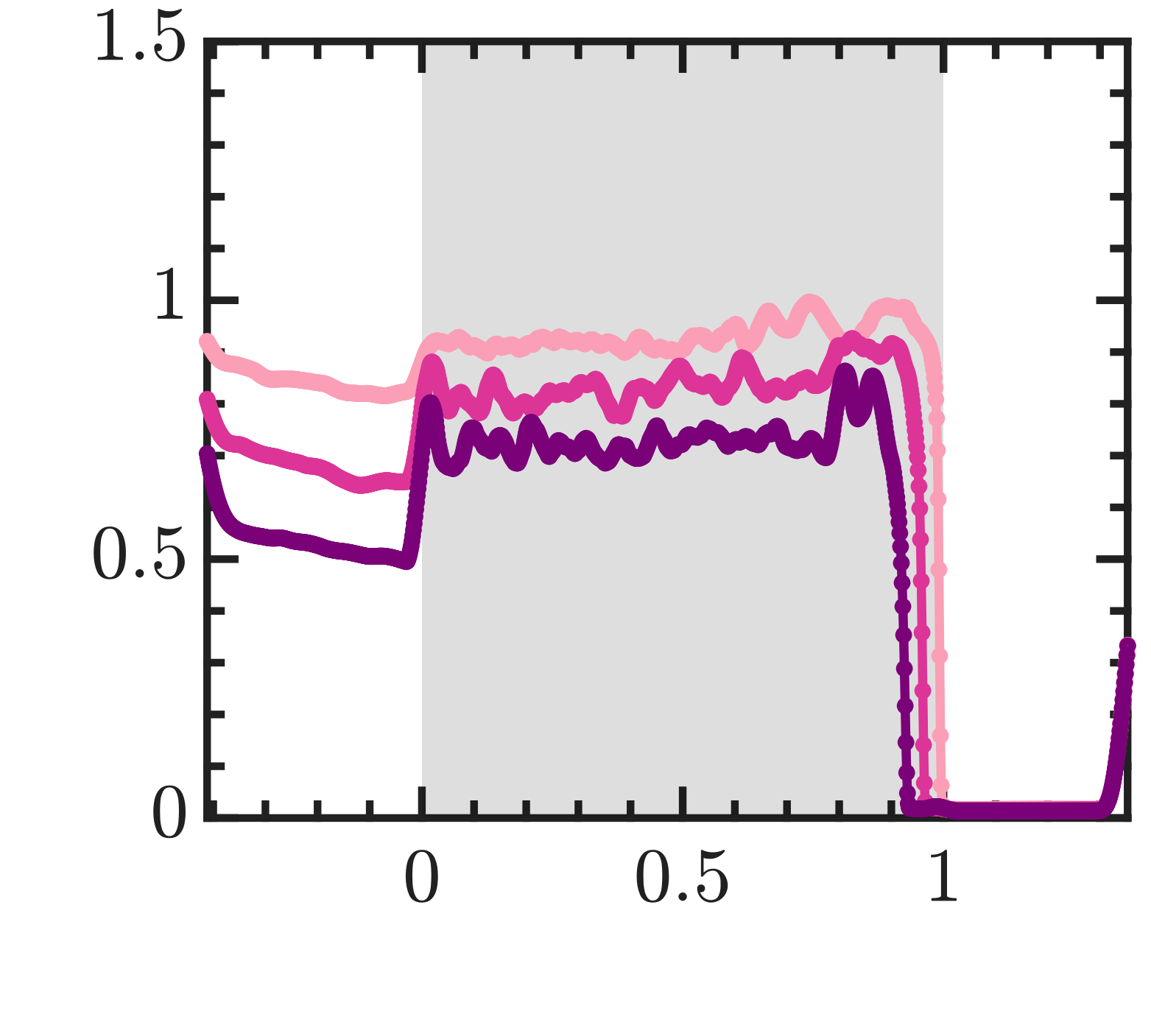}
      } 
      \subfloat{
          \includegraphics[width=0.28\textwidth]{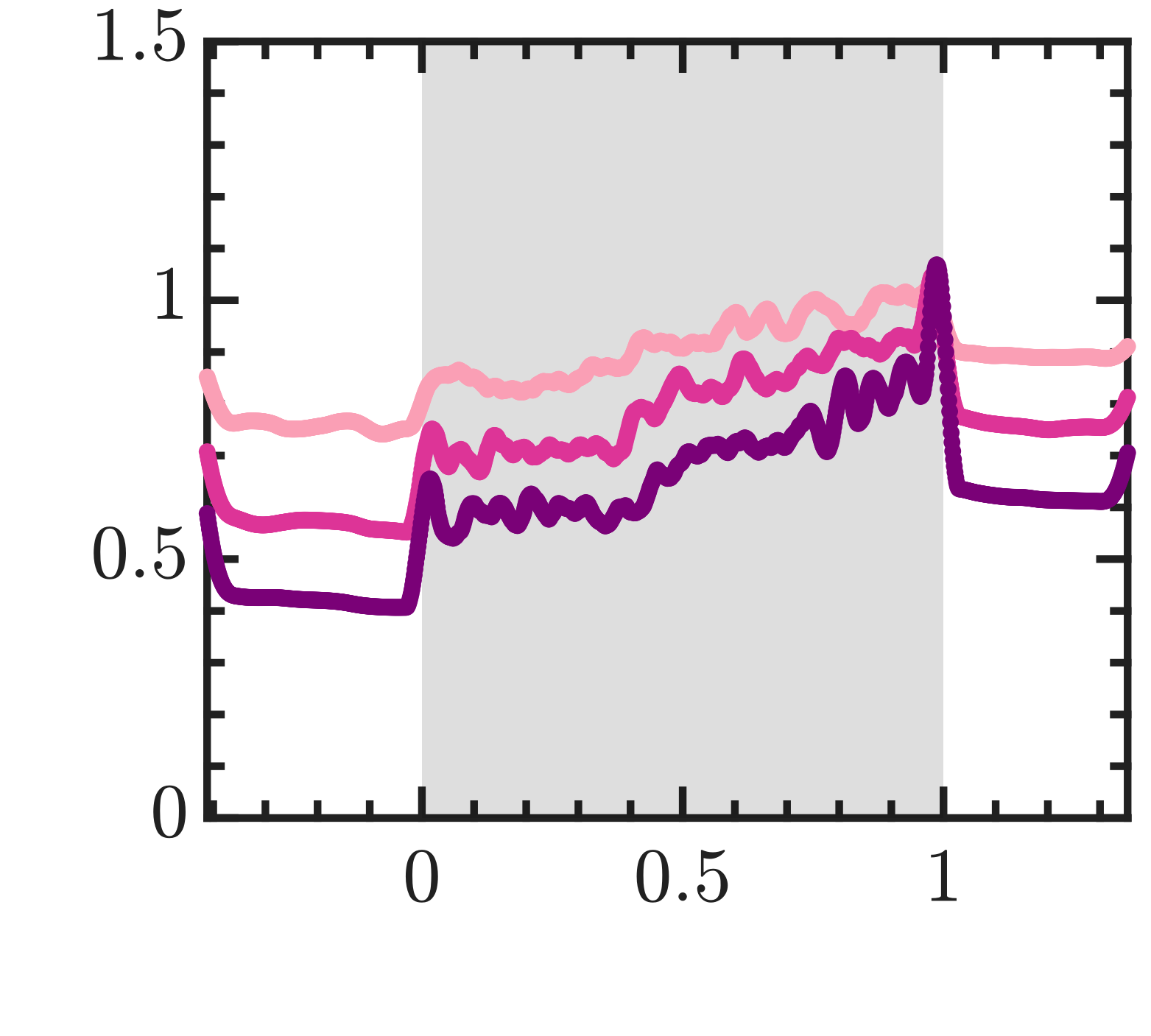}
      } \vspace{-10pt} \\
      \subfloat{
          \includegraphics[width=0.28\textwidth]{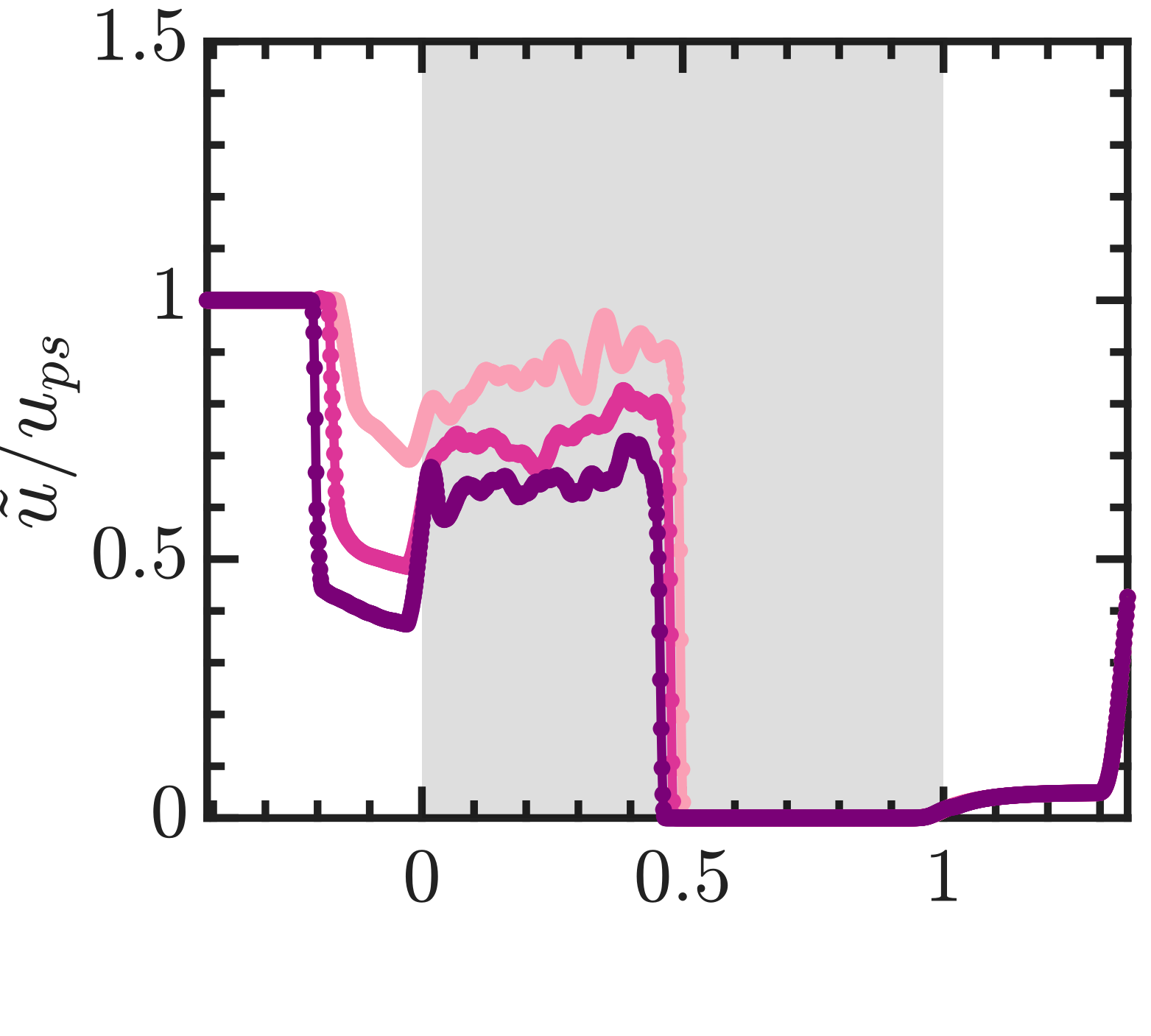}
      } 
      \subfloat{
          \includegraphics[width=0.28\textwidth]{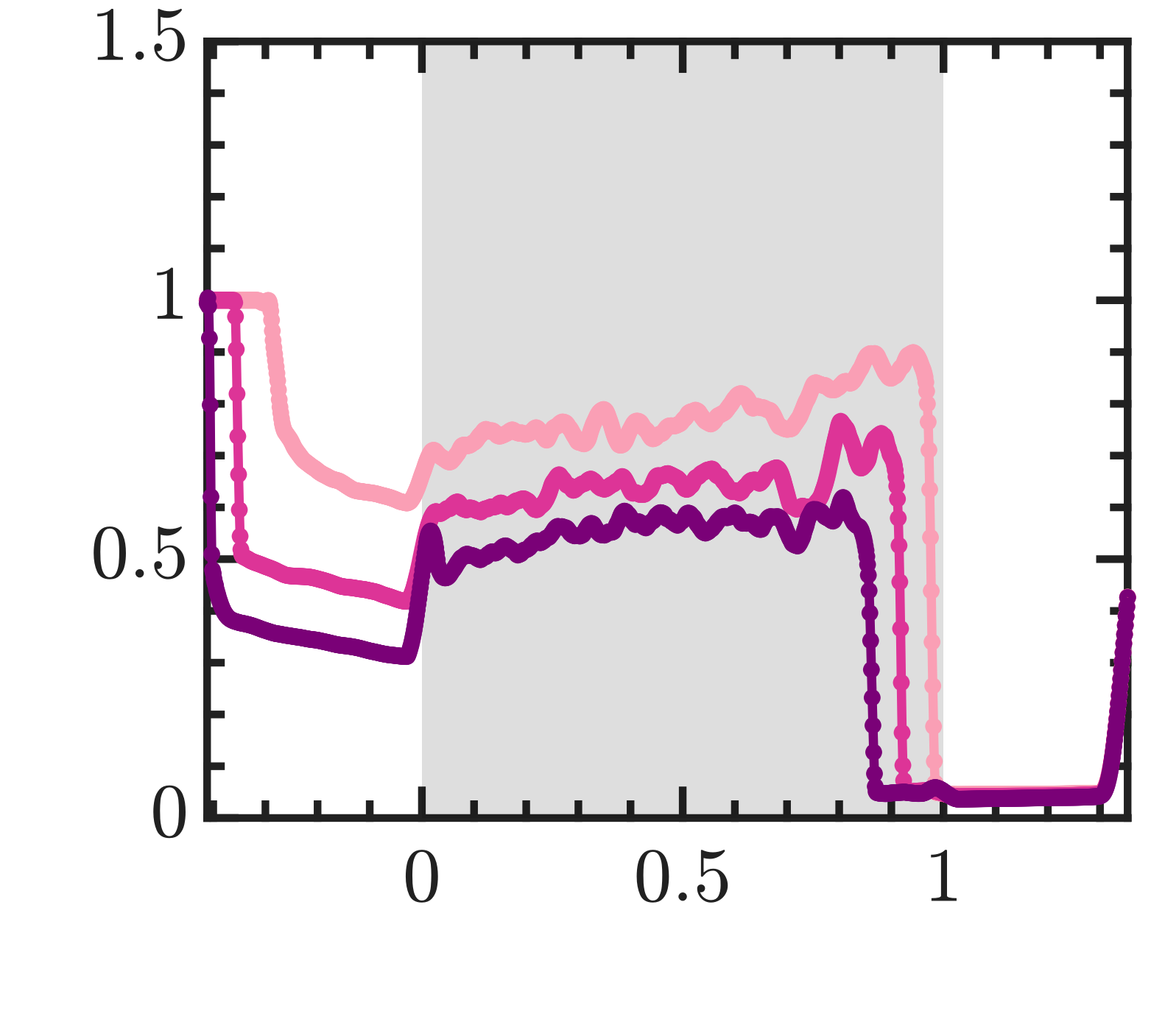}
      } 
      \subfloat{
          \includegraphics[width=0.28\textwidth]{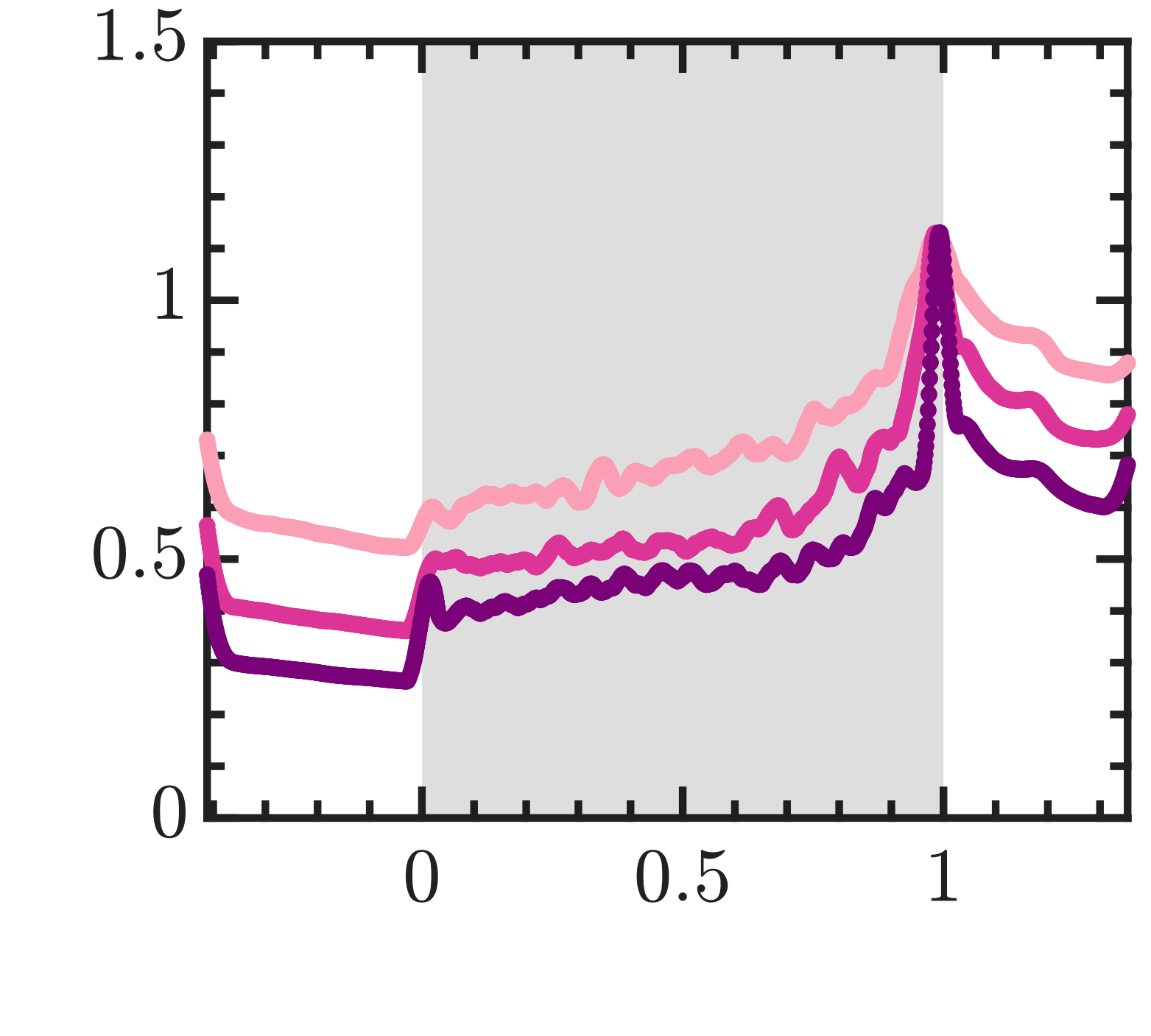}
      } \vspace{-10pt} \\
      \subfloat{
          \includegraphics[width=0.28\textwidth]{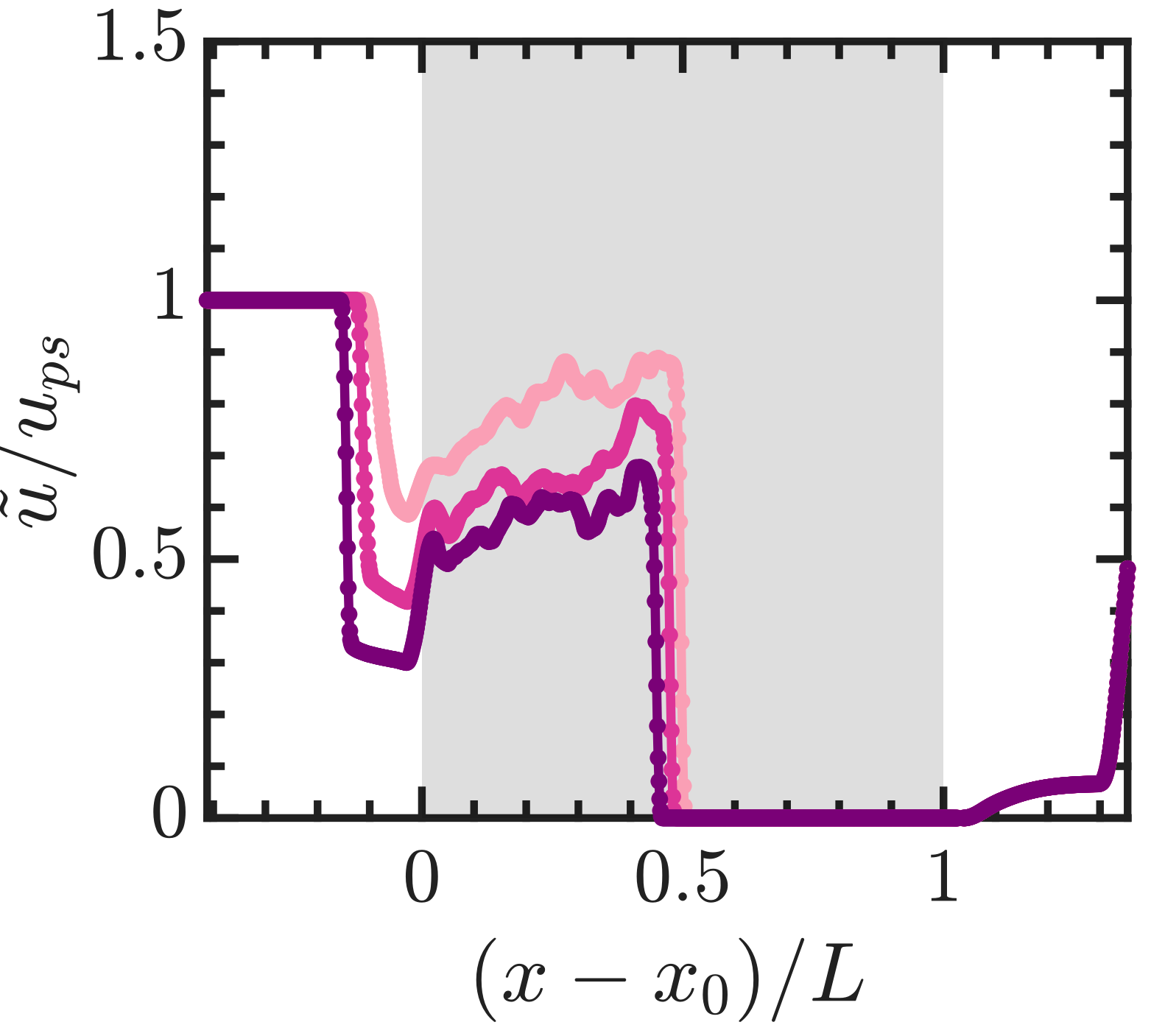}
      }
      \subfloat{
          \includegraphics[width=0.28\textwidth]{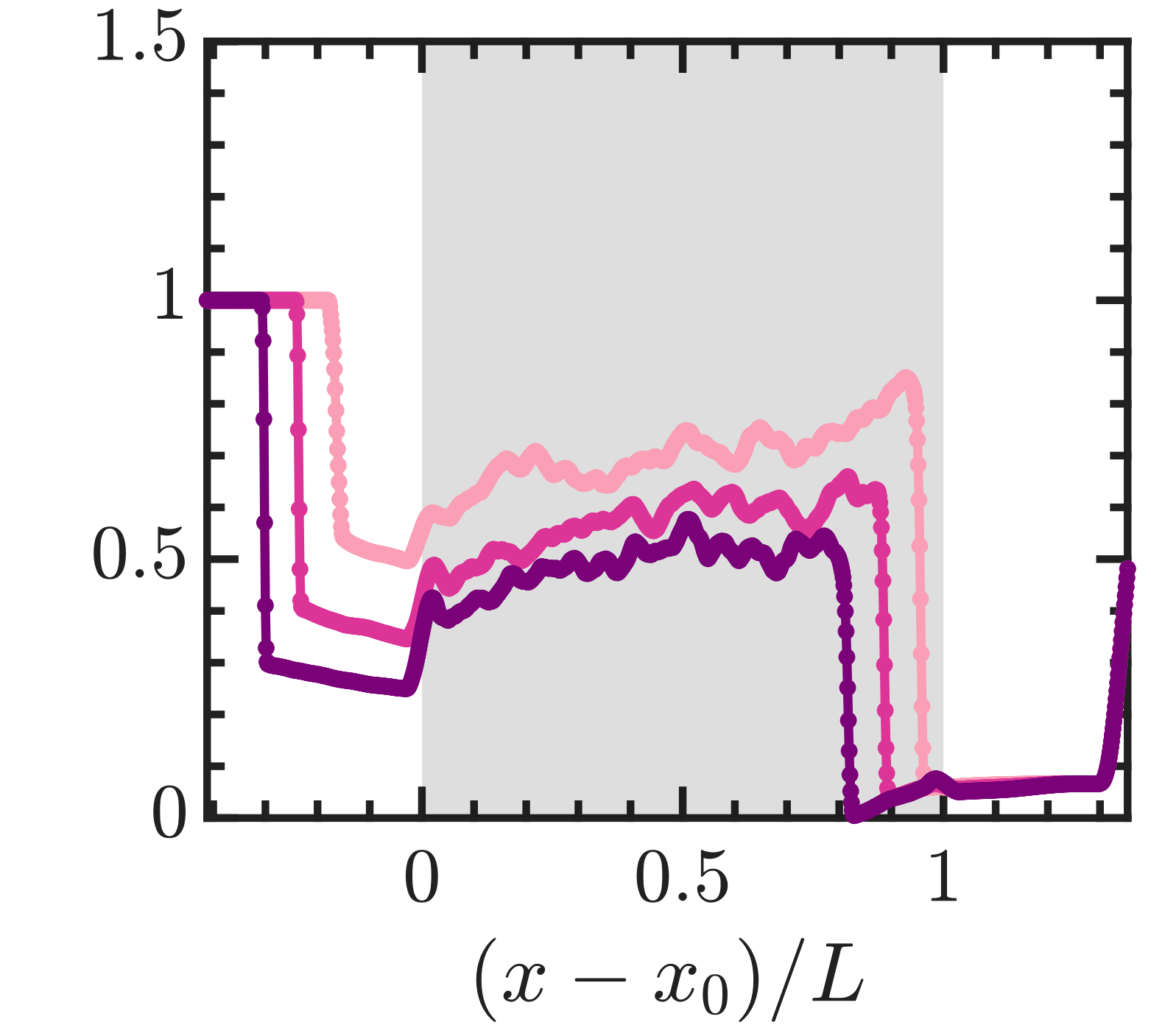}
      }
      \subfloat{
          \includegraphics[width=0.28\textwidth]{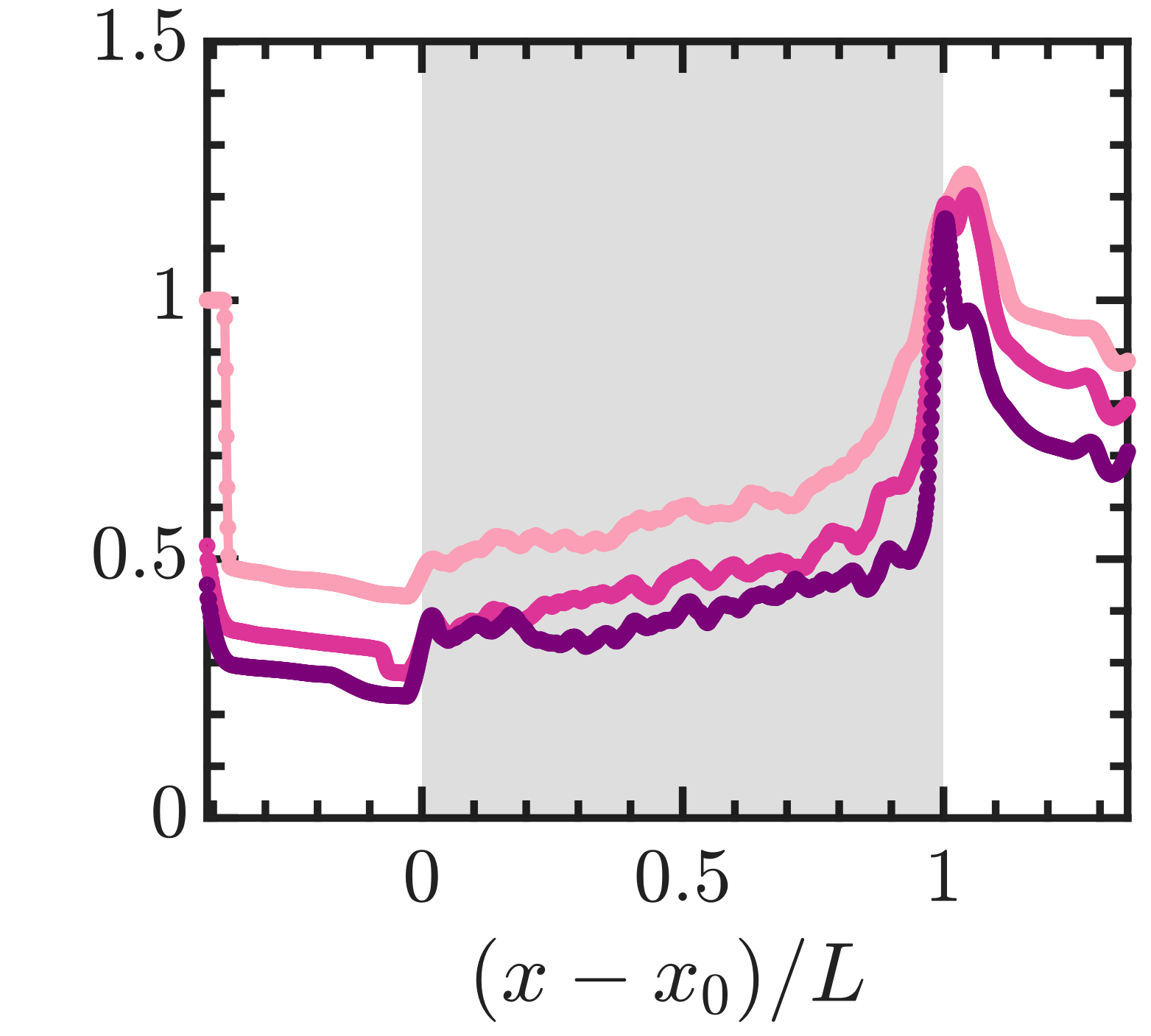}
      }
  \end{tabular}
  \begin{tikzpicture}[overlay, remember picture]
    \node at (-10.8,4.6) {$(a)$};
    \node at ([xshift=0.29\linewidth]-10.8,4.6) {$(b)$};
    \node at ([xshift=0.58\linewidth]-10.8,4.6) {$(c)$};
    \node at (-10.8,1.4) {$(d)$};
    \node at ([xshift=0.29\linewidth]-10.8,1.4) {$(e)$};
    \node at ([xshift=0.58\linewidth]-10.8,1.4) {$(f)$};
    \node at (-10.8,-1.8) {$(g)$};
    \node at ([xshift=0.29\linewidth]-10.8,-1.8) {$(h)$};
    \node at ([xshift=0.58\linewidth]-10.8,-1.8) {$(i)$};  
    
    
\end{tikzpicture}
  \caption{Mean gas-phase velocity profiles. Darker lines indicate higher volume fractions: $\Phi_p=0.1$ (light pink), $\Phi_p=0.2$ (pink), $\Phi_p=0.3$ (purple). $t/\tau_L=0.5$ (left), $t/\tau_L=1$ (middle) and $t/\tau_L=2$ (right). $(a)$--$(c)$ $M_s=1.2$, $(d)$--$(f)$ $M_s=1.66$, $(g)$--$(i)$ $M_s=2.1$. The gray-shaded region indicates the location of the particle curtain.}
  \label{fig:meanvel}
\end{figure}

The Favre-averaged gas-phase velocity, $\widetilde{u}$, as a function of the streamwise direction at three different time instances ($t/\tau_L = 0.5$, $1$, and $2$) is shown in figure~\ref{fig:meanvel}. The abrupt drop in velocity observed at early times ($t/\tau_L = 0.5$ and $1$) marks the location of the transmitted shock. The flow decelerates significantly as it approaches the particle curtain due to drag, with greater reduction in velocity relative to the post-shock velocity at higher volume fractions. The flow then accelerates as it traverses the curtain. At the latest time ($t/\tau_L = 2$), a sharp increase in $\widetilde{u}$ at the downstream edge of the curtain is seen across all cases, indicating a region of choked flow transitioning to supersonic velocities. Similar trends in the velocity field have been reported previously \citep[e.g.][]{theofanous2018shock,mehta2018propagation,osnes2019computational}. \citet{mehta2018propagation} obtained an analytical solution of the Riemann problem for a duct with a sudden change in cross-sectional area as a simpler means of predicting the flow through a particle curtain. The solution was found to compare well with inviscid simulations of shock--particle interactions, though it is unable to predict the choking behaviour leading to supersonic velocities observed here.

The amplitude and speed of the reflected shocks, indicated by the abrupt increase in velocity upstream of the particle curtain, increase with $\Phi_p$. The transmitted shock travels faster through the curtain at lower $\Phi_p$ where the flow is less obstructed. For a given volume fraction, the magnitude of $\widetilde{u}$ decreases with increasing $M_s$, and the flow-expansion region at the downstream edge rises sharply with increasing $M_s$.

\begin{figure}
  \centering
  \begin{tabular}{cc}
      \subfloat{
          \includegraphics[width=0.48\textwidth]{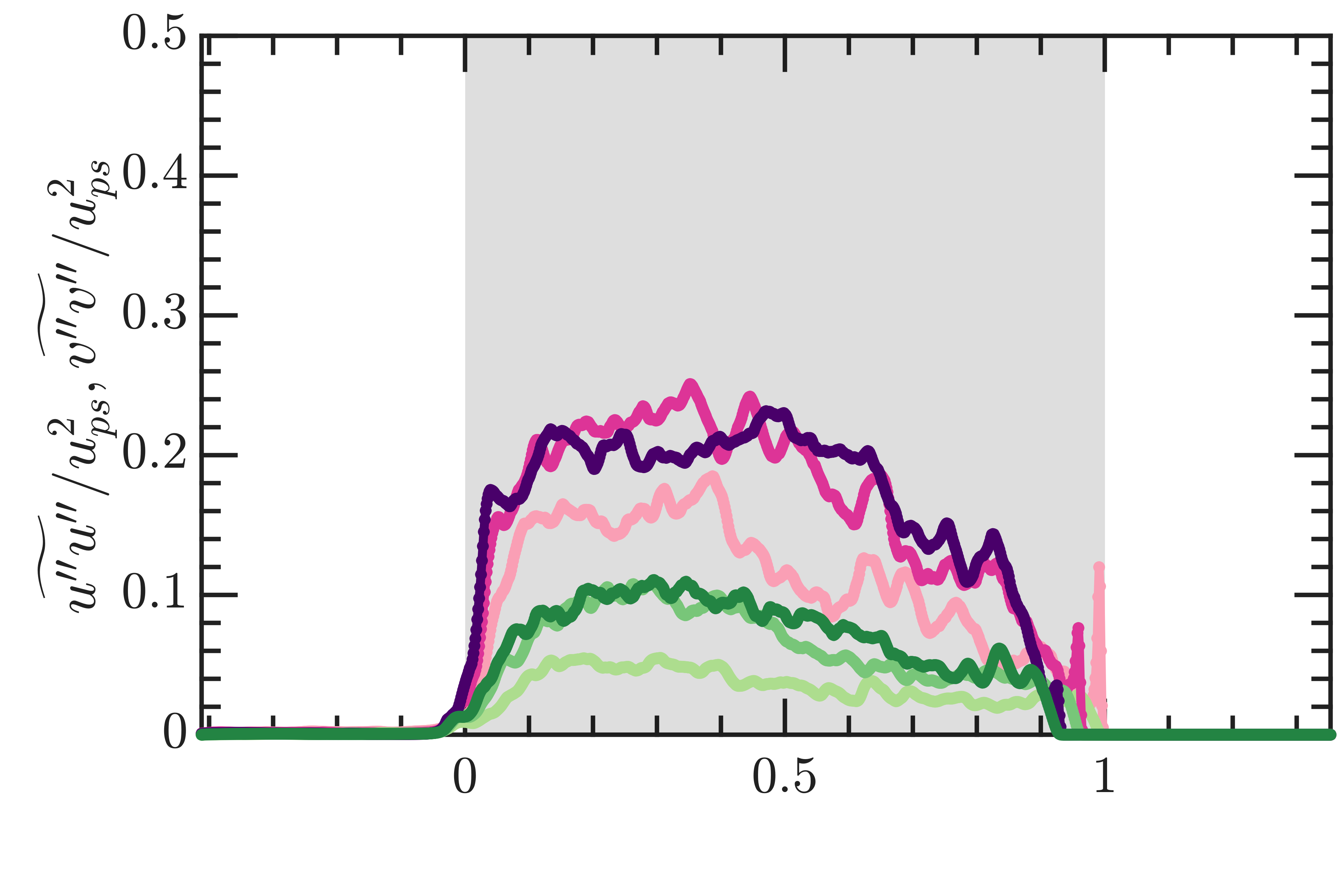}
      } 
      \subfloat{
          \includegraphics[width=0.48\textwidth]{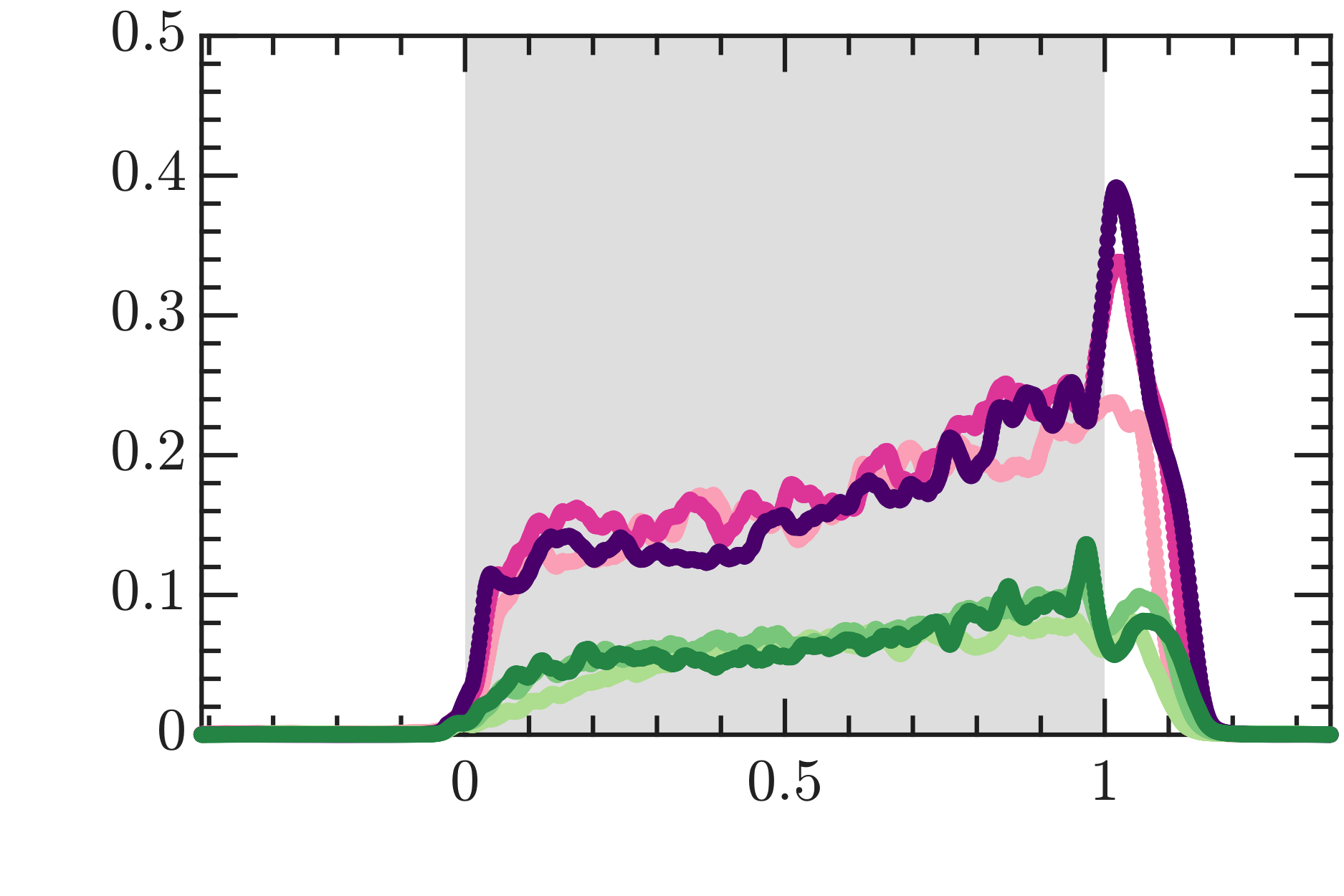}
      } \vspace{-10pt} \\
      \subfloat{
          \includegraphics[width=0.48\textwidth]{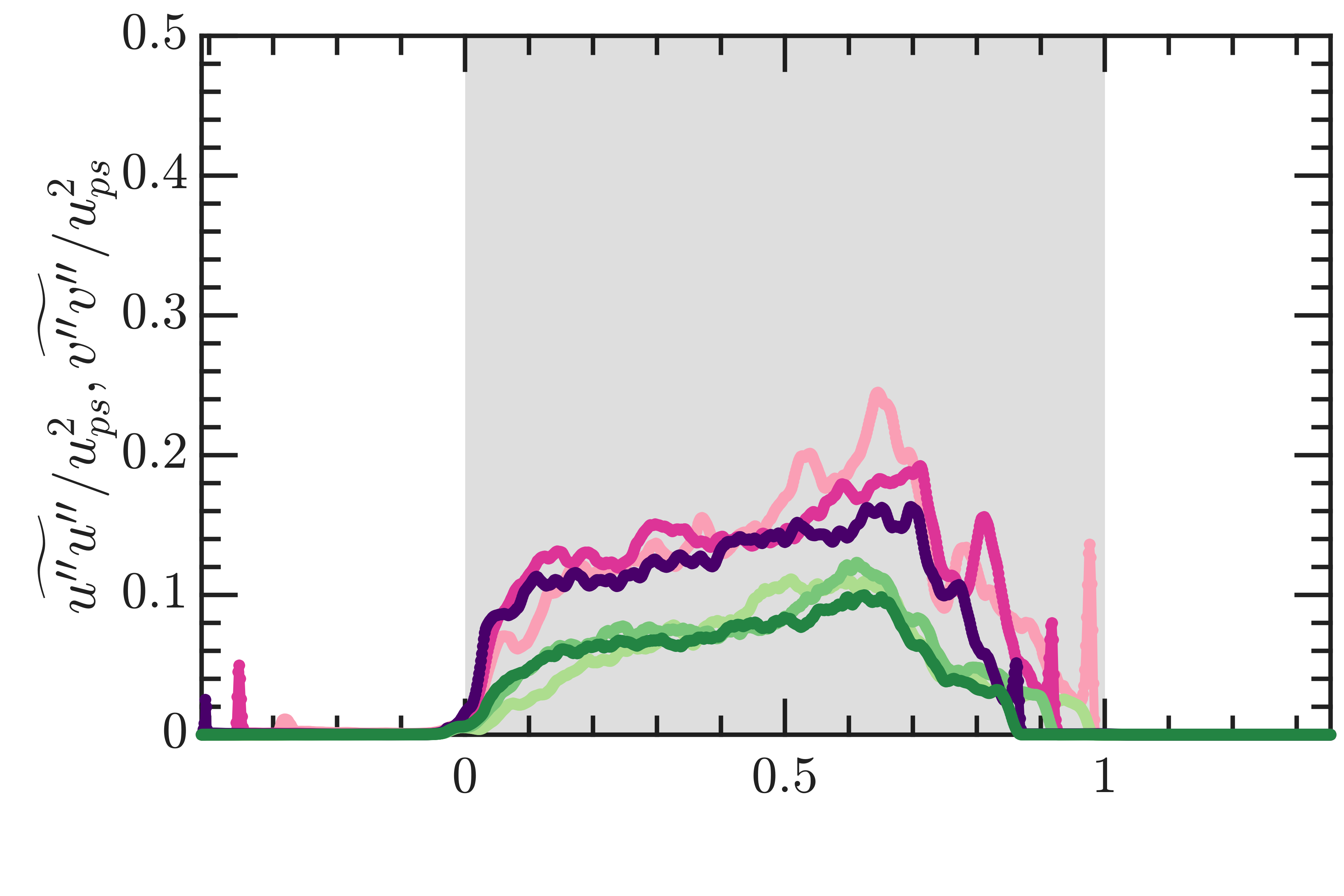}
      } 
      \subfloat{
          \includegraphics[width=0.48\textwidth]{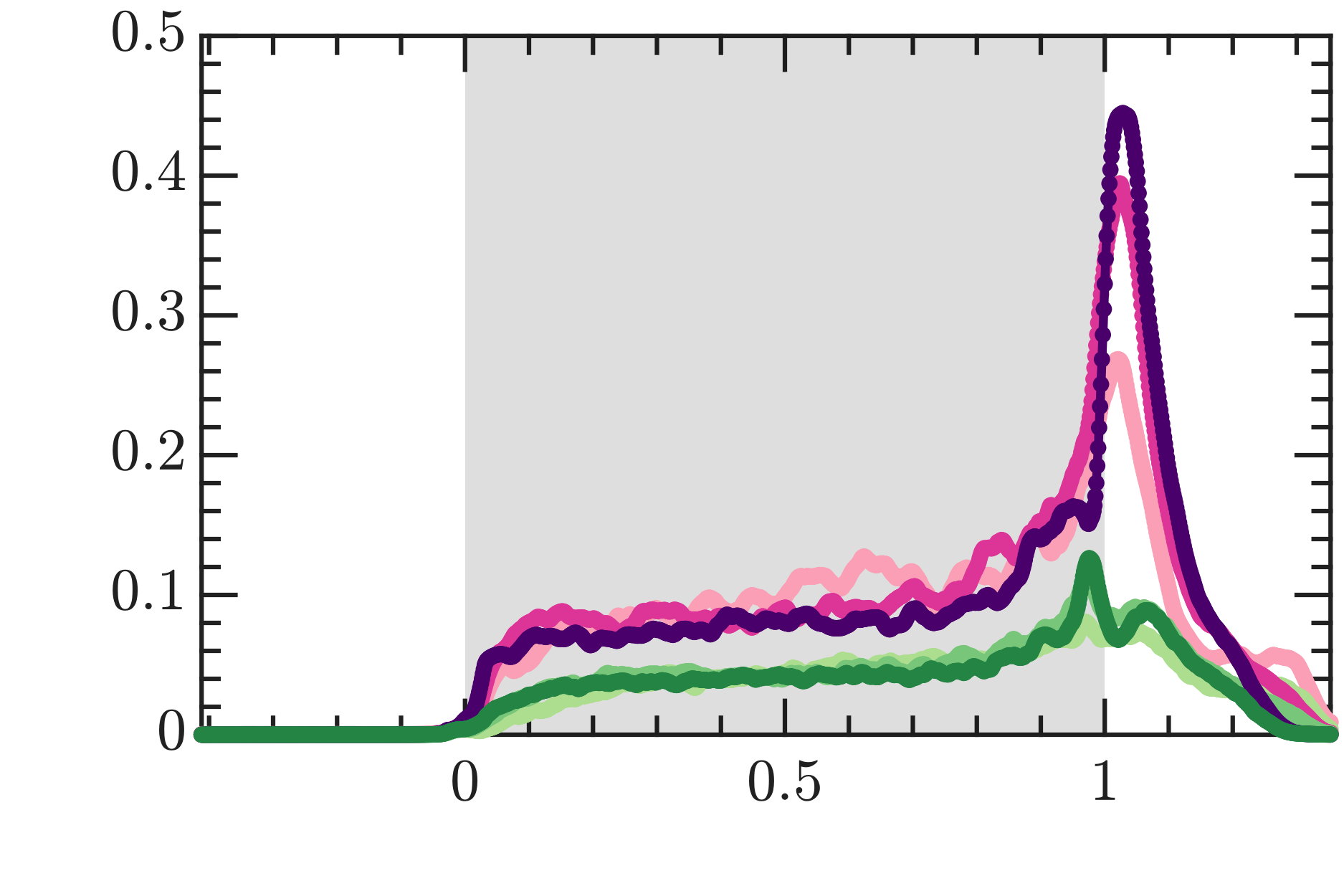}
      } \vspace{-10pt} \\
      \subfloat{
          \includegraphics[width=0.48\textwidth]{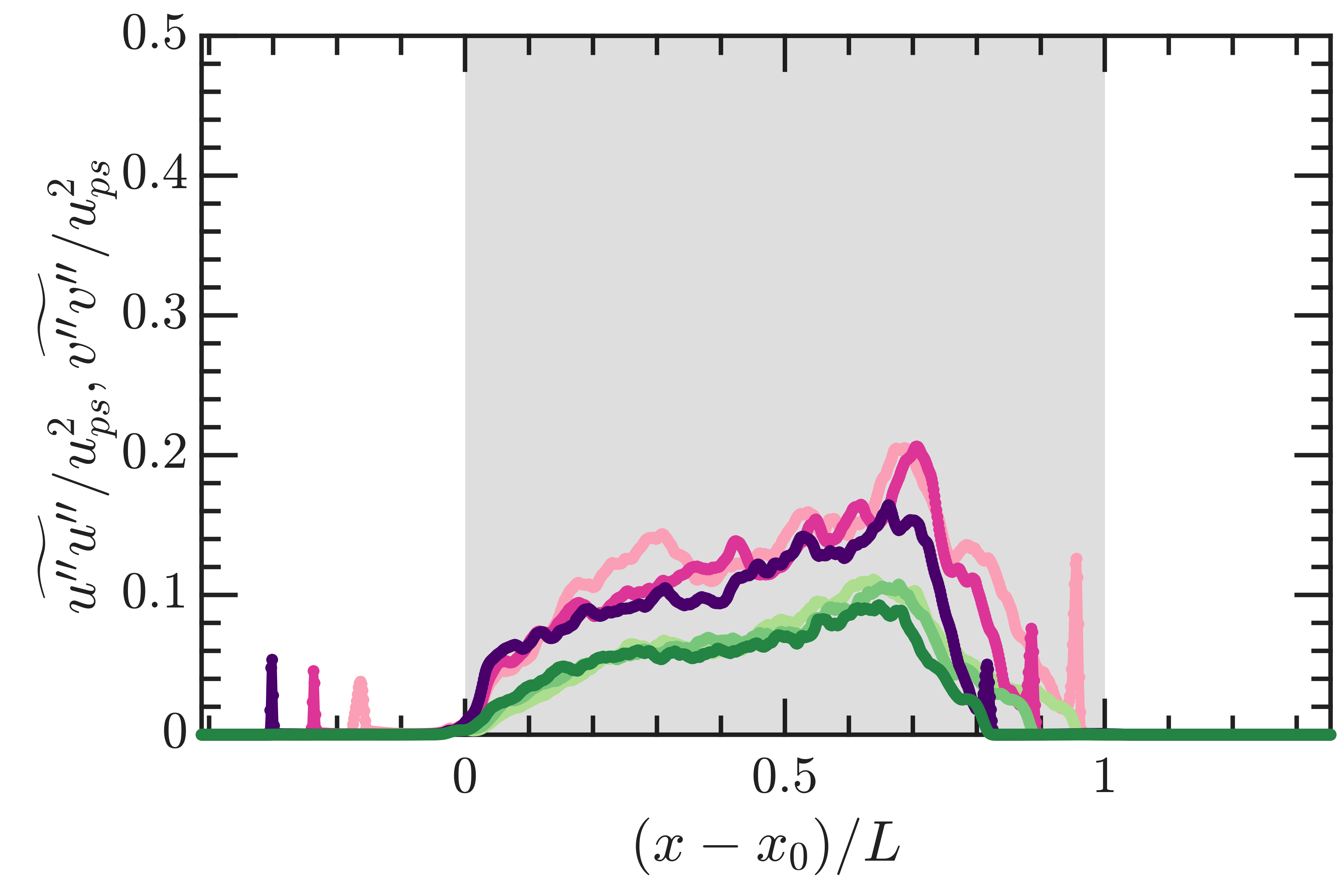}
      }
      \subfloat{
          \includegraphics[width=0.48\textwidth]{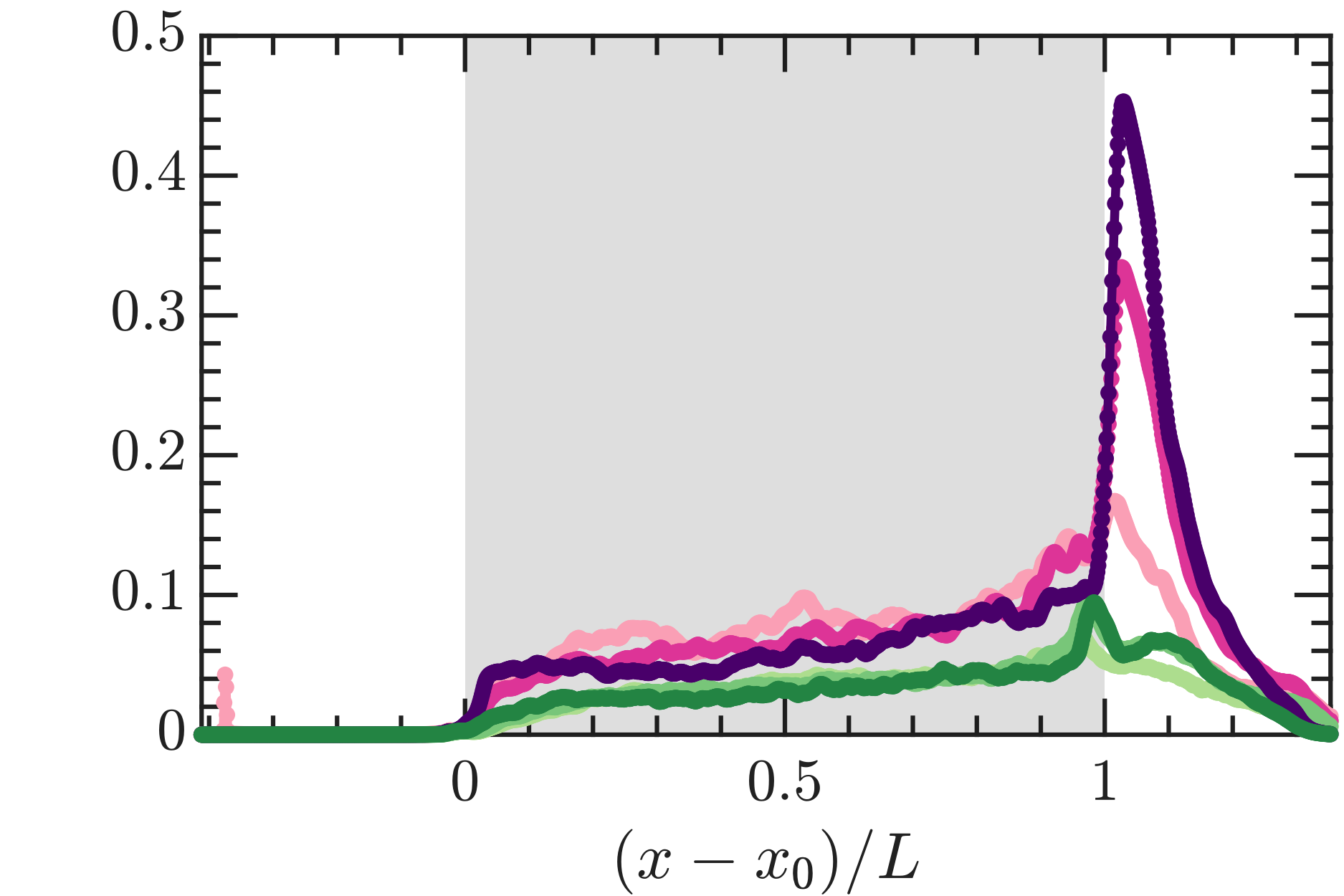}
      }
  \end{tabular}
  \begin{tikzpicture}[overlay, remember picture]
      \node at (-5.1,12.2) {$(a)$};
      \node at ([xshift=0.49\linewidth]-5.1,12.2) {$(b)$};
      \node at (-5.1,8.0) {$(c)$};
      \node at ([xshift=0.49\linewidth]-5.1,8.0) {$(d)$};
      \node at (-5.1,3.9) {$(e)$};
      \node at ([xshift=0.49\linewidth]-5.1,3.9) {$(f)$};
  \end{tikzpicture}
    \caption{Velocity fluctuations at $t/\tau=1$ (left) and $t/\tau_L=2$ (right). $(a,b)$ $M_s=1.2$, $(c,d)$ $M_s=1.66$ and $(e,f)$ $M_s=2.1$. Streamwise fluctuations $u_{\rm rms}^2$ (pink/purple), spanwise fluctuations $v_{\rm rms}^2$ (shades of green). $\Phi_p=0.1$ (light shade), $\Phi_p=0.1$ (intermediate shade), $\Phi_p=0.3$ (dark shade).}
    \label{fig:uuvv}
\end{figure}
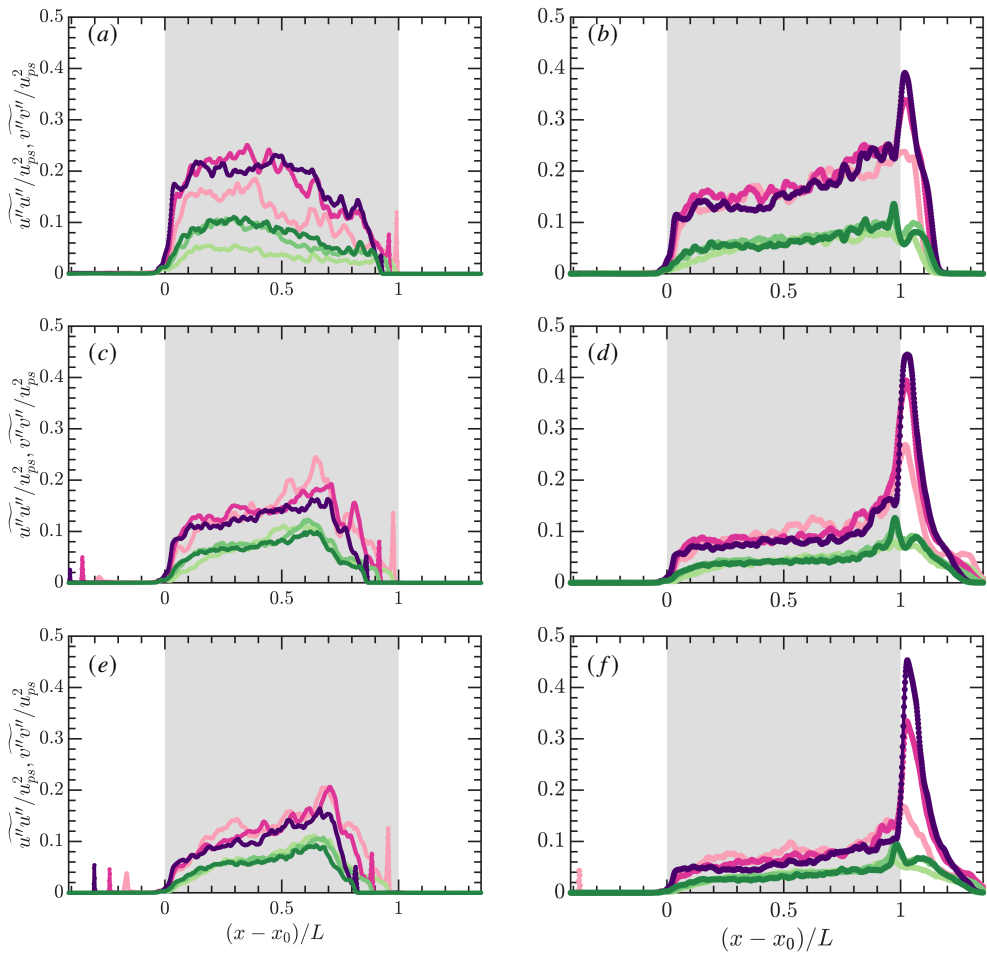

The root-mean-square (rms) gas-phase velocity fluctuations in the streamwise direction is defined as $u_{\rm rms}^2 = \widetilde{u''u''}$. Due to symmetry, the spanwise fluctuations are taken as $v_{\rm rms}^2 = (\widetilde{v''v''}+\widetilde{w''w''})/2$. Figure~\ref{fig:uuvv} shows these components at $t/\tau_L=1$ and $2$. All values are normalized by the post-shock kinetic energy, $u_{ps}^2$. Velocity fluctuations originate almost immediately within the particle curtain. The magnitude of the streamwise fluctuations are nearly twice the spanwise components. The fluctuations are higher at initial times, shortly after the shock passes over the particles. The maximum velocity fluctuations occur at the downstream edge where the flow chokes. Overall, the fluctuations decrease in magnitude with increasing $M_s$. This reduction can be attributed to an increase in compressibility effects with higher $M_s$. The precise dissipation mechanisms will be quantified in \S~\ref{sec:budget}, where individual terms of the PTKE budget are reported.

It is interesting to note that the normalized fluctuations are nearly invariant with volume fraction except for the lowest shock Mach number case at early times (see figure~\ref{fig:uuvv}$(a)$). Previous studies by \citet{mehta2020pseudo} observed an increase in velocity fluctuations with $\Phi_p$. However, we only observe significant variation due to $\Phi_p$ at the downstream edge of the curtain. In this region, the streamwise fluctuations increase by approximately a factor of four, yet the spanwise fluctuations remain unaffected. 

To better quantify the level of anisotropy, we define the gas-phase anisotropy tensor as
\begin{equation}
    b_{ij} = \frac{R_{ij}}{2k_g} - \frac{\delta_{ij}}{3},
\end{equation}
where $R_{ij}=\widetilde{u_i''u_j''}$ is the pseudo-turbulent Reynolds stress, $k_g=\widetilde{u_i''u_i''}/2$ (repeated indices imply summation) is the PTKE and $\delta_{ij}$ is the Dirac delta function. The streamwise component $b_{11}$ is dominant compared with the components perpendicular to the flow direction $b_{22}$ and $b_{33}$. The cross-correlation of velocity fluctuations, $b_{12}$, is often negligible in gas--solid flows~\citep{mehrabadi2015pseudo}. Due to symmetry in the flow, only $b_{11}$ and $b_{22}$ are reported.

\begin{figure}
    \centerline{\includegraphics[width=0.6\linewidth]{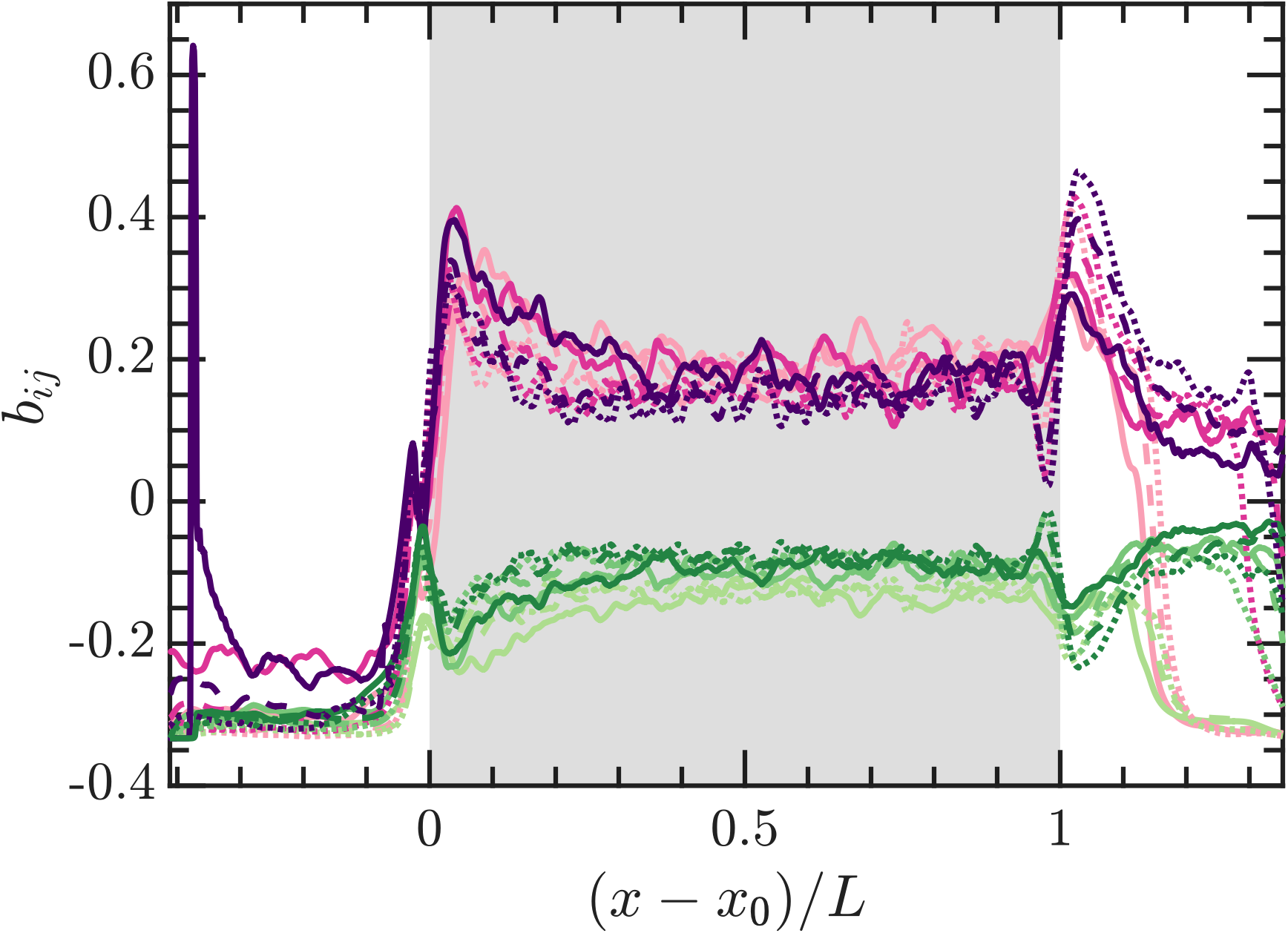}}
    \caption{Components of the Reynolds stress anisotropy tensor for cases 1--9 at $t/\tau_L=2$. $\Phi_p=0.1$ (\full), $\Phi_p=0.2$ (\dashed), and $\Phi_p=0.3$ (\dotted). Parallel component $b_{11}$ (light pink, pink, purple) and perpendicular component $b_{22}$ (light green, green, dark green) for $M_s=1.2$, $1.66$ and $2.1$ (light to dark).}
    \label{fig:anisotropy}
\end{figure}

All nine cases are overlaid in figure~\ref{fig:anisotropy} at $t/\tau_L=2$ with each line style representing a distinct volume fraction and each shade of colour representing a distinct shock Mach number. A strong degree of anisotropy is observed. Interestingly, the level of anisotropy remains approximately constant across the curtain, with $b_{11}\approx0.2$ and $b_{22}\approx-0.1$ for all cases, regardless of $\Phi_p$ and $M_s$. Variations are noted only near the curtain edges, where the streamwise component becomes even more dominant. Upstream of the curtain, the anisotropy tensor becomes ill-defined as $k_g$ is zero due to lack of a turbulence production mechanism. The level of anisotropy suggests that approximately 50\% of PTKE is partitioned in $\widetilde{u''u''}$ and $25\%$ in $\widetilde{v''v''}$ and $\widetilde{w''w''}$.

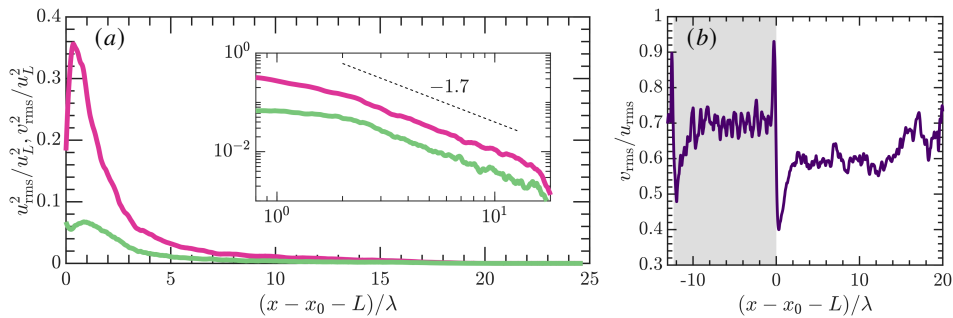
\begin{figure}
    \centering
    \begin{tabular}{ccc}
    \centering
    \subfloat{\includegraphics[height=0.3\textwidth]{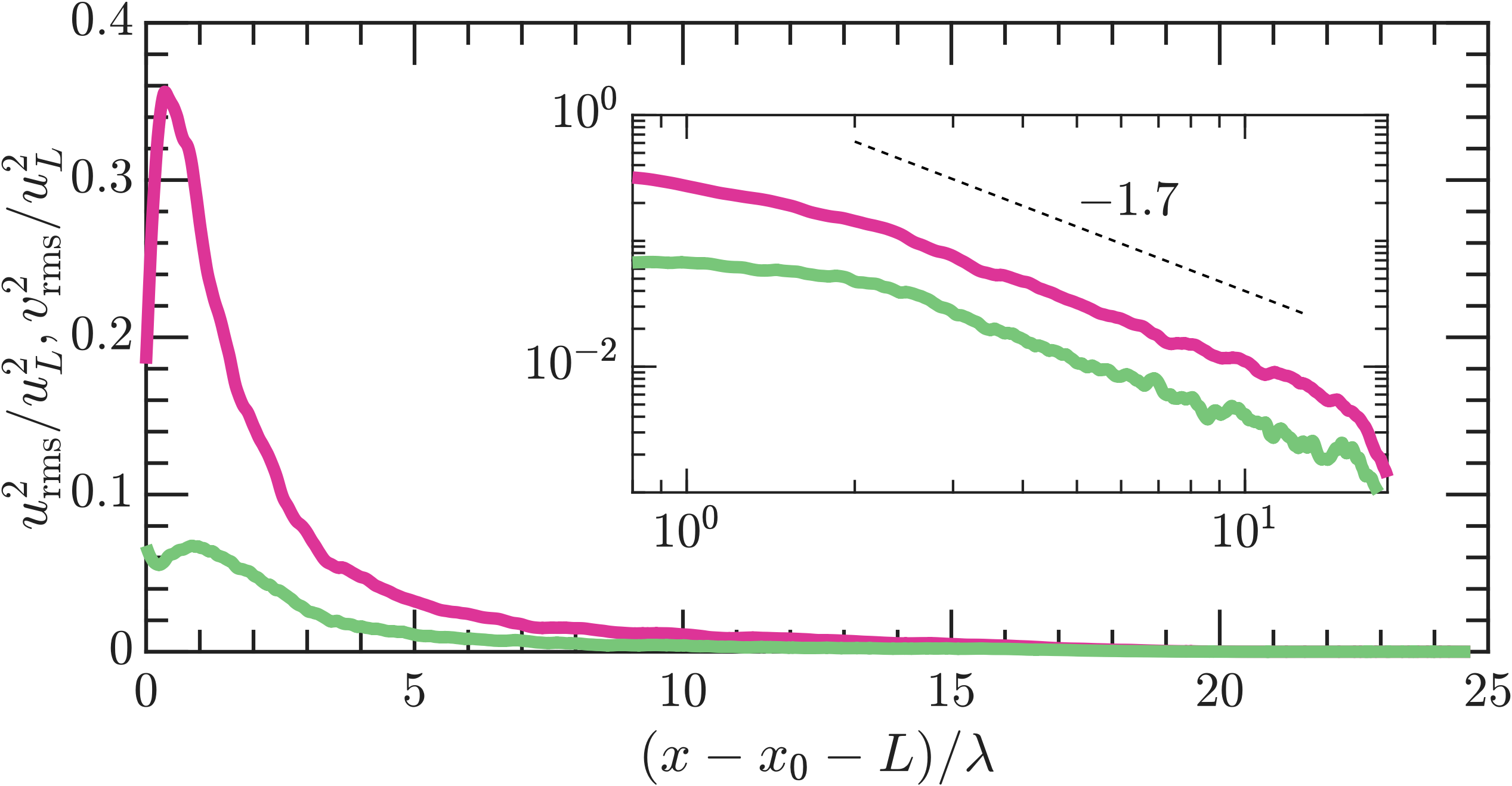}} &     
    \subfloat{\includegraphics[height=0.3\textwidth]{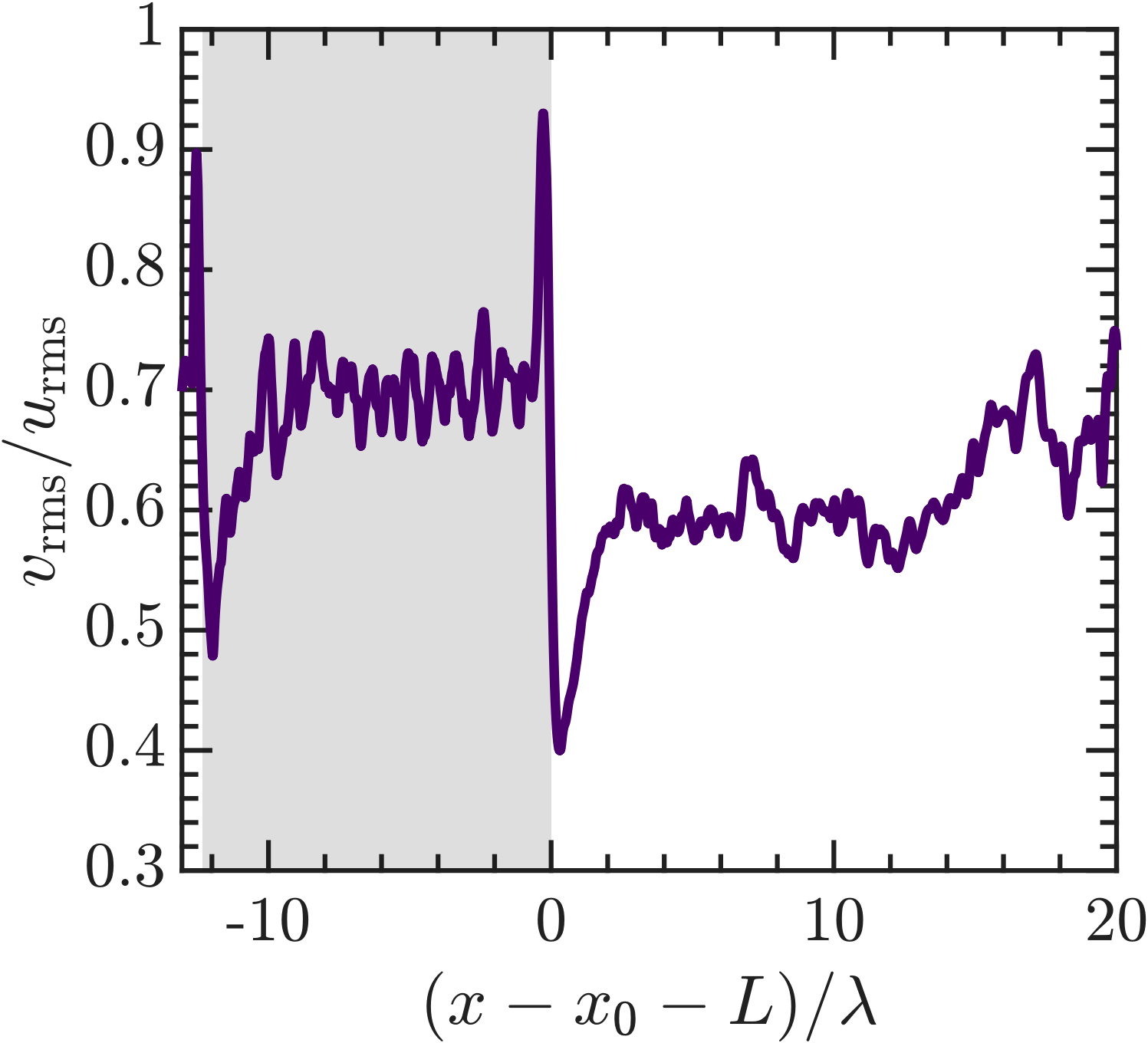}}
    \end{tabular}
    \caption{$(a)$ Components of rms velocities ${u_{\rm rms}^2}$ (pink) and $v_{\rm rms}^2$ (green) for Case 10 downstream of the particle curtain at $t/\tau_L=5$. The inset shows the components in log-scale. $(b)$ Ratio of the spanwise to streamwise rms velocities as a function of streamwise distance normalized by interparticle spacing $\lambda$.}
    \label{fig:turbdecay}
    \begin{tikzpicture}[overlay, remember picture]
        \node at (-4.9,5.8) {$(a)$};
        \node at ([xshift=0.58\linewidth]-4.9,5.8) {$(b)$};
    \end{tikzpicture}
\end{figure}

PTKE advects and decays downstream of the particle curtain. Case $10$ extends $3L$ downstream of the particle curtain to examine this behaviour in greater detail. Figure~\ref{fig:turbdecay} shows the rms velocity components at $t/\tau_L=5$, where the flow reaches a steady state. It can be seen that the flow remains anisotropic beyond the curtain and eventually the fluctuations completely decay. This is analogous to grid-generated turbulence~\citep{batchelor1948decay,mohamed1990decay,kurian2009grid}. According to \cite{batchelor1948decay}, the decay of turbulence intensity downstream of a grid (or screen) with mesh width $M$ follows a power law, given by
\begin{equation}
    \Bigg (\frac{u_{\rm rms}}{u_0}\Bigg)^2 = A \Bigg ( \frac{x-x_L}{M} \Bigg )^{n}
\end{equation}
where $u_0$ is the velocity of the gas phase at a point of virtual origin of turbulence $x_0$ and $A$ is an empirical constant. An analogy can be drawn to our shock-particle configuration by setting the mesh width to the average interparticle spacing, $\lambda$, which can be defined as
\begin{equation}
    \lambda = D \left(\frac{\pi}{6 \Phi_p} \right)^{1/3}.
\end{equation}
Additionally, we set the point of origin of turbulence decay to the location of the downstream curtain edge, $x_L=x_0+L$, and consider the velocity at this point $u_L=\tilde{u}(x{=}x_L;t{=}5\tau_L)$ when normalizing the turbulence intensity.

Figure~\ref{fig:turbdecay}$(a)$ shows the decay of streamwise and spanwise components of rms velocities as a function of the downstream distance normalized by $\lambda$. The inset illustrates this decay in log scale, from which we conclude that the decay does indeed follow a power-law behaviour with an exponent $n=-1.7$. This value is slightly higher than the reported values for $n$ in incompressible, single-phase grid-generated turbulence reported in the literature, which range from $-1.13$ to $-1.6$ \citep{mohamed1990decay,kurian2009grid}. The ratio $v_{\rm rms}/u_{\rm rms}$ shown in figure~\ref{fig:turbdecay}$(b)$  highlights significant anisotropy of approximately $0.7$, while downstream it reduces to $\approx0.5$ suggesting that the flow remains anisotropic even at later time periods. 

\begin{figure}
  \centerline{\includegraphics[width=0.5\linewidth,keepaspectratio]{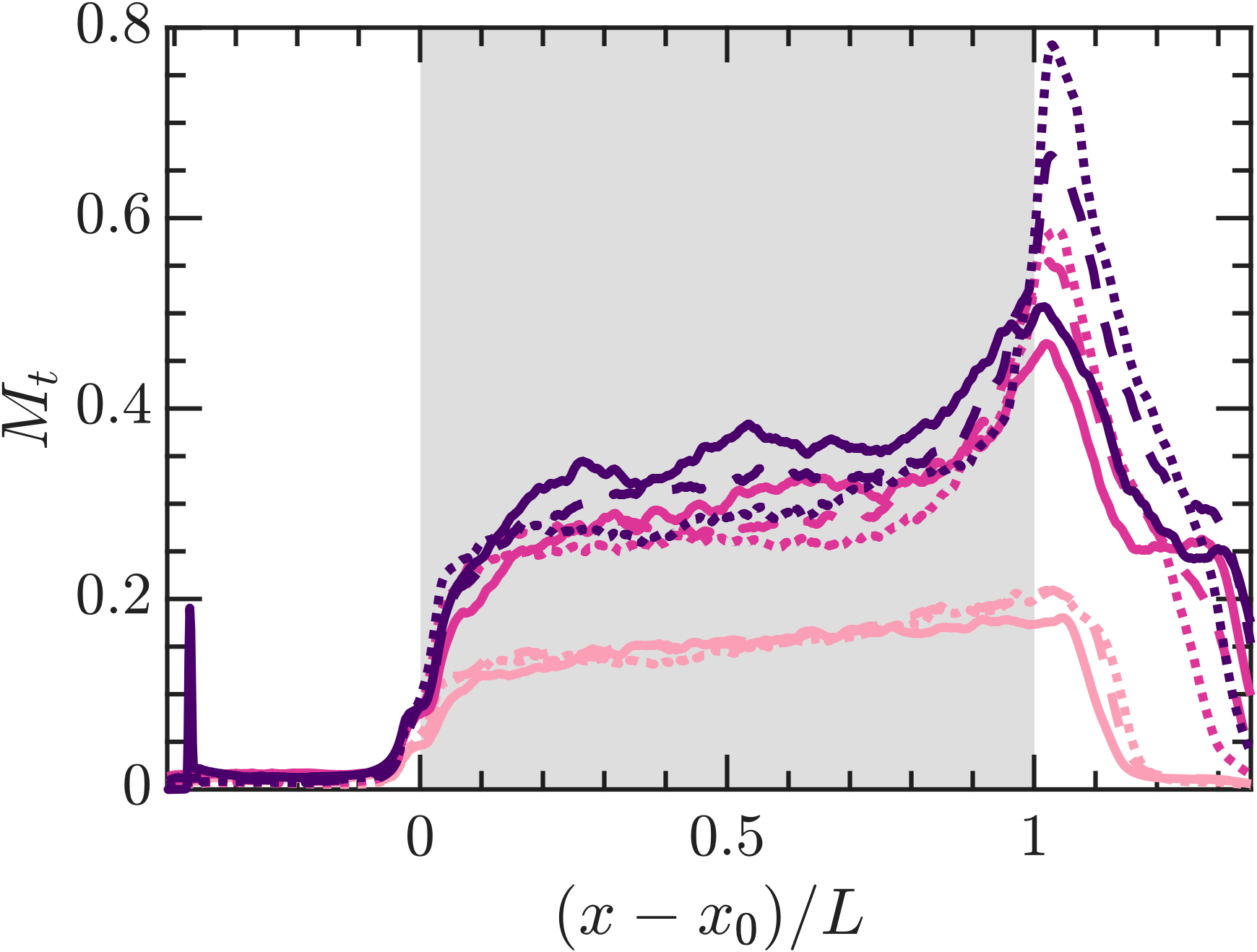}}
  \caption{Turbulent Mach number for cases $1-9$ at $t/\tau_L=2$. Same legend as $b_{11}$ in figure~\ref{fig:anisotropy}.}
  \label{fig:Mt}
\end{figure}
    
The turbulent Mach number, defined as $M_t = \sqrt{2k_g}/\overline{c}$, is shown at $t/\tau_L=2$ for cases $1-9$ in figure~\ref{fig:Mt}. $M_t$ tends to increase rapidly at the upstream edge of the curtain where turbulence is first generated, then gradually increases throughout the curtain and peaks at the downstream edge where the flow chokes. The turbulent Mach number increases monotonically with the incident shock speed. Within the curtain, $M_t$ is relatively independent of $\Phi_p$, but increases with increasing $\Phi_p$ at the downstream edge. For the cases with the lowest shock Mach number ($M_s=1.2$), $M_t\approx 0.2$, which falls in the quasi-isentropic regime, as classified by \citet{sagaut2008homogeneous}, where pressure fluctuations are not significant. These cases are distinct from the higher $M_s$ cases in that the velocity does not rapidly increase at the downstream edge of the curtain (see figure~\ref{fig:meanvel}$(c)$) and the mean sound speed remains relatively constant (not shown) and thus the trends in $M_t$  are qualitatively different from the two higher $M_s$ cases. For the two higher shock Mach number cases, $M_t$ varies between $0.3$ and $0.8$, placing them in the nonlinear subsonic regime where dilatational fluctuations are expected to be important.

\subsection{Budget of pseudo-turbulent kinetic energy}\label{sec:budget}

The  presence of particles in the flow generates local gas-phase velocity fluctuations characterized by the pseudo-turbulent kinetic energy (PTKE), defined as $k_g = (\widetilde{{u}_i'' {u}_i''})/2$. Reynolds-averaged transport equations for compressible flows have previously been derived by \citet{sarkar1991analysis}, among others. In this study, the transport equation for PTKE is derived in a similar manner, but the presence of particles is accounted for by including the indicator function in the averaging process as defined in \S~\ref{sec:avg}. Multiplying through the Navier--Stokes equations in \S~\ref{sec:eqs} by the indicator function and averaging over the homogeneous $y$- and $z$-directions yields a one-dimensional, time-dependent transport equation for PTKE (a similar derivation is given by \citet{vartdal2018using}), given by
\begin{equation}\label{eq:budget}
    \begin{aligned}
        \frac{\partial}{\partial t}(\alpha_g \, \overline{\rho} \, k_g)  + \frac{\partial}{\partial x}(\alpha_g \, \overline{\rho} \, \widetilde{u} \, k_g) &= \mathcal{P}_S + \mathcal{T} + \Pi + \alpha_g \rho \epsilon + \mathcal{M} + \mathcal{P}_D.
    \end{aligned}
\end{equation}
The terms on the right-hand side represent various mechanisms for producing, dissipating, and transporting PTKE. $\mathcal{P}_{s}$ is production due to mean shear, $\mathcal{T}$ is a term akin to diffusive transport, ${\Pi}$ is the pressure-dilatation correlation term and $\epsilon$ is the viscous dissipation tensor. The trailing terms arising from the averaging procedure are lumped into $\mathcal{M}$. $\mathcal{P}_D=\mathcal{P}_D^P+\mathcal{P}_D^V$ is production due to drag that contains contributions from pressure and viscous stresses, respectively. These terms are each defined as
\begin{align}
     \mathcal{P}_S &= - \alpha_g \overline{\rho}  \widetilde{u'' u''} \frac{\partial \widetilde{u}}{\partial x}, \\
     \mathcal{T}&= -\frac{1}{2} \frac{\partial}{\partial x}(\alpha_g \overline{\rho {u_i'' u_i'' u''}} ) - \frac{\partial}{\partial x}(\alpha_g \overline{p'u''}) + \frac{\partial}{\partial x}(\alpha_g \overline{u_i'' \sigma_{i1}'}), \\  
     {\Pi} &= \alpha_g \ \overline{ p' \frac{\partial u_i''}{\partial x_i} }, \\
     \alpha_g \rho \epsilon & = - \alpha_g \ \overline{ \sigma_{ik}' \frac{\partial u_i''}{\partial x_k} }, \\ 
     \mathcal{M} &= -\frac{\partial} {\partial x_i}(\alpha_g \overline{p} \overline{u_i''}) + \alpha_g \overline{p} \frac{\partial \overline{u_i''}}{\partial x_i}  + \frac{\partial } {\partial x_i} (\alpha_g \overline{\sigma_{11}}\overline{u_i''}) - \alpha_g \overline{\sigma_{11}} \frac{\partial \overline{u_i''}}{\partial x_i}, \\
     \mathcal{P}_D &=   \mathcal{P}_D^P + \mathcal{P}_D^V = \overline{p' u_i'' \frac{\partial \mathcal{I}}{\partial x_i}} -  \overline{\sigma_{ik}' u_i'' \frac{\partial \mathcal{I}}{\partial x_k}}.
\end{align}

\begin{figure}
  \centering
  \begin{tabular}{cc}
      \subfloat{
          \includegraphics[width=0.48\textwidth]{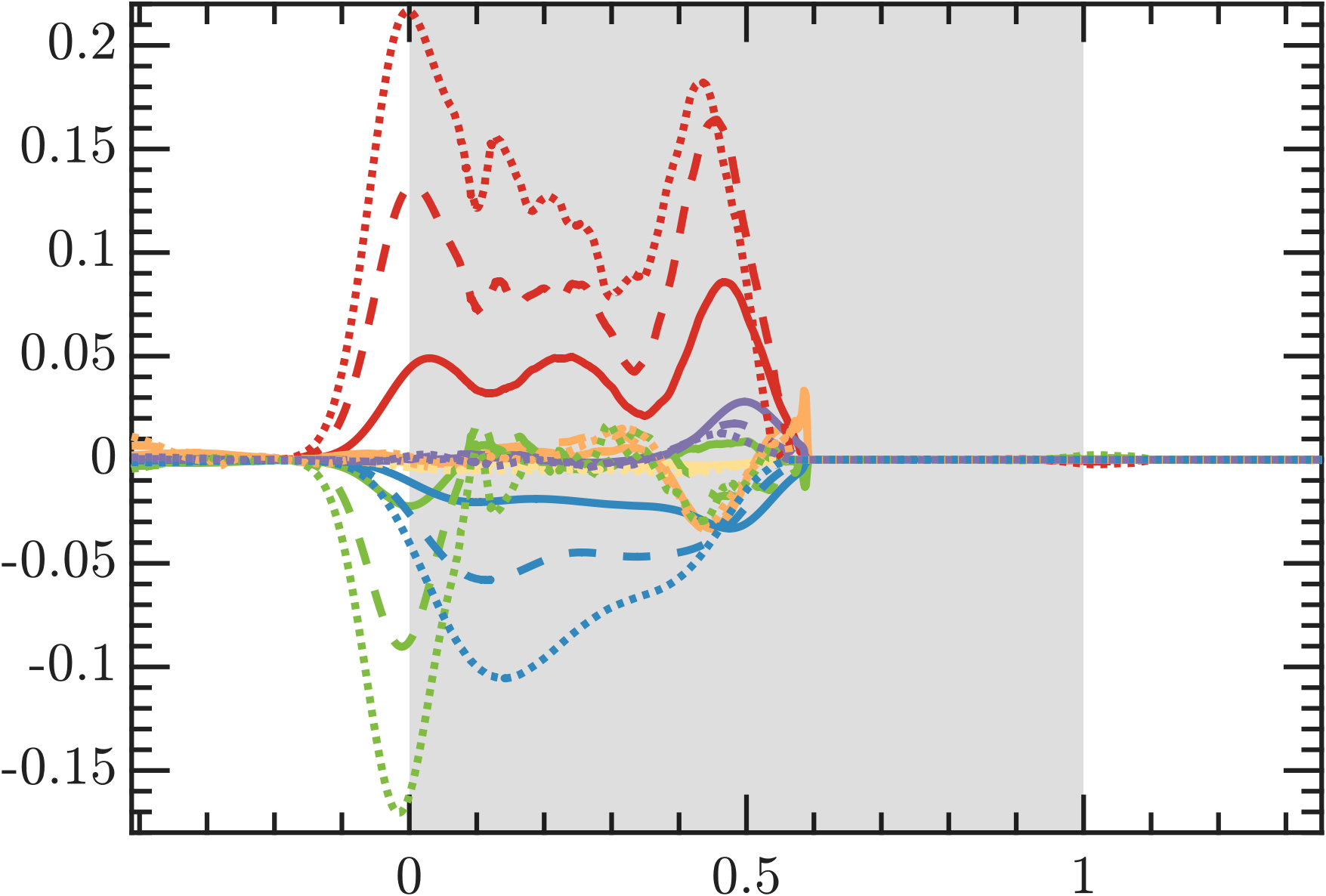}
      } &
      \subfloat{
          \includegraphics[width=0.48\textwidth]{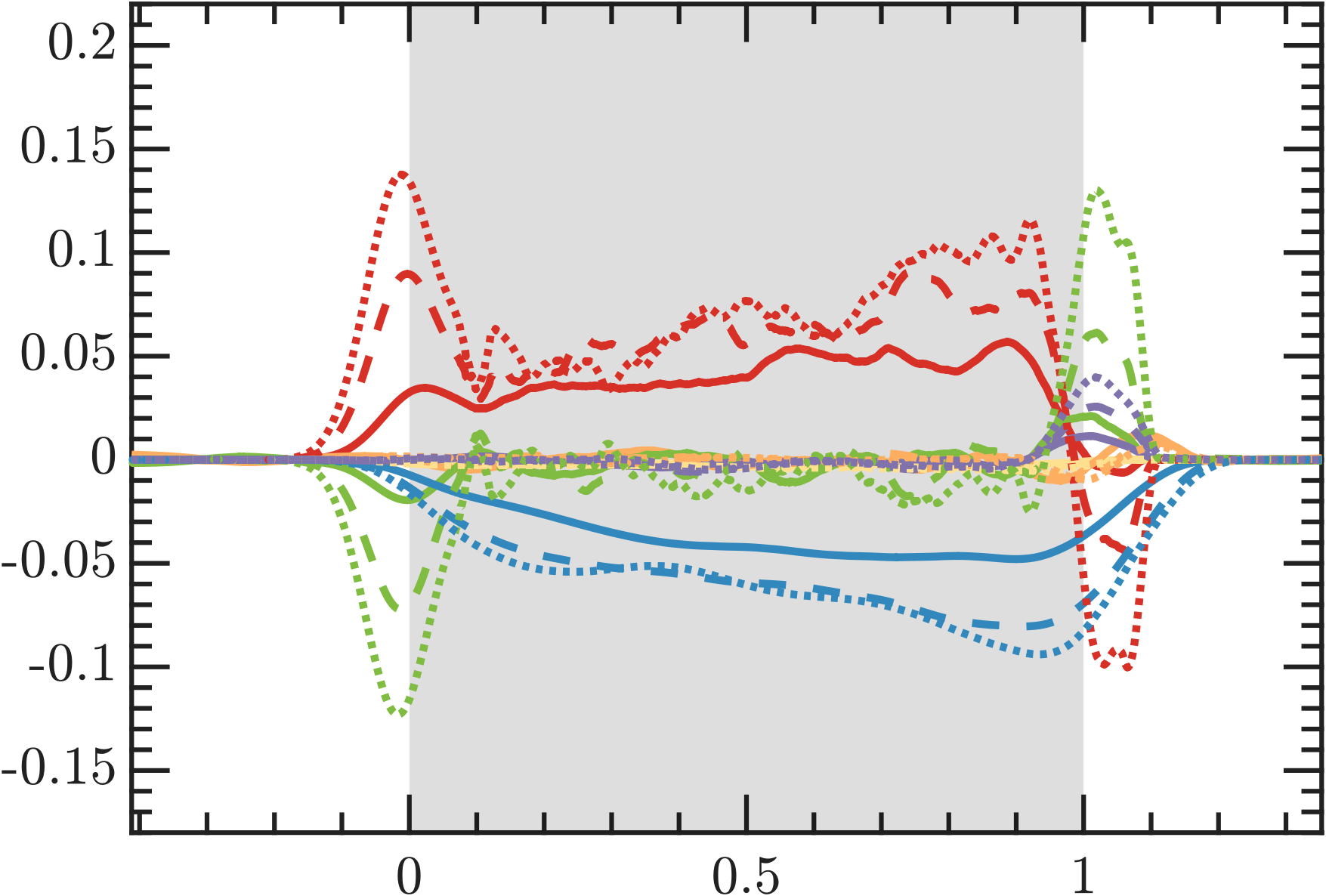}
      }  \\
      \subfloat{
          \includegraphics[width=0.48\textwidth]{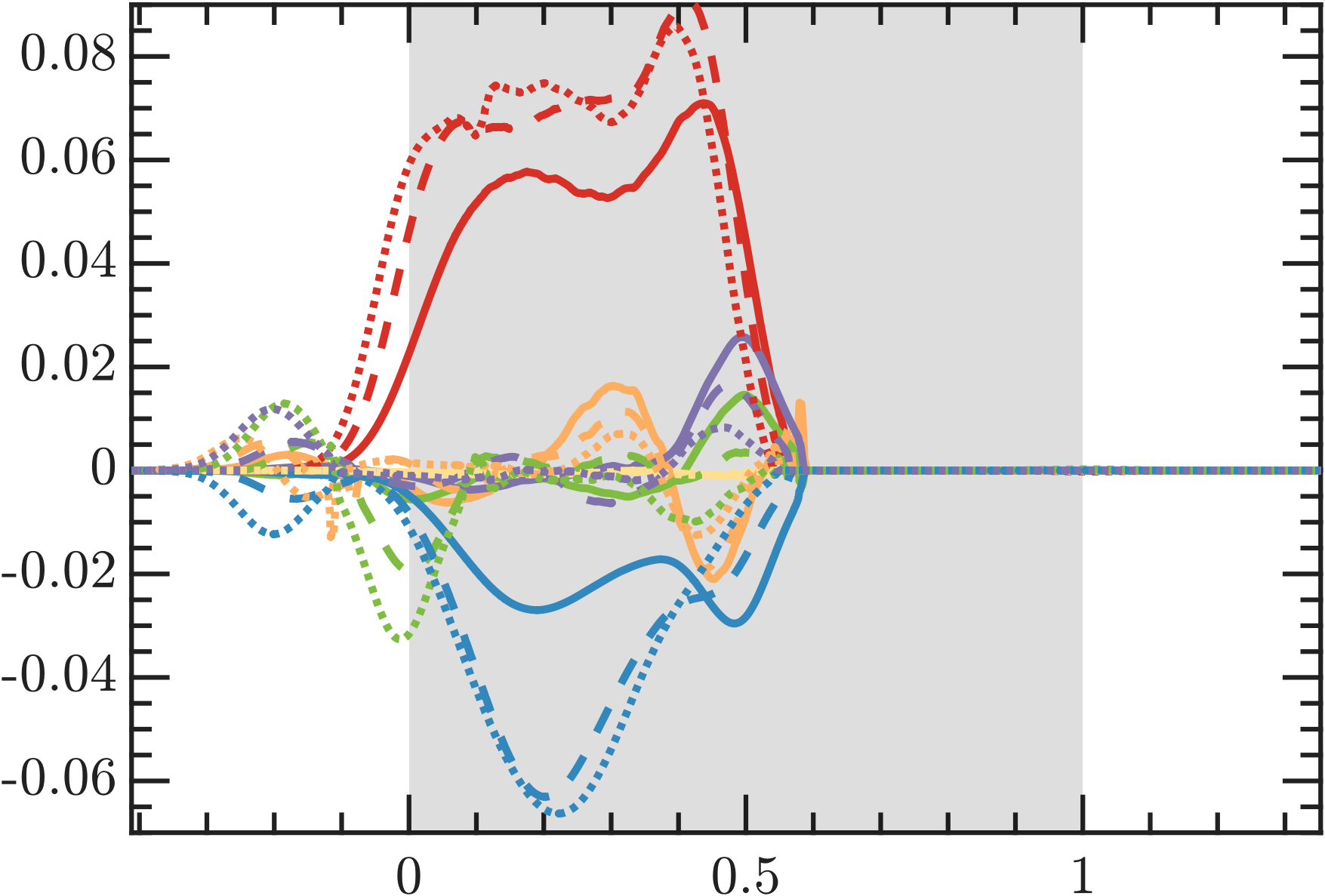}
      } &
      \subfloat{
          \includegraphics[width=0.48\textwidth]{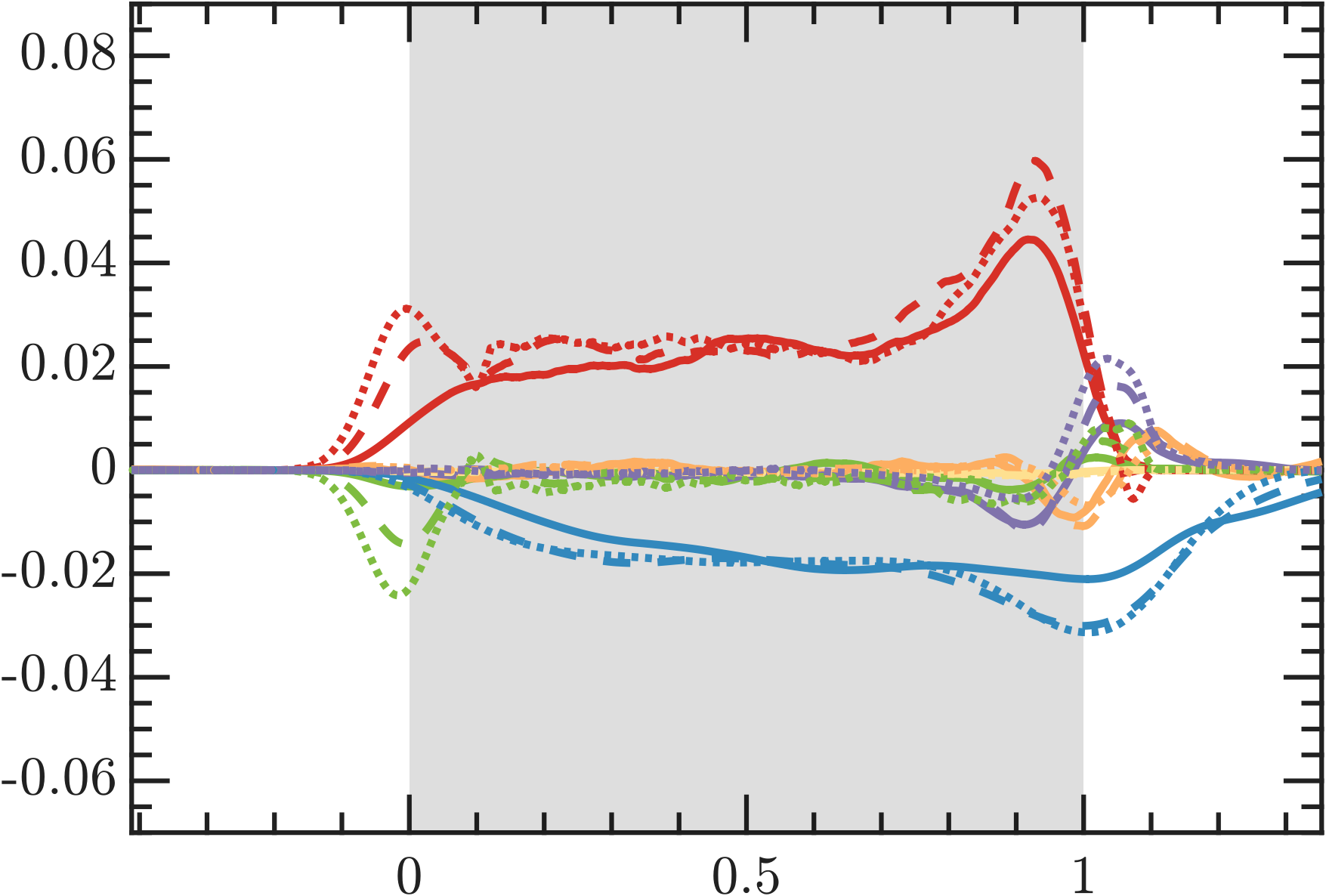}
      }  \\
      \subfloat{
          \includegraphics[width=0.48\textwidth]{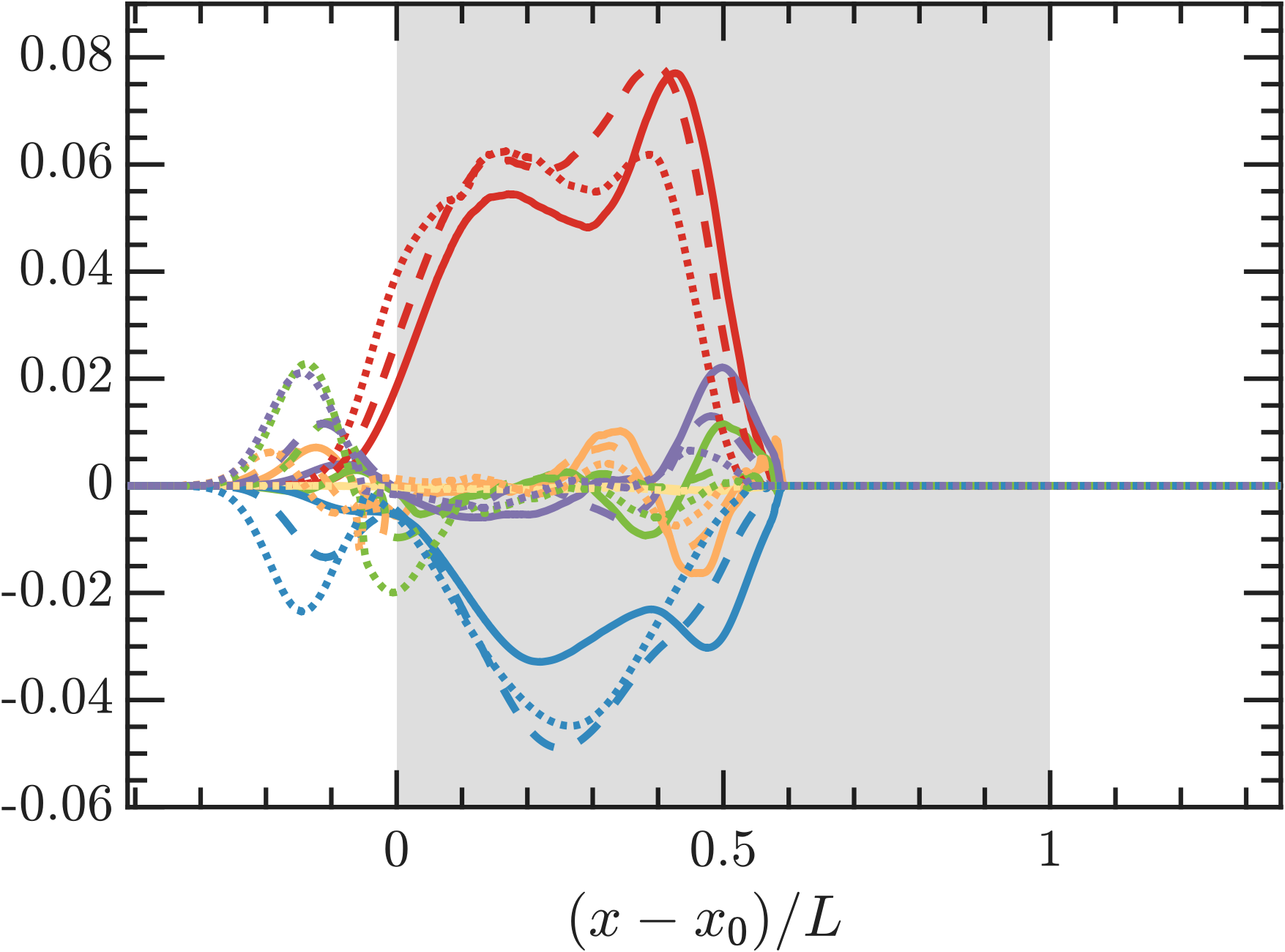}
      } &
      \subfloat{
          \includegraphics[width=0.48\textwidth]{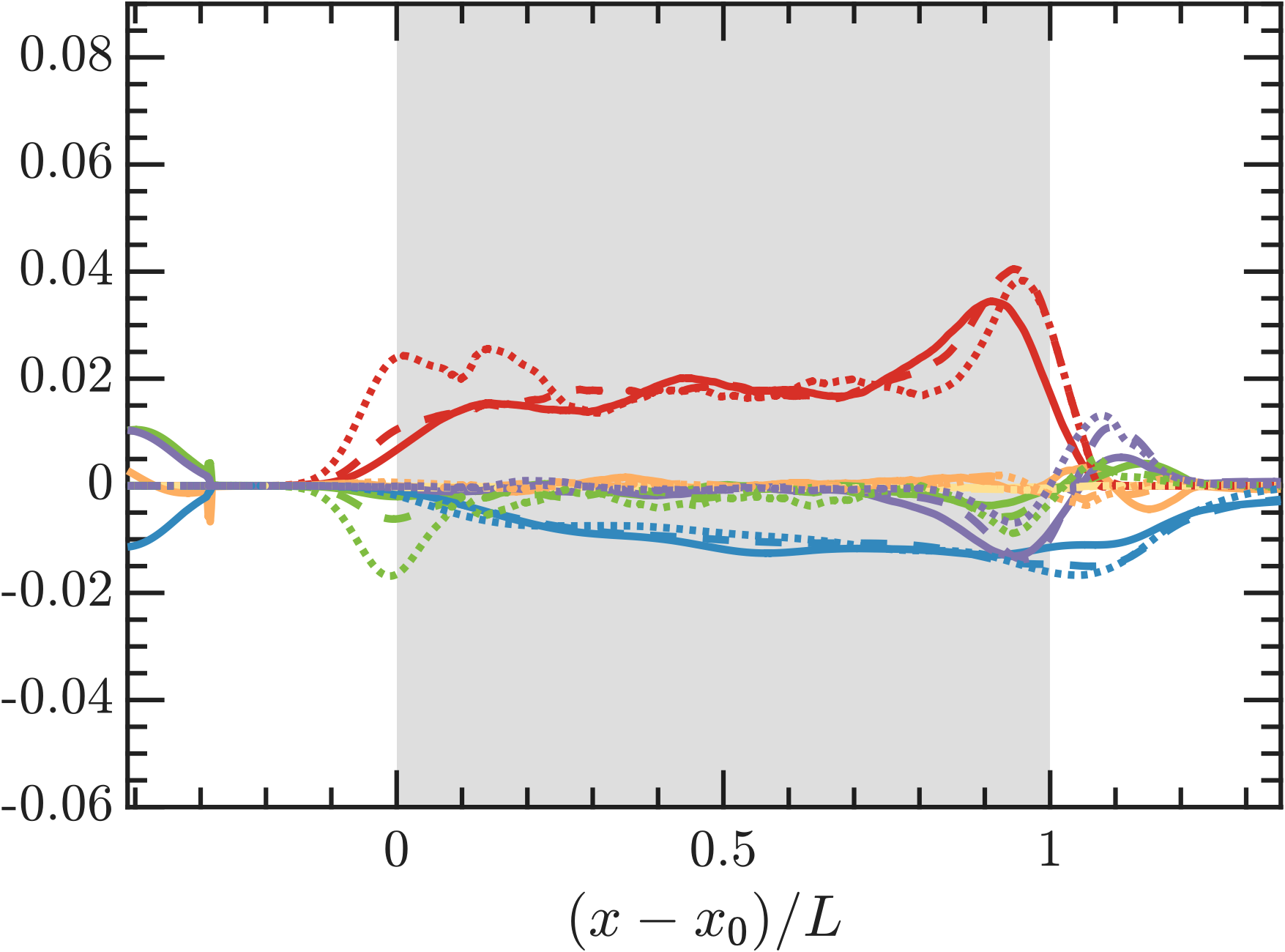}
      }
  \end{tabular}
  \begin{tikzpicture}[overlay, remember picture]
      \node at (-5.3,13.6) {$(a)$};
      \node at ([xshift=0.5\linewidth]-5.3,13.6) {$(b)$};
      \node at (-5.3,9.1) {$(c)$};
      \node at ([xshift=0.5\linewidth]-5.3,9.1) {$(d)$};
      \node at (-5.3,4.5) {$(e)$};
      \node at ([xshift=0.5\linewidth]-5.3,4.5) {$(h)$};
      
      \draw [line width=2, pd ] (-4.5,-0.5) -- (-3.5,-0.5) node [right, text=black] 
      {$\mathcal{P}_D^P$};
      \draw [line width=2, pdv ] ([xshift=0.2\linewidth]-4.5,-0.5) -- ([xshift=0.2\linewidth]-3.5,-0.5) node [right, text=black] 
      {$\mathcal{P}_D^V$};
      \draw [line width=2, transport ] ([xshift=0.4\linewidth]-4.5,-0.5) -- ([xshift=0.4\linewidth]-3.5,-0.5) node [right, text=black] 
      {$\mathcal{T}$};
      \draw [line width=2, press-dil] ([xshift=0.6\linewidth]-4.5,-0.5) -- ([xshift=0.6\linewidth]-3.5,-0.5) node [right, text=black] 
      {$\Pi$};
      \draw [line width=2, viscdiss] ([xshift=0.2\linewidth]-4.5,-1.0) -- ([xshift=0.2\linewidth]-3.5,-1.0) node [right, text=black] {$\alpha_g \rho_g \epsilon_g$};
      \draw [line width=2, ps] ([xshift=0.4\linewidth]-4.5,-1.0) -- ([xshift=0.4\linewidth]-3.5,-1.0) node [right, text=black] {$\mathcal{P}_s$};
  \end{tikzpicture}
  \vspace{40pt}
    \caption{Budgets of PTKE at $t/\tau_L=0.5$ (left) and $t/\tau_L=2$ (right). $(a,b)$ $M_s=1.2$, $(c,d)$ $M_s=1.66$ and $(e,f)$ $M_s=2.1$. $\Phi_p=0.1$ (\full), $\Phi_p=0.2$ (\dashed) and $\Phi_p=0.3$ (\dotted).}.
    \label{fig:budget}
\end{figure}
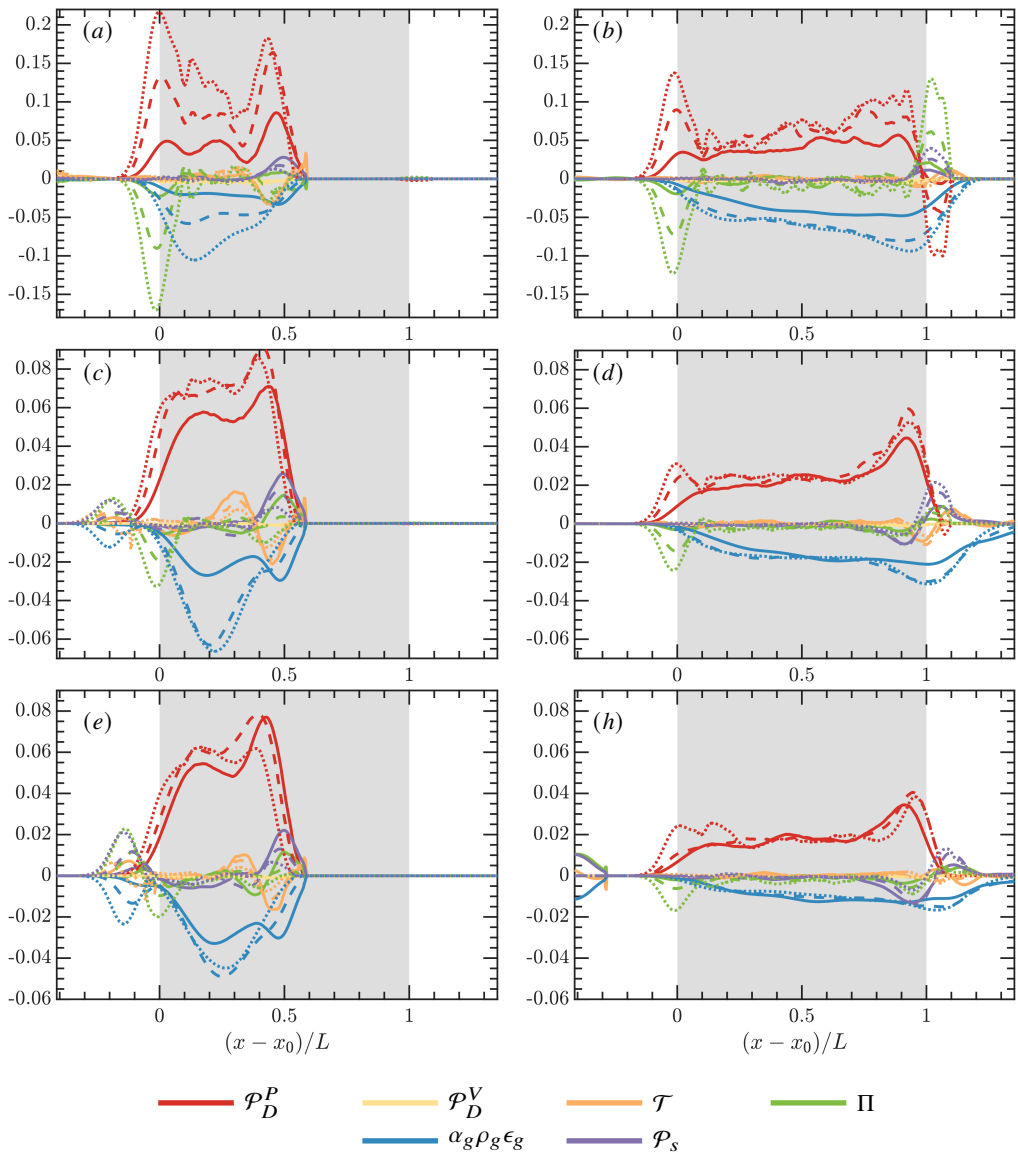

Figure~\ref{fig:budget} shows the budget of PTKE at $t/\tau_L=0.5$ and $2$ for different $M_s$ and $\Phi_p$. The terms are normalized by post-shock quantities: $\rho_{ps}u_{ps}^3/D$. The statistics from the particle-resolved simulations are noisy due to the indicator function and to provide reliable data, a low-pass (Gaussian) filter is applied in the streamwise direction with filter size of standard deviation $3D$ after averaging in the periodic directions. It should be noted that most coarse-grain simulations of particle-laden flows use grid spacing larger than $D$. Also, the resulting profiles were found to be insensitive to a wide range of filter sizes. Note in figures~\ref{fig:budget}$(c),(e),(h)$, small oscillations upstream of the curtain indicate the location of the reflected shock.

The majority of PTKE is generated via drag production, which is balanced by viscous dissipation. The remaining terms are negligible except for shear production, $\mathcal{P}_s$, and the pressure-dilatation correlation term $\Pi$ near the shock and at the edge of the curtain where the volume fraction gradient is large. $\mathcal{M}_s$ is omitted from the plots since it was found to be negligible. At later times after the shock has passed through the curtain, mean-shear production and the pressure-strain correlation act as the dominant production terms at the downstream edge of the curtain. Downstream of the curtain, there are no production mechanisms and viscous dissipation dominates.

The magnitude of the terms in the budget are observed to increase with increasing $\Phi_p$ and decrease with increasing shock Mach number. This reduction at higher Mach number is not due to enhanced dilatational dissipation, but rather a reduction of all terms, similar to what has been observed in single-phase compressible shear layers~\citep{sarkar_stabilizing_1995,pantano2002study}. 


  
  



\subsection{Energy spectra}\label{sec:spectra}
Two-dimensional energy spectra of the phase-averaged streamwise velocity fluctuations are computed at different locations along the curtain. Special care is taken to account for the presence of particles. At each location along the $x$-axis, the instantaneous energy spectrum is defined as
\begin{equation}\label{eq:Euu}
        E_{uu}(x,t) = \widehat{ \sqrt{\mathcal{I} \rho} u''} \widehat{ \sqrt{\mathcal{I} \rho} u''}^*,
\end{equation}
where the $\widehat{(\cdot)}$ notation denotes the two-dimensional Fourier transform and $^*$ indicates its complex conjugate. The integration of $E_{uu}$ at each streamwise location is taken over a circular shell in the $[\kappa_y \times \kappa_z]$ space, where $\kappa$ represents the wave number. This definition of the Fourier coefficient is consistent with classic compressible turbulence literature~\citep{kida_energy_1990,lele1992compact}, extended to include the indicator function to account for particles. 

\begin{figure}
  \centerline{\includegraphics*[width=\linewidth]{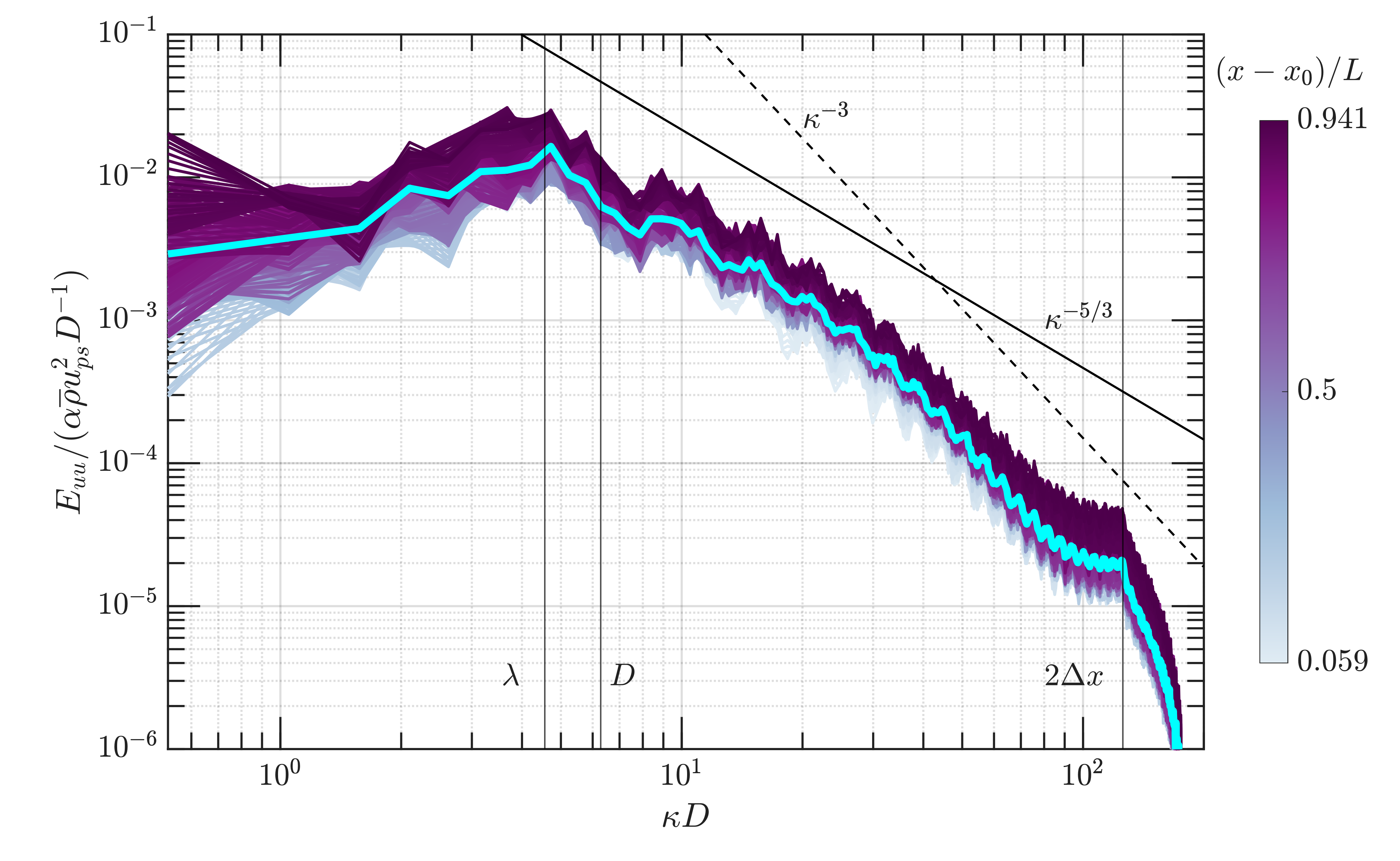}}
  \caption{One-dimensional spectra of streamwise velocity fluctuations for $M_s=1.66$ and $\Phi_p=0.2$ at $t/\tau_L=2$. The colour bar corresponds to different locations in the particle curtain. Ensemble average of all the spectra within the curtain (thick cyan line). Vertical lines indicate relevant length scales in the flow. Solid and dashed lines correspond to slopes of $-5/3$ and $-3$, respectively.}
  \label{fig:spec_1.66_0.2}
\end{figure}

Figure~\ref{fig:spec_1.66_0.2} shows the energy spectra for Case 5 ($M_s=1.66$, $\Phi_p=0.2$) at various $x$ locations within the particle curtain at $t/\tau_L=2$. The spectra for the initial 40 grid points ($x-x_0<D$) are excluded because the turbulence is not fully developed in this region. It is evident that the spectra remain relatively consistent across the streamwise positions, exhibiting minimal variation from the ensemble average of all spectra. Thus, although the flow is inhomogeneous in $x$, the turbulence is relatively homogeneous in the majority of the curtain. Consequently, subsequent figures will only display the ensemble average.

The inclusion of the discontinuous indicator function in \eqref{eq:Euu} introduces oscillations throughout the spectrum, known as a `ringing' artifact. While the ringing can be mitigated by applying a Butterworth filter or similar methods, such filtering was not employed to avoid the introduction of ad-hoc user-defined parameters.

Most of the energy resides at length scales that coincide with the mean interparticle spacing, $\lambda$. The interparticle spacing is found to differentiate the energy-containing range from the inertial subrange, indicating that wakes in the interstitial spaces between particles are responsible for the generation of PTKE. An inertial subrange is evident at scales smaller than $\lambda$, characterized by an energy spectrum that follows a $-5/3$ power law before transitioning to a steeper $-3$ power law at higher wavenumbers. The energy diminishes rapidly at scales below $2\Delta x$, which is attributed to numerical dissipation. Interestingly, part of the inertial subrange aligns with characteristics of homogeneous single-phase turbulence, displaying a $-5/3$ power law, while the smaller scales align with bubble-induced turbulence, evidenced by a $-3$ power law. However, the presence of noise in the spectra makes it challenging to draw definitive conclusions.

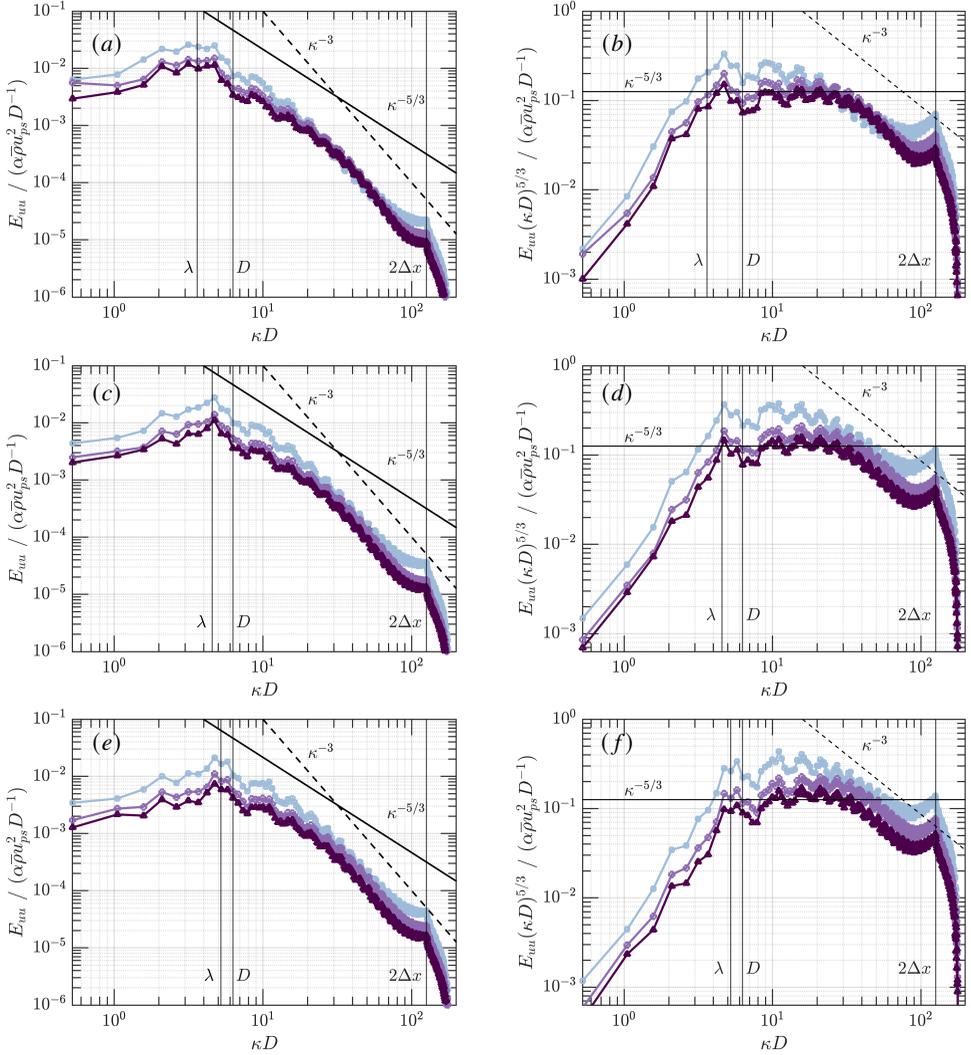
\begin{figure}
  \centering
  \begin{tabular}{ccc}
      \subfloat{
          \includegraphics[width=0.45\textwidth]{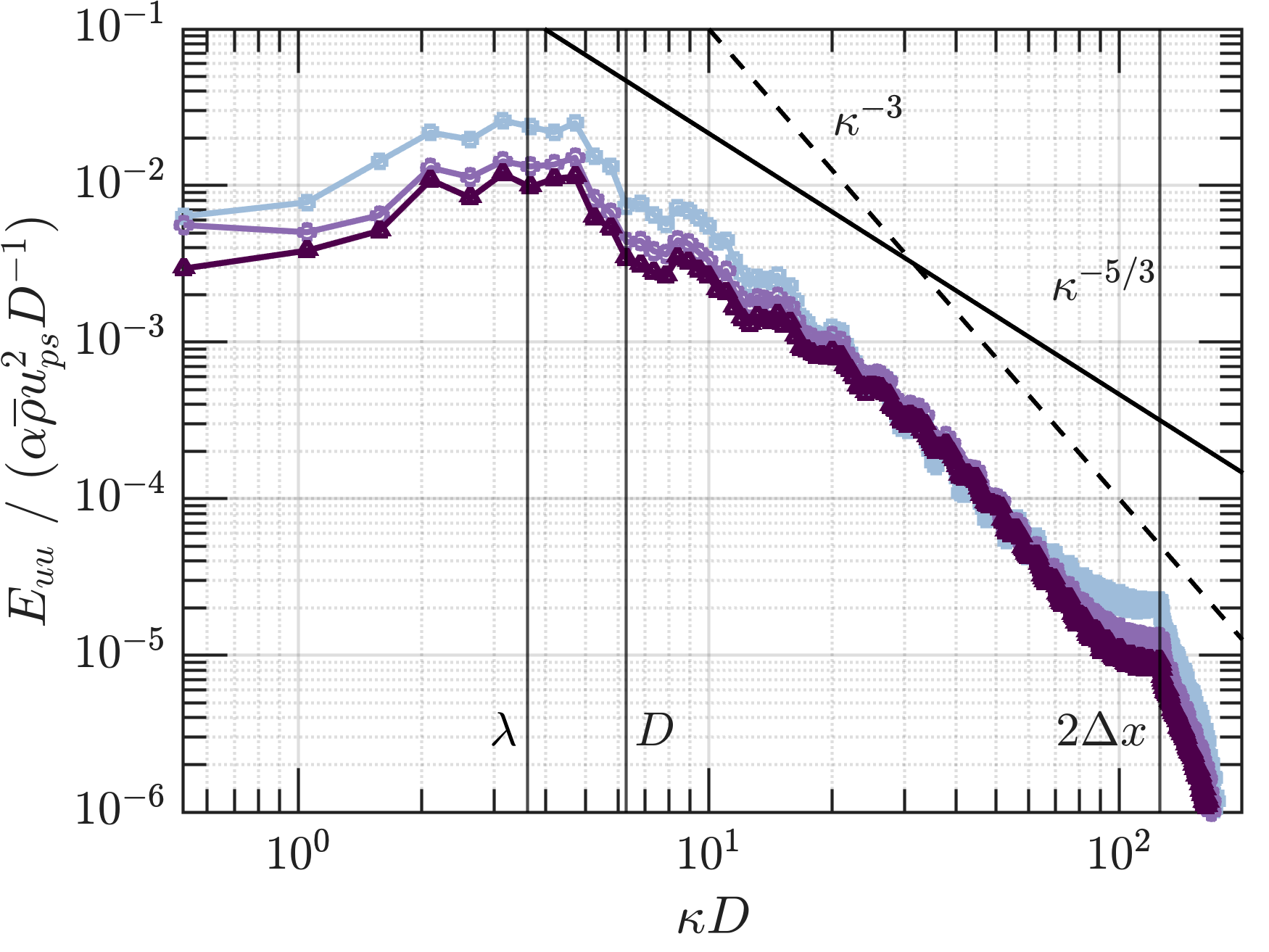}
      } \hspace{12pt}
      \subfloat{
          \includegraphics[width=0.45\textwidth]{{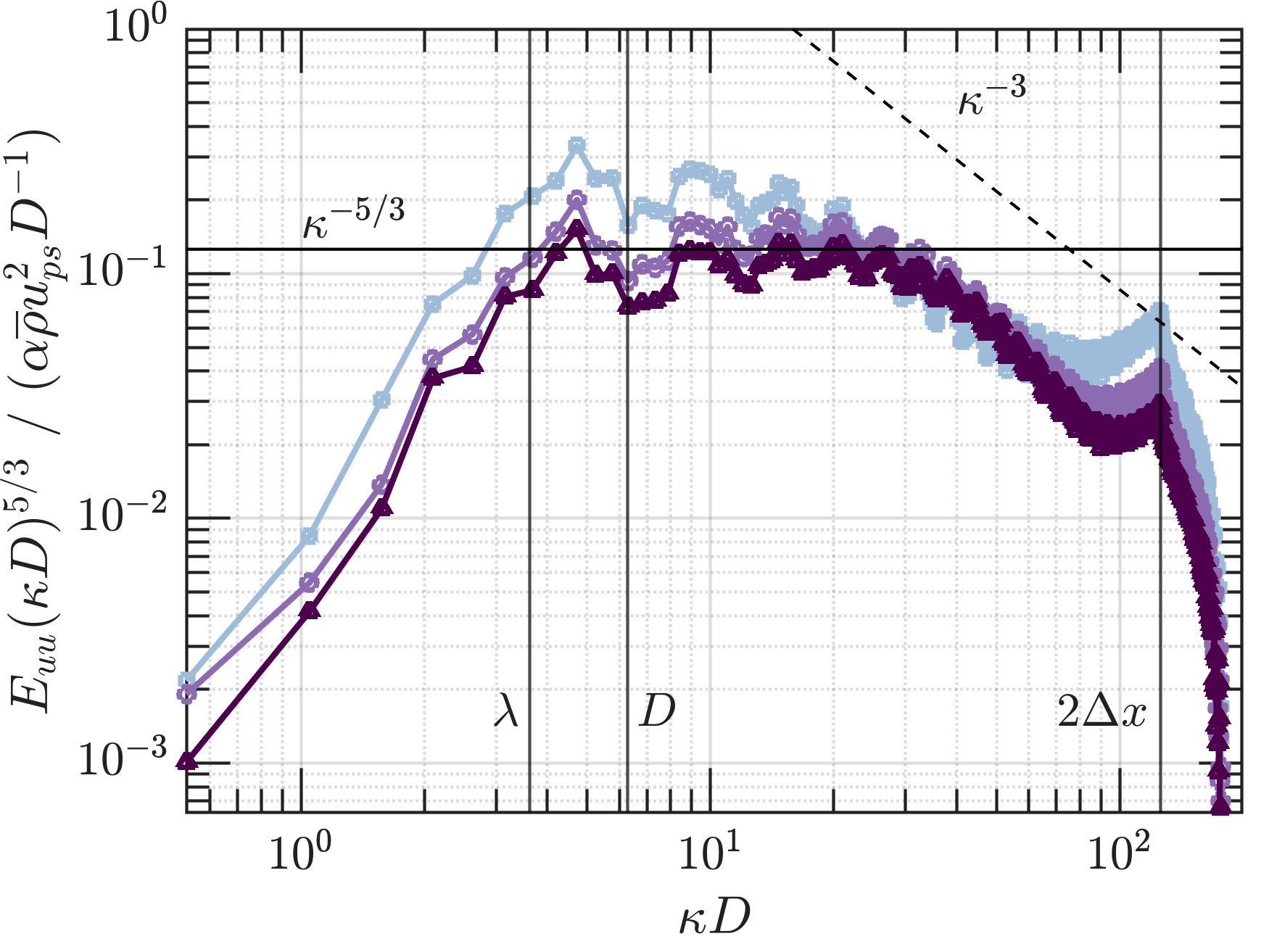}}
      } \\
      \subfloat{
          \includegraphics[width=0.45\textwidth]{{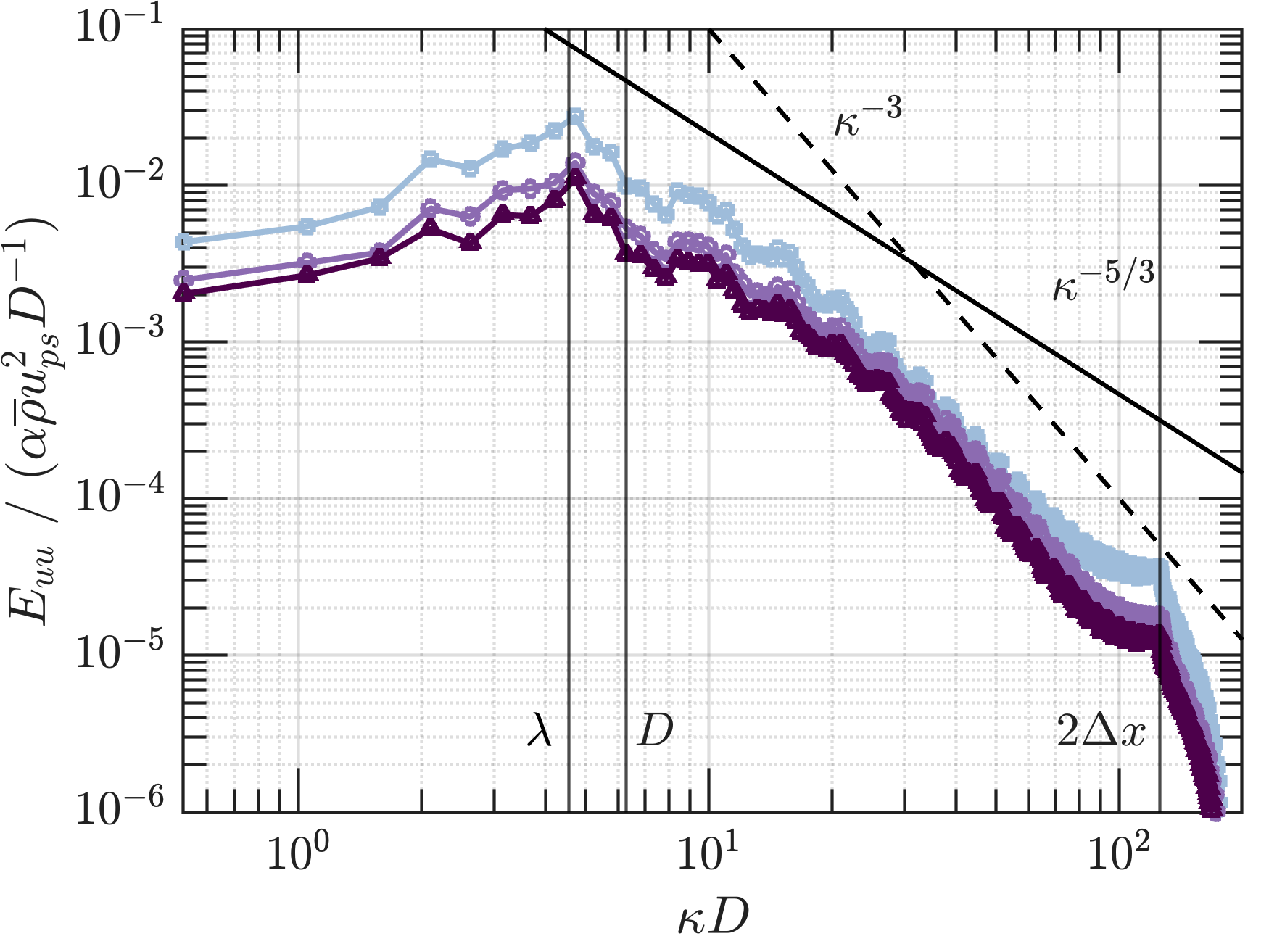}}
      } \hspace{12pt}
      \subfloat{
          \includegraphics[width=0.45\textwidth]{{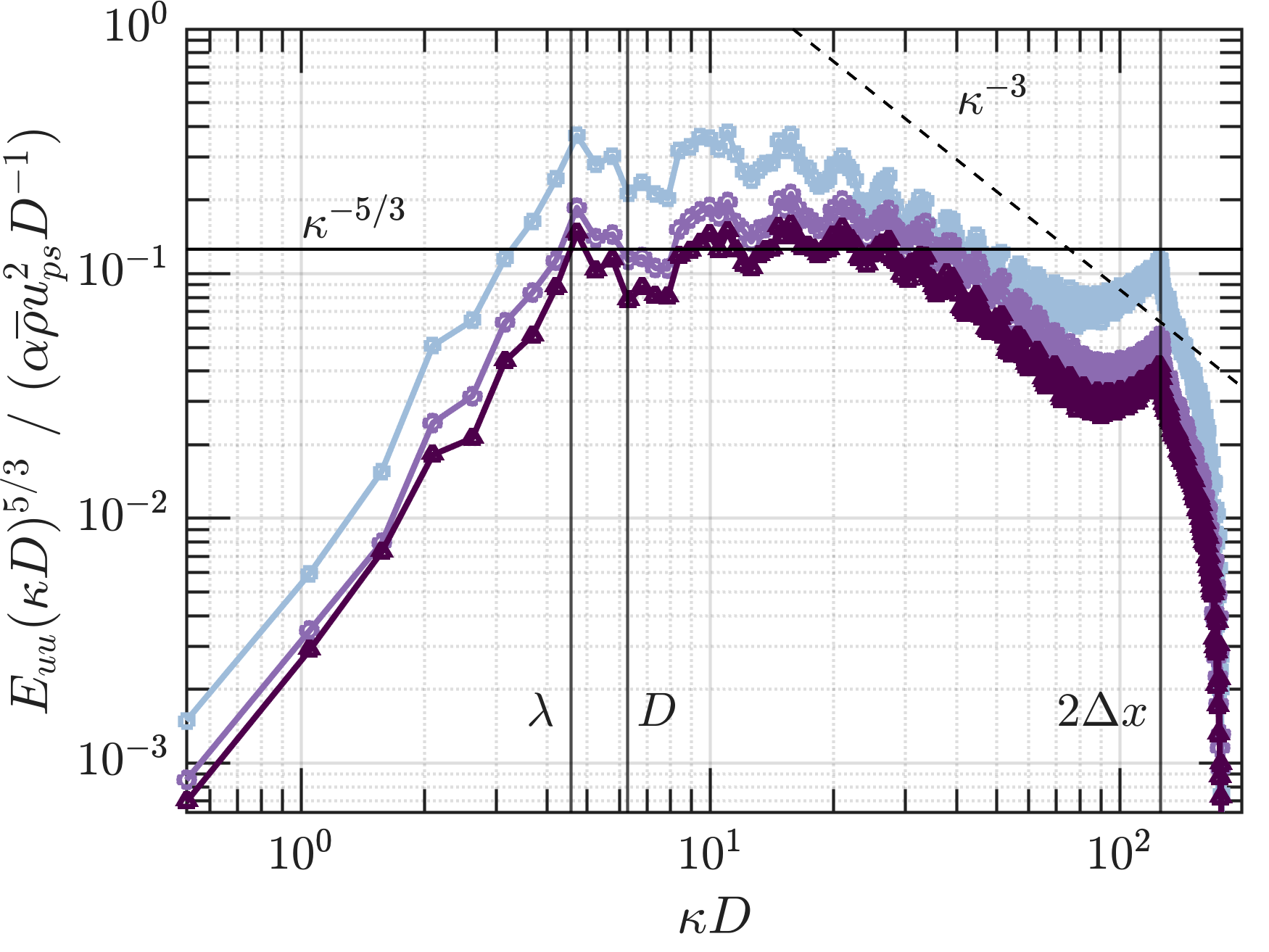}}
      } \\ 
      \subfloat{
          \includegraphics[width=0.45\textwidth]{{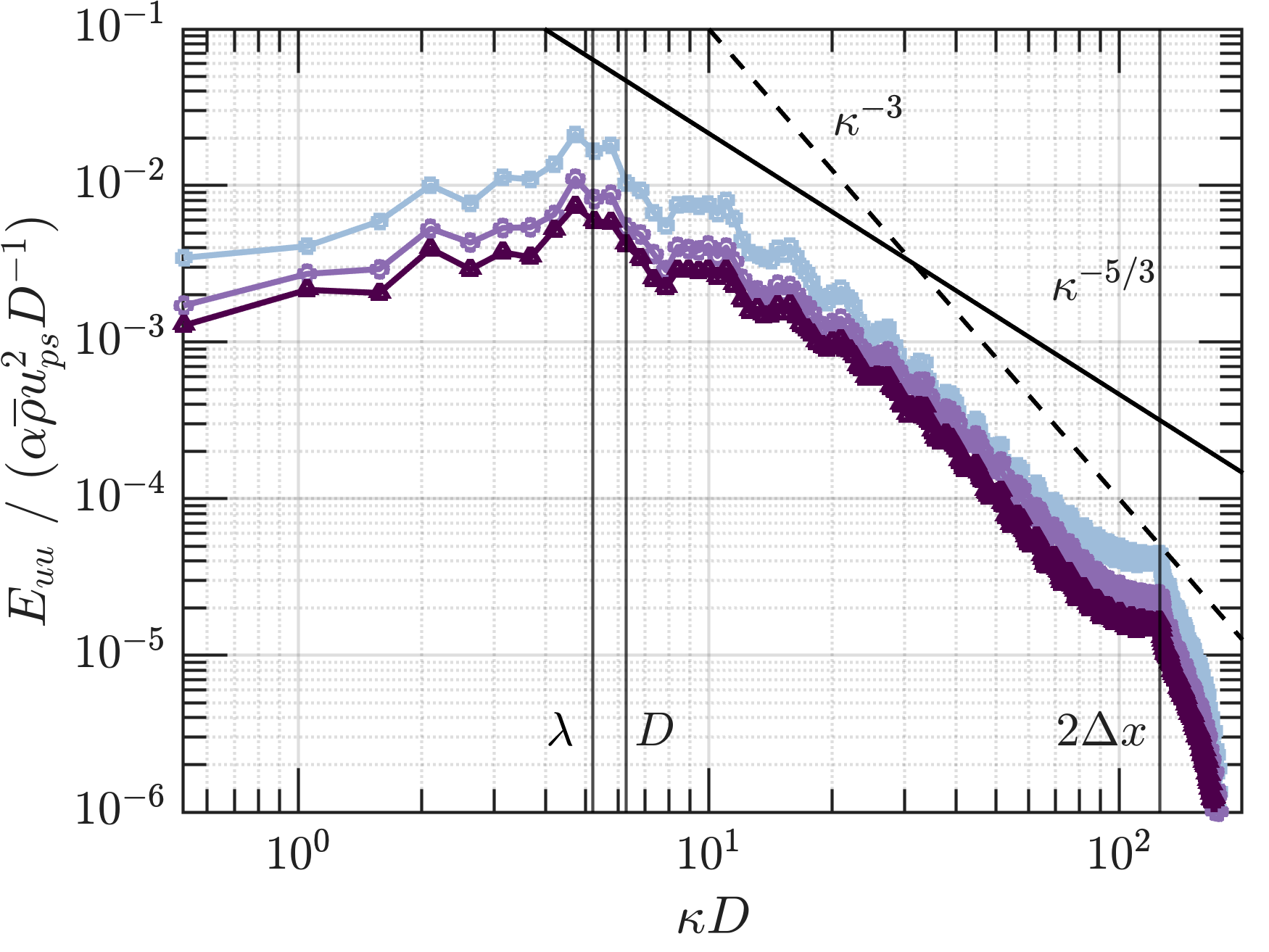}}
      } \hspace{12pt}
      \subfloat{
          \includegraphics[width=0.45\textwidth]{{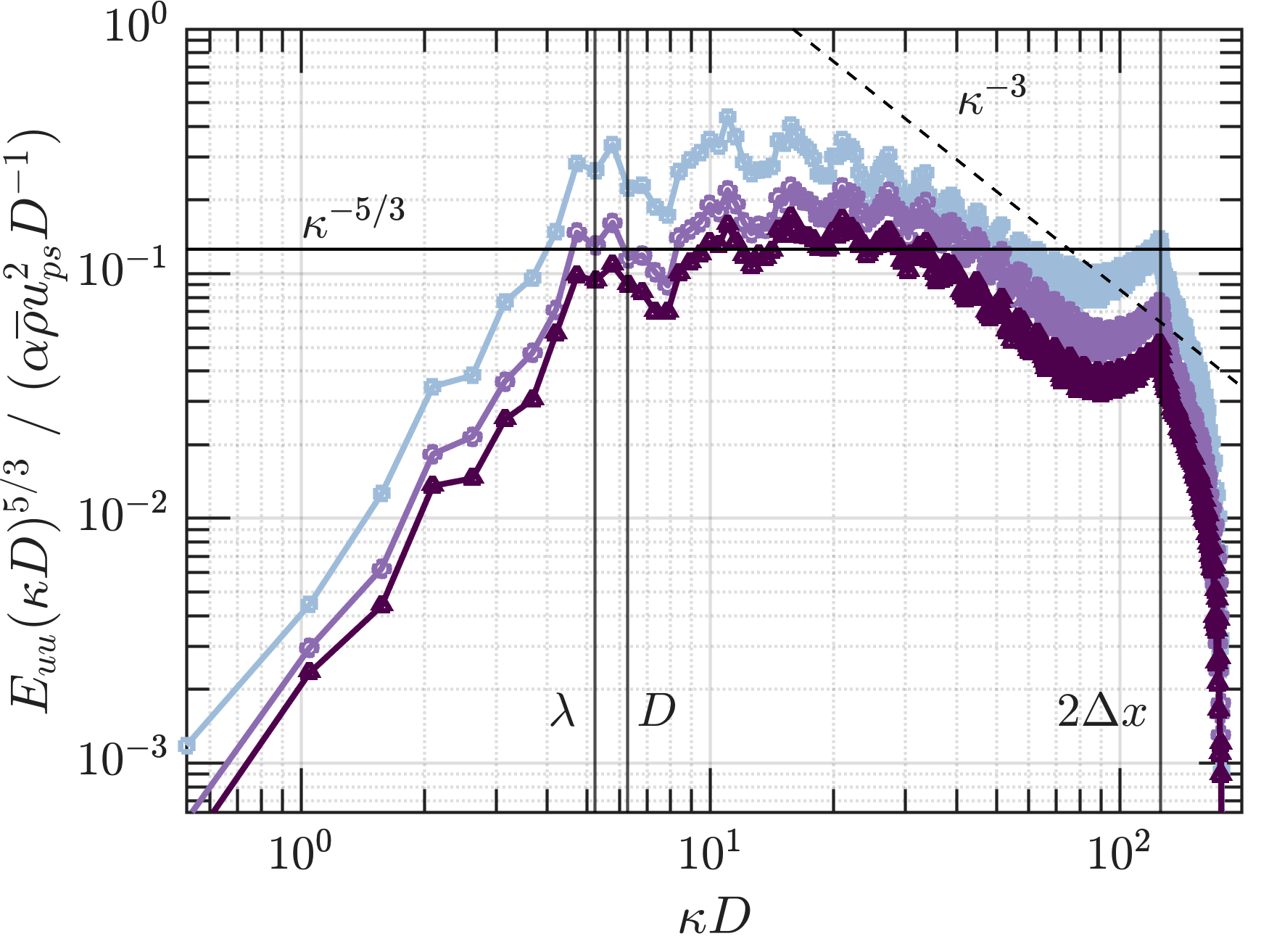}}
      }
  \end{tabular}
    \begin{tikzpicture}[overlay, remember picture]
      \node at (-11.7,6.5) {$(a)$};
      \node at ([xshift=0.5\linewidth]-11.7,6.5) {$(b)$};
      \node at (-11.7,1.9) {$(c)$};
      \node at ([xshift=0.5\linewidth]-11.7,1.9) {$(d)$};
      \node at (-11.7,-2.7) {$(e)$};
      \node at ([xshift=0.5\linewidth]-11.7,-2.7) {$(f)$};         
    \end{tikzpicture}
  \caption{Mean (left) and compensated (right) energy spectra of streamwise velocity fluctuations within the particle curtain at $t/\tau_L=2$ for $(a,b)$ $\Phi_p=0.1$, $(c,d)$ $\Phi_p=0.2$ and $(e,f)$ $\Phi_p=0.3$. $M_s=1.2$ (light blue, square), $M_s=1.66$ (lavender, circle), and $M_s=2.1$ (purple, triangle).}
  \label{fig:spectrend}
\end{figure}

In Figure~\ref{fig:spectrend}, the ensemble-averaged spectra are compared across different cases at $t/\tau_L = 2$. A broadband reduction in $E_{uu}$ is observed with increasing $M_s$, which is consistent with the observations made in the PTKE budget. As before, the turbulence levels are largely invariant with $\Phi_p$. For each case, the mean interparticle spacing is found to delineate the inertial subrange. Compensated spectra are also shown to better identify the power-law scaling, which appears consistent in each case. The spectrum decays with a $-5/3$ law at wave numbers $\mathcal{O}(D)$, while at higher wave numbers there is a steeper $-3$ decay. It remains unclear whether this steepening is due to gas-phase compressibility, interphase exchange with particles, or both. The following section decomposes the turbulent velocity field into solenoidal and dilatational components to gain further insight.

\subsubsection{Helmholtz decomposition}\label{sec:helm} 
A Helmholtz decomposition of the velocity field is performed to analyse the solenoidal (divergence-free) and dilatational (curl-free) components separately, according to~\citep{kida_energy_1990,yu_genuine_2019}
\begin{equation}
    \bm{u} = \bm{u}_{sol} +\bm{u}_{dil},
\end{equation}
where $\u_{sol} = \nabla \times \bm{A}$ and $\u_{dil} = \nabla \varphi$. Here $\bm{A}$ is the vector potential satisfying $\nabla^2 \bm{A} = - \bm{\omega}$, where $\bm{\omega}=\nabla\times\u$ is the local vorticity. The velocity potential $\varphi$ satisfies  $\nabla^2 \varphi = \nabla \cdot \u$.

\begin{figure}
\centering
    \hspace*{-36pt}
    \begin{tabular}{c}
      \subfloat{
          \includegraphics[width=0.8\linewidth] {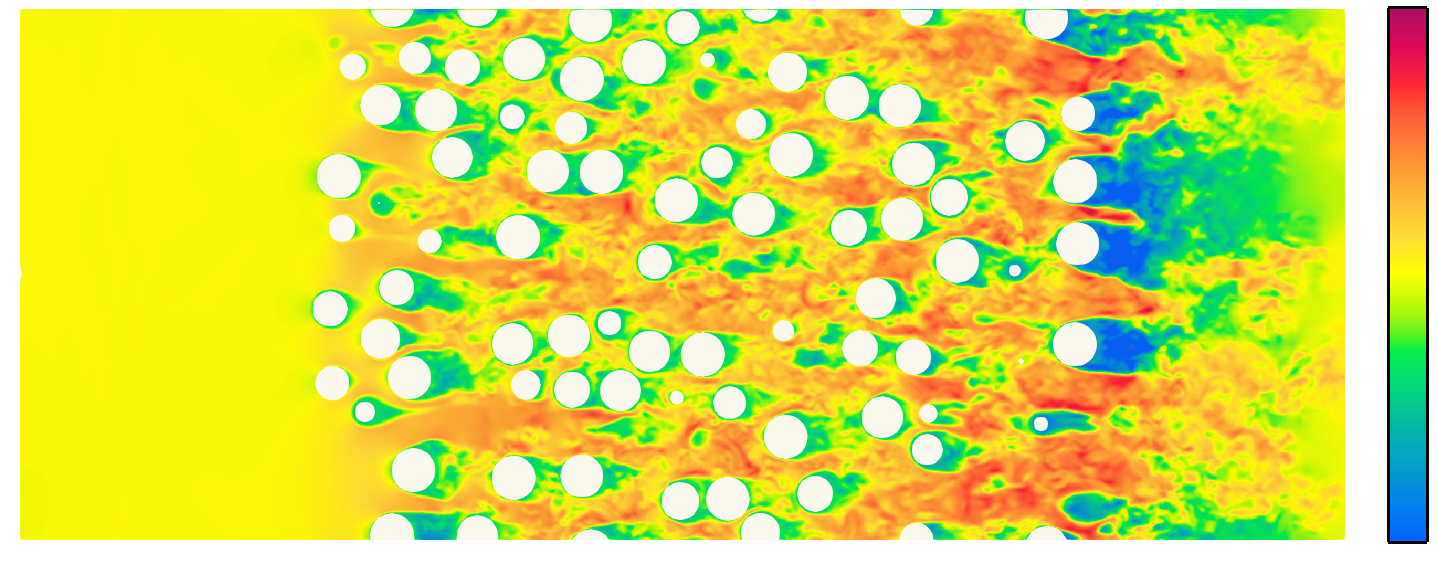}
      } \\\\
      \subfloat{
          \includegraphics[width=0.8\linewidth]{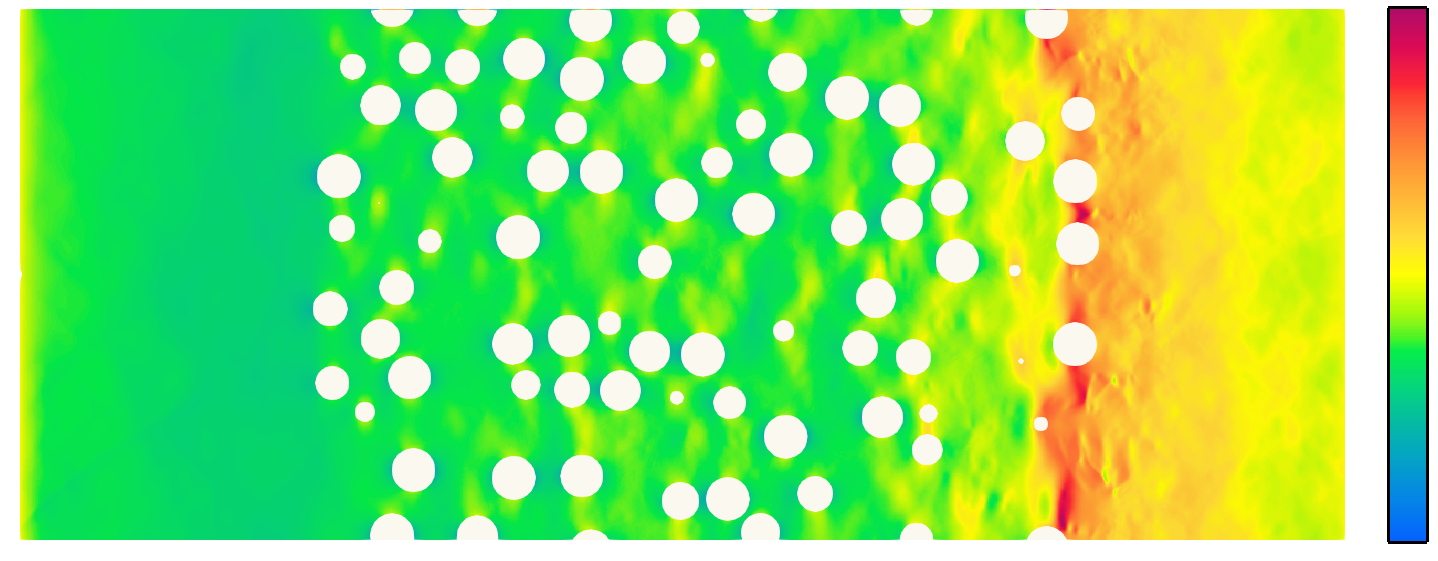}
      } 
      \end{tabular}
    \caption{A two-dimensional slice of the $(a)$ solenoidal and $(b)$ dilatational streamwise velocity fields at $t/\tau_L=2$ for Case $5$.
    }
    \begin{tikzpicture}[overlay, remember picture]
        \node at (-5.3,10.4) {$(a)$};
        \node at (-5.3,5.8) {$(b)$};
        \node at ([xshift=0.78\linewidth]-5.3,10.2) {$1.0$};
        \node at ([xshift=0.77\linewidth]-5.3,8.25) {$0$};
        \node at ([xshift=0.82\linewidth]-5.3,8.25) {$\frac{u_{sol}}{u_{ps}}$};
        \node at ([xshift=0.78\linewidth]-5.3,6.3) {$-1.0$};
        \node at ([xshift=0.78\linewidth]-5.3,5.5) {$0.5$};
        \node at ([xshift=0.77\linewidth]-5.3,3.55) {$0$};
        \node at ([xshift=0.82\linewidth]-5.3,3.55) {$\frac{u_{dil}}{u_{ps}}$};
        \node at ([xshift=0.78\linewidth]-5.3,1.6){$-0.5$};
    \end{tikzpicture}
    \label{fig:soldilcontour}
\end{figure}

Figure~\ref{fig:soldilcontour} shows two-dimensional slices of the instantaneous streamwise velocity components. The solenoidal component exhibits significant fluctuations throughout the curtain, capturing particle wakes. In contrast, the dilatational velocity field remains relatively small within the curtain and increases sharply at the downstream edge, where the flow chokes. This indicates that the majority of PTKE is concentrated in the solenoidal portion, with compressibility playing a minor role except near large volume fraction gradients.

\begin{figure}
  \centering
  \begin{tabular}{cc}
      \subfloat{
          \includegraphics[width=0.45\textwidth]{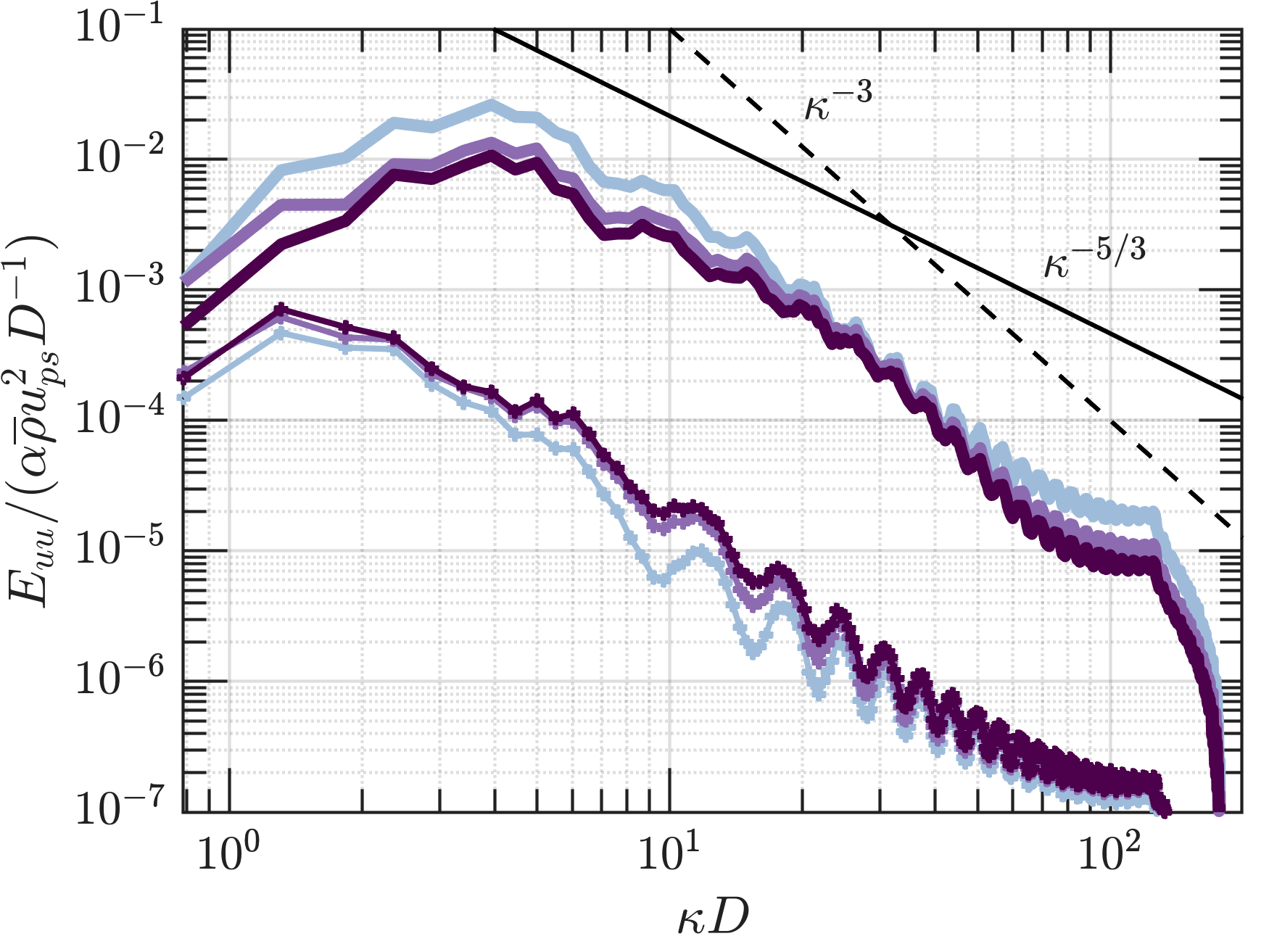}
      } \hspace{6pt}
      \subfloat{
          \includegraphics[width=0.45\textwidth]{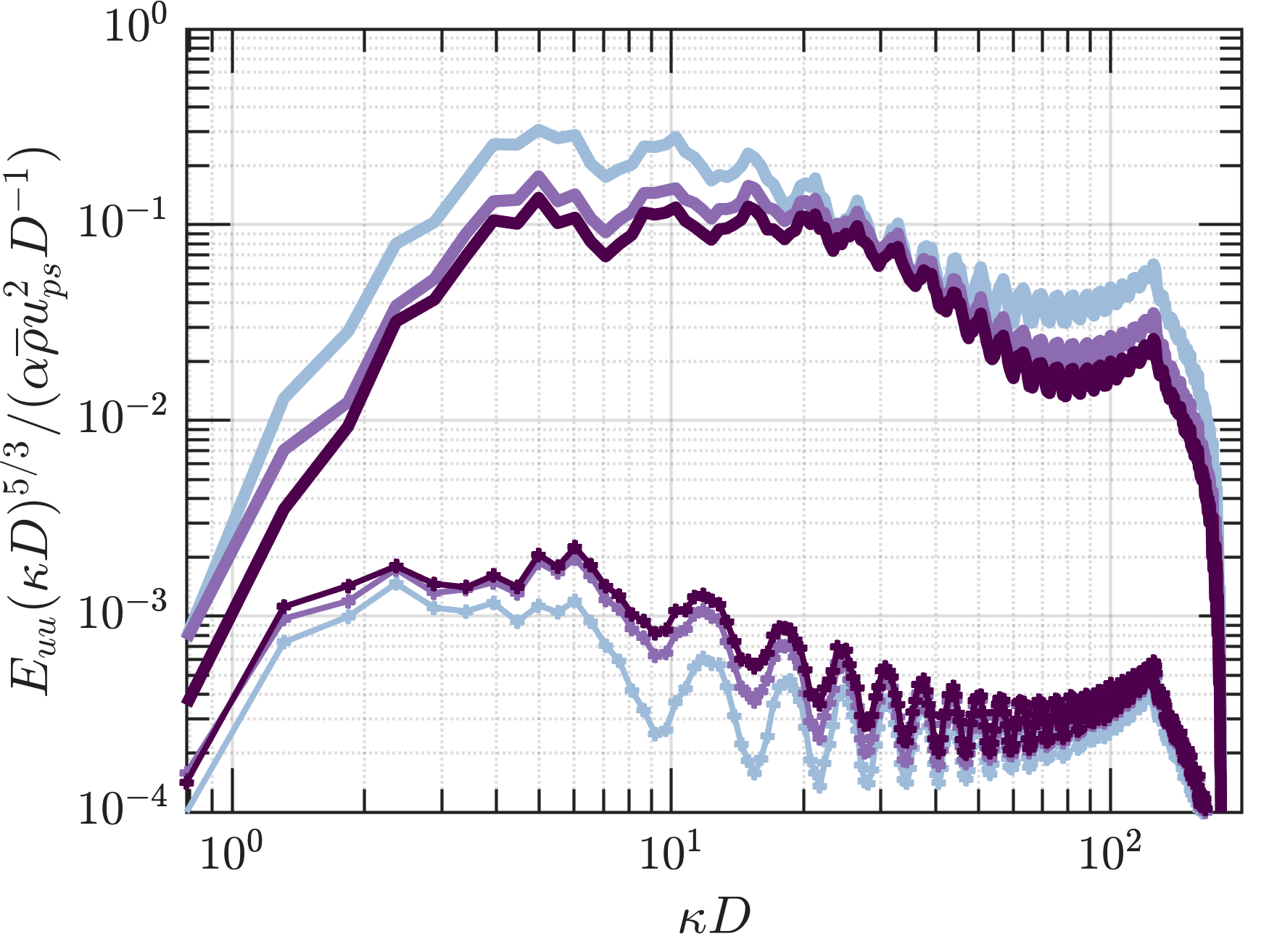}
      } \\
      \subfloat{
          \includegraphics[width=0.45\textwidth]{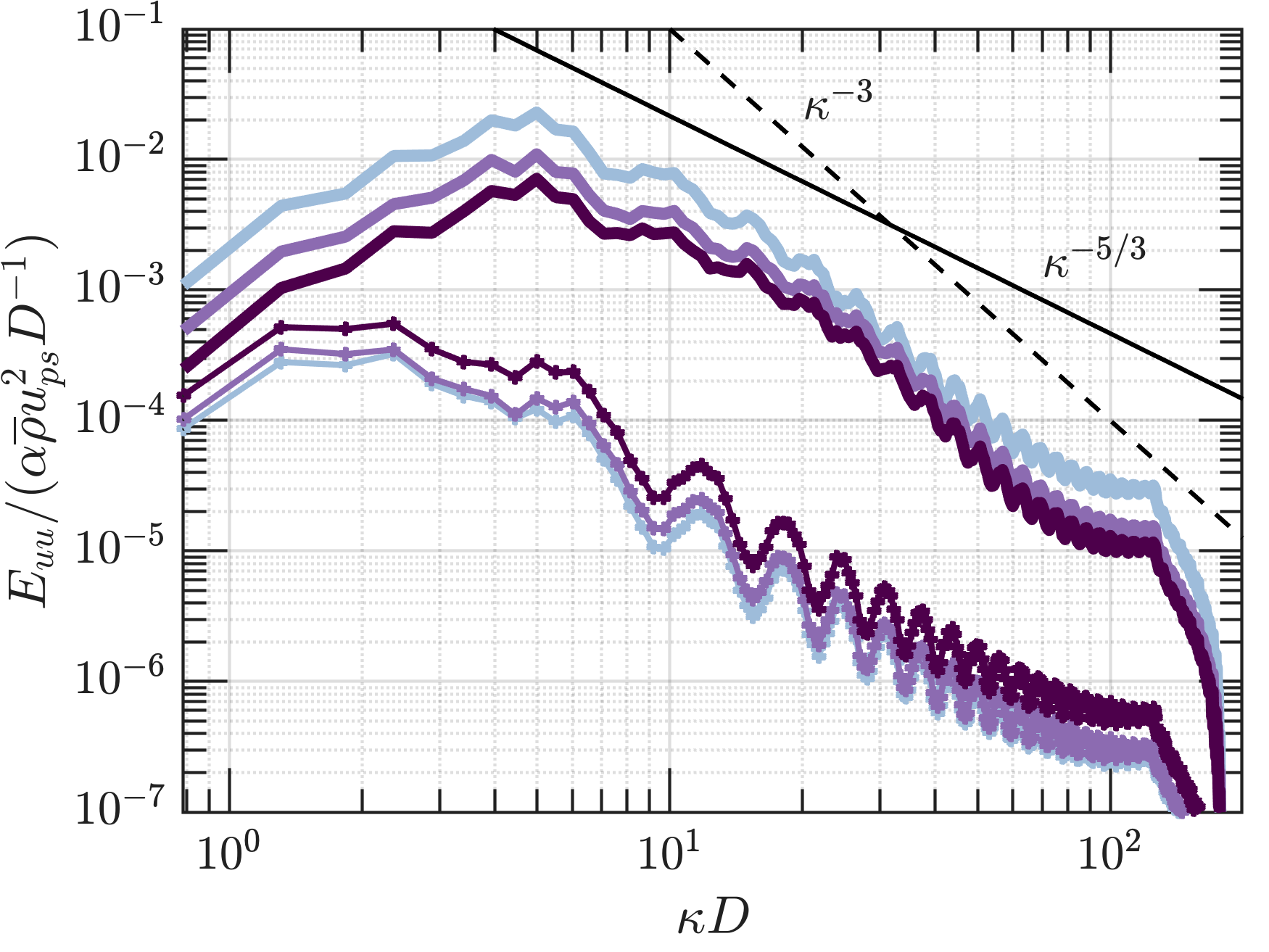}
      } \hspace{6pt}
      \subfloat{
          \includegraphics[width=0.45\textwidth]{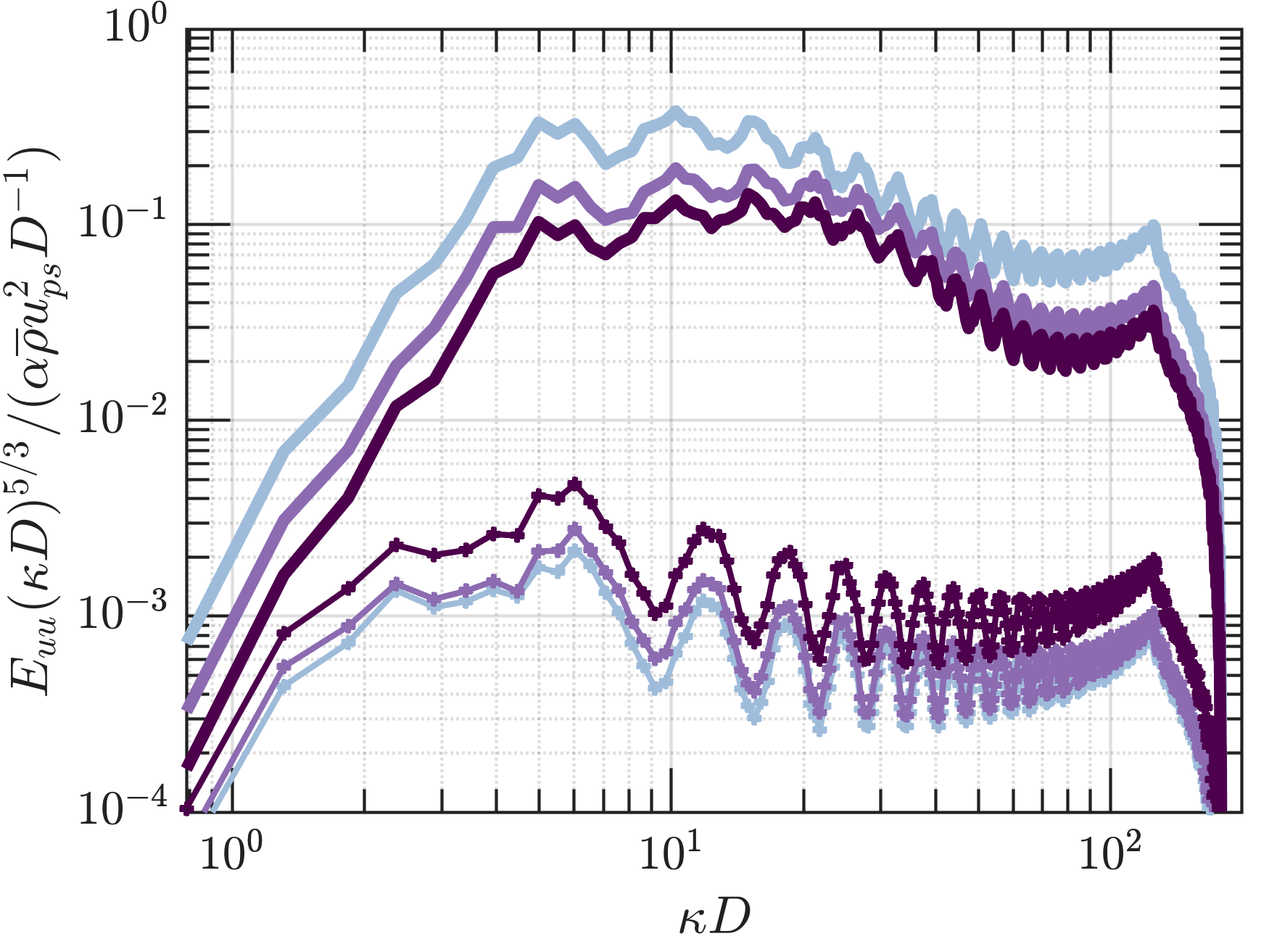}
      } \\ 
      \subfloat{
          \includegraphics[width=0.45\textwidth]{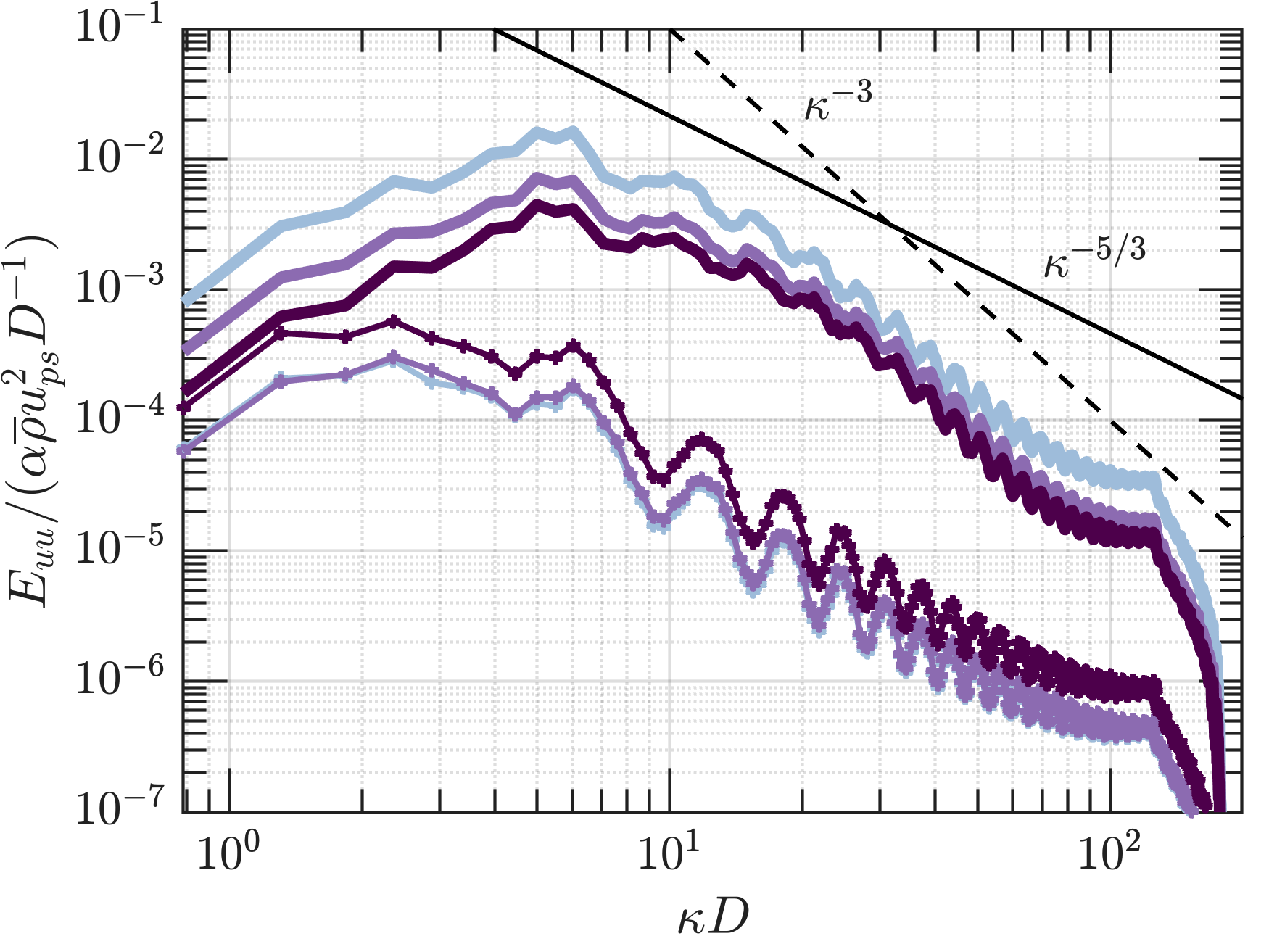}
      } \hspace{6pt}
      \subfloat{
          \includegraphics[width=0.45\textwidth]{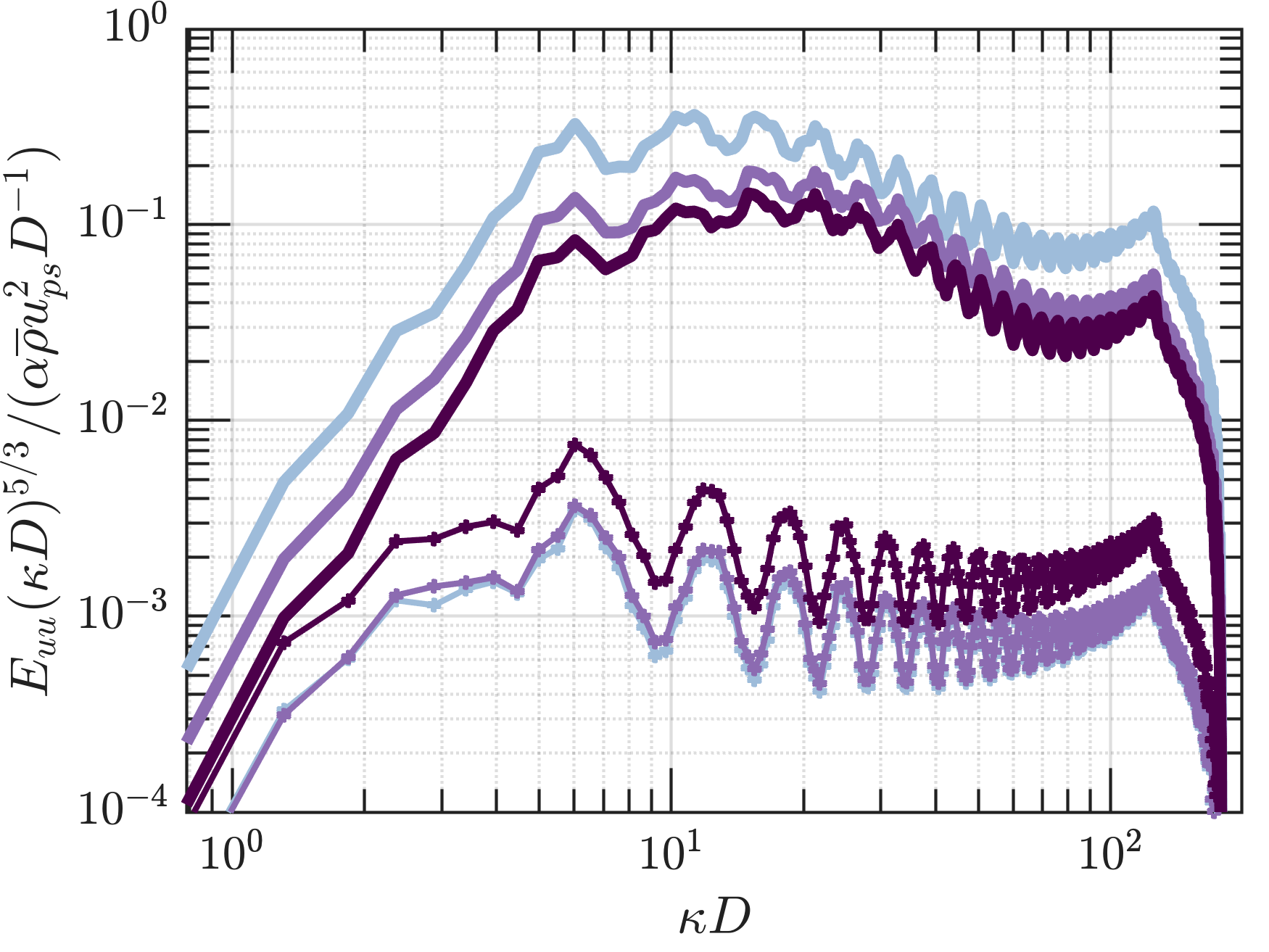}
      }
  \end{tabular}
    \begin{tikzpicture}[overlay, remember picture]
      \node at (-11.6,6.5) {$(a)$};
      \node at ([xshift=0.49\linewidth]-11.6,6.5) {$(b)$};
      \node at (-11.6,1.9) {$(c)$};
      \node at ([xshift=0.49\linewidth]-11.6,1.9) {$(d)$};
      \node at (-11.6,-2.7) {$(e)$};
      \node at ([xshift=0.49\linewidth]-11.6,-2.7) {$(f)$};         
    \end{tikzpicture}
  \caption{Mean (left) and compensated (right) spectra of the streamwise velocity fluctuations computed using solenoidal (\full) and dilatational (\mar) velocity fields at $t/\tau_L=2$ for $(a,b)$ $\Phi_p=0.1$, $(c,d)$ $\Phi_p=0.2$ and $(e,f)$ $\Phi_p=0.3$. Colour scheme same as figure~\ref{fig:spectrend}.}
  \label{fig:soldil}
\end{figure}

Figure~\ref{fig:soldil} shows energy spectra of the streamwise solenoidal and dilatational velocity components at $t/\tau_L=2$. The solenoidal energy spectrum is approximately two orders of magnitude larger than the dilatational component across all wavenumbers and tends to decrease with increasing $M_s$, while the dilatational component increases with increasing Mach number. These findings align with observations from direct numerical simulations of compressible homogeneous isotropic turbulence \citep{donzis_fluctuations_2013}. Interestingly, only the solenoidal spectrum demonstrates a $-3$ power law decay, while the dilatational component maintains an approximate $-5/3$ scaling throughout the inertial subrange. Consequently, the $-3$ power law decay may be attributed to incompressible wakes rather than compressible effects.      

\section{Two-fluid turbulence model}\label{sec:model}
In this section, we propose a two-equation model for PTKE and its dissipation. This turbulence model is integrated into a one-dimensional Eulerian-based two-fluid framework. The hyperbolic equations for particle-laden compressible flows include added mass and internal energy contributions, derived from kinetic theory based on the recent work of \citet{fox2019kinetic,fox2020hyperbolic}. The section ends with an a-posteriori analysis of the turbulence model and comparisons are made against the particle-resolved simulations.

\subsection{A kinetic-based hyperbolic two-fluid model}
Particle-resolved simulations require grid spacing significantly smaller than the particle diameter to adequately resolve boundary layers and capture relevant aerodynamic interactions. Eulerian-based two-fluid models are a widely used coarse-grained modelling approach that assume the properties of both solid and fluid phases can be expressed as interpenetrating continua interacting through interphase drag terms. Unlike particle-resolved simulations, the computational cost of modelling the particle phase scales with the number of grid cells rather than the number of particles, making it a more efficient option for simulating systems with a large number of particles.

The added mass is included in the mass, momentum, and energy balances, augmented to account for particle wakes. These equations are fully hyperbolic and avoid the ill-posedness common in conventional compressible two-fluid models with two-way coupling~\citep{fox2020hyperbolic}. To match the conditions used in the particle-resolved simulations, stationary monodisperse particles are considered (i.e. the particle velocity $\bm{u}_p=0$, granular temperature $\varTheta_p =0$ and $\alpha_p \rho_p$ is constant in the curtain, where $\alpha_p=1-\alpha_g$ is the particle volume fraction and $\rho_p$ is the particle density). Heat transfer between the phases is neglected. For brevity, brackets and tildes are omitted and it is implied that the equations are written in terms of Favre- and phase-averaged quantities.

The governing equation for mass balance (added mass, gas phase) in one spatial dimension are given by
\begin{equation}
    \begin{aligned}
        \frac{\partial}{\partial t} (\alpha_a \rho_a ) &= S_a, \\
        \frac{\partial}{\partial t} (\alpha_g^\star \rho ) + \frac{\partial}{\partial x} (\alpha_g^\star \rho   u) & = -S_a. 
    \end{aligned}
\end{equation}
The gas-phase momentum balance is
\begin{equation}
    \begin{aligned}
        \frac{\partial}{\partial t} (\alpha_g^\star \rho u) + \frac{\partial}{\partial x} (\alpha_g^\star \rho u^2 + \hat{p} + \alpha_p^\star \alpha_g^\star \rho u^2 ) &= -\frac{\alpha_p^\star \rho }{\tau_p} u + \alpha_p^\star \Big (\frac{\partial}{\partial x} \hat{p} + F_{pg} \Big) - S_{gp}, \\ 
    \end{aligned}    
\end{equation}
and the gas-phase total energy balance is
\begin{equation}
    \begin{gathered}
        \frac{\partial}{\partial t} (\alpha_g^\star \rho E ) + \frac{\partial}{\partial x} (\alpha_g^\star \rho u E + \alpha_g^\star u \hat{p} ) = - S_E.\\
    \end{gathered}
\end{equation}
The added-mass internal energy balance is
\begin{equation}
    \begin{gathered}
        \frac{\partial}{\partial t} (\alpha_a \rho_a  e_a) = S_E.
    \end{gathered}
\end{equation}
Here, $\alpha_a$ is the volume fraction of the added-mass phase and $\rho_a$ is its density.  The gas-phase volume fraction is replaced by $\alpha_g^\star=\alpha_g-\alpha_a$, $\alpha_p^\star = \alpha_p + \alpha_a$, $\alpha_g^\star = 1 - \alpha_p^\star$  and $e_a$ is the specific internal energy of the added mass. The gas- and added-mass phases have the same pressure $p$, but different temperatures $T$ and $T_a$, found from $e$ and $e_a$, respectively. $S_a$ represents mass exchange between the two phases through added mass, leading to momentum $S_{gp}$ and energy $S_E$ exchange. The particle response timescale $\tau_p=4\rho D^2/(3\mu C_D \Rey_p)$ 
depends on the drag coefficient $C_D$ modelled using the drag law from \citet{osnes2023comprehensive}. This model takes into account the effects of local volume fraction, the particle Reynolds number $\Rey_p=\rho |u| D/\mu$ and particle Mach number $M_p=|u|/c$ based on slip velocity $|u|$ ($|u| = |u - u_p|$, $u_p=0$).  The remaining parameters are provided in Appendix~\ref{appB}. Note that PTKE contributes to the modified pressure $\hat{p}$.

The equations are solved using a standard finite-volume method implemented in \texttt{MATLAB}. A HLLC scheme \citep{toro1994restoration} is employed to solve the hyperbolic part of the system. Further details on the implementation and discretization of the one-dimensional two-fluid model can be found in \citet{boniou2023shock}.

\subsection{Two-equation model for PTKE}\label{sec:two-eqn}
To capture PTKE in the Eulerian framework, a two-equation $k_g$--$\epsilon$ model is proposed that retains only the significant source terms from the budget:
\begin{equation}\label{eqn:2eqnmodel}
  \begin{aligned}
    &\frac{\partial}{\partial t}(\alpha_g^\star  \rho k_g) + \frac{\partial}{\partial x}(\alpha_g^\star \rho k_g u \ ) = \mathcal{P}_s + \mathcal{P}_D - (1+M_t^2) {\alpha_g^\star \rho} \epsilon,\\
    & \frac{\partial}{\partial t}(\alpha_g^\star \rho \epsilon) + \frac{\partial}{\partial x}(\alpha_g^\star \rho \epsilon u \ ) = C_{\epsilon,1} \frac{\epsilon}{k_g} \mathcal{P}_S + \frac{C_{\epsilon,D}}{\tau_D}\mathcal{P}_D - C_{\epsilon,2} \ \alpha_g^\star \rho  \frac{\epsilon^2}{k_g},
  \end{aligned}
\end{equation}
where  $C_{\epsilon,1}=1.44$ and $C_{\epsilon,2}=1.92$  are constants from single-phase turbulence modelling. The mean-shear production term is modelled as $\mathcal{P}_s = - \alpha_g^\star \rho \widetilde{u''u''} (\partial u / \partial x)$. The interphase PTKE exchange term due to drag induced by particles is $\mathcal{P}_D = \alpha_p^\star \rho u^2/\tau_p$. 
The compressibility correction in the PTKE equation is written in terms of the turbulent Mach number $M_t=\sqrt{2k_g}/c$, where $c=\sqrt{\gamma p/\rho}$ is the local speed of sound~\citep{sarkar1991analysis}. 
$\tau_D$ is rate of drag dissipation and is modelled using slip velocity as $\tau_D= D /|u|$.

The mean-shear production term, $\mathcal{P}_s$, includes the streamwise component of the Reynolds stress, $\widetilde{u''u''}$. Based on the findings from \S~\ref{sec:1dstats}, the anisotropy was found to be relatively constant across the curtain and independent of volume fraction and shock Mach number (see figure~\ref{fig:anisotropy}). The streamwise component of Reynolds stress is therefore given by
\begin{equation}\label{eq:Rij}
\begin{aligned}
\widetilde{u''u''} =  2 \left( b_{11} + \frac{1}{3} \right) k_g,\\
\widetilde{v''v''} = \widetilde{w''w''} =  2 \left( b_{22} + \frac{1}{3} \right) k_g,
\end{aligned}
\end{equation}
with $b_{11}=0.2$ and $b_{22}=-0.1$.

In the two-equation model \eqref{eqn:2eqnmodel}, the only remaining term requiring closure is $C_{\epsilon,D}$, a model coefficient that controls the portion of PTKE produced through drag that ultimately gets dissipated. In the limits of homogeneity and steady state with $M_t=0$, \eqref{eqn:2eqnmodel} reduces to
\begin{equation}
    \frac{k_g}{u^2} = \frac{C_{\epsilon,2}}{C_{\epsilon,D} }\frac{\alpha_p^\star \rho}{\alpha_g^\star \rho} \frac{\tau_D}{\tau_p}.
\end{equation}
Rearranging for the unclosed model parameter yields
\begin{equation}\label{eq:CepsD}
    C_{\epsilon,D} =  \frac{3}{4} \frac{\alpha_p^\star}{\alpha_g^\star}\frac{u^2}{k_g} C_{\epsilon,2}  C_D.
\end{equation}
\citet{mehrabadi2015pseudo} proposed the following algebraic model for PTKE for homogeneous particle suspensions valid for $ \alpha_g\Rey_p <300$, $M_p=0$ and $\alpha_p \ge 0.1$:
\begin{equation}
    \frac{k_g}{u^2} = \alpha_p \left(1+1.25\alpha_g^3\exp(-\alpha_p \sqrt{\alpha_g\Rey_p}) \right).
\end{equation}
Plugging in this expression into \eqref{eq:CepsD} provides closure for $C_{\epsilon,D}$ and ensures the model returns the correct level of PTKE in the limit of incompressible, homogeneous, steady flow. Because $\alpha_p^\star \to \alpha_p$ when $\alpha_p \to 0$, $C_{\epsilon,D}$ remains finite outside the particle curtain.

\subsection{A-posteriori analysis}
\begin{figure}    \centerline{\includegraphics[height=0.2\textheight,width=0.6\linewidth,keepaspectratio]{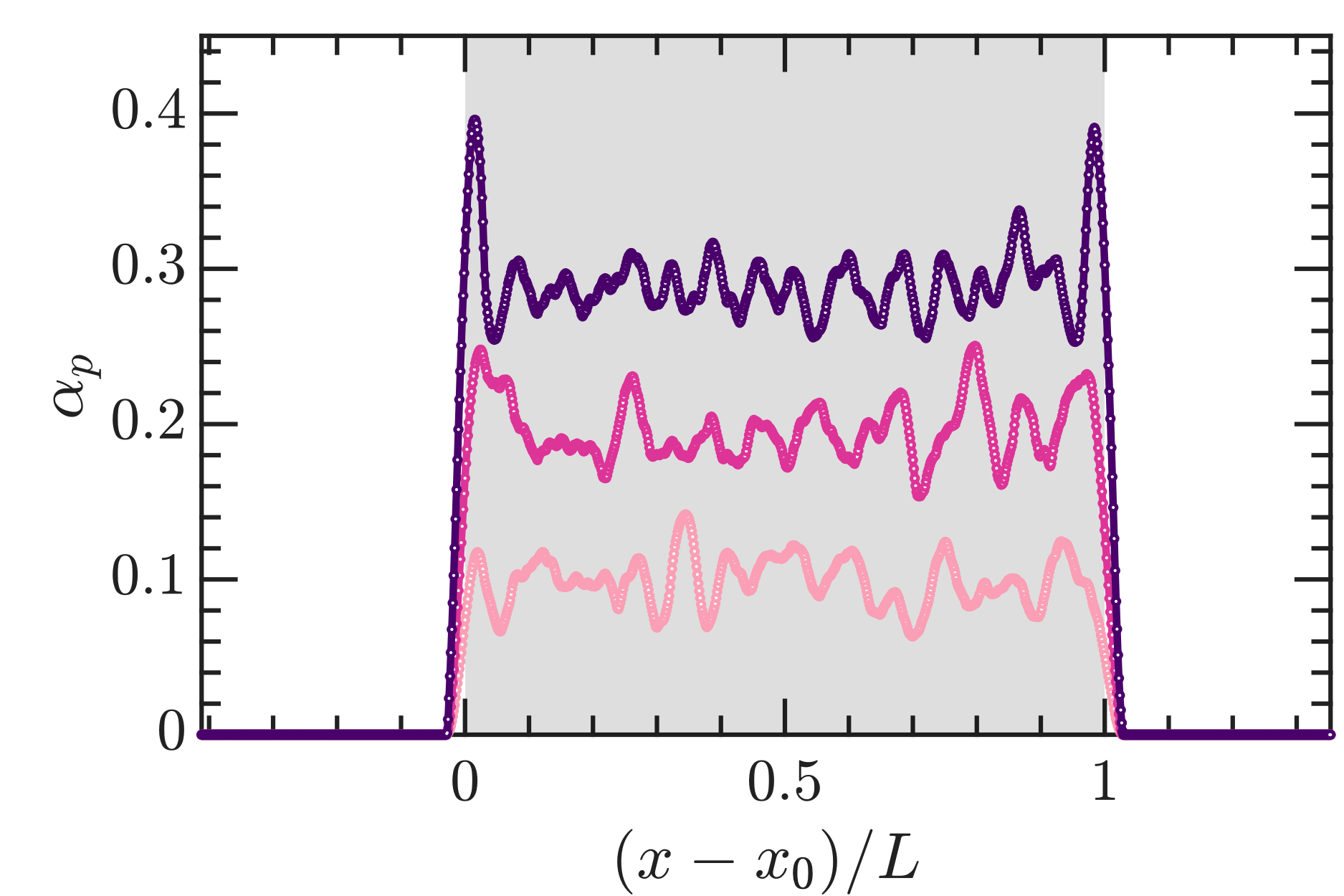}}
    \caption{One-dimensional particle volume fraction profiles obtained from the particle-resolved simulations for $\Phi_p=0.1$ (light pink), $\Phi_p=0.2$ (pink) and $\Phi_p=0.3$ (purple).}
    \label{fig:volfrac}
\end{figure}

One-dimensional shock--particle interactions are simulated using the two-fluid model detailed above with the parameters used in the particle-resolved simulations. It should be noted that the results will depend significantly on the volume fraction profile. To ensure a fair comparison, one-dimensional volume fraction profiles are extracted from the particle-resolved simulations and used in the model (see figure~\ref{fig:volfrac}).

\begin{figure}
  
  \begin{tabular}{ccc}
      \subfloat{
          \includegraphics[width=0.31\textwidth]{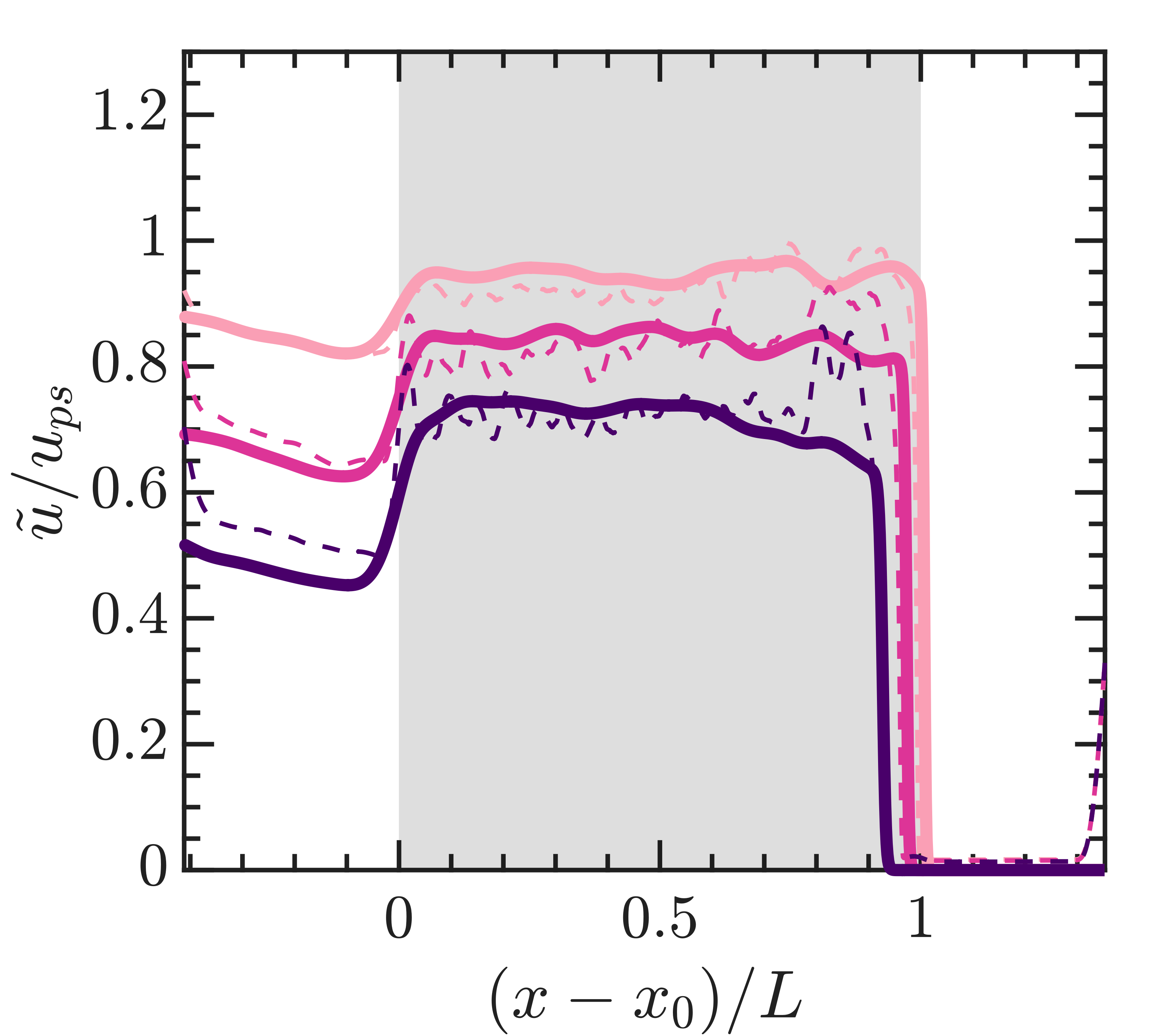}
      } 
      \subfloat{
          \includegraphics[width=0.31\textwidth]{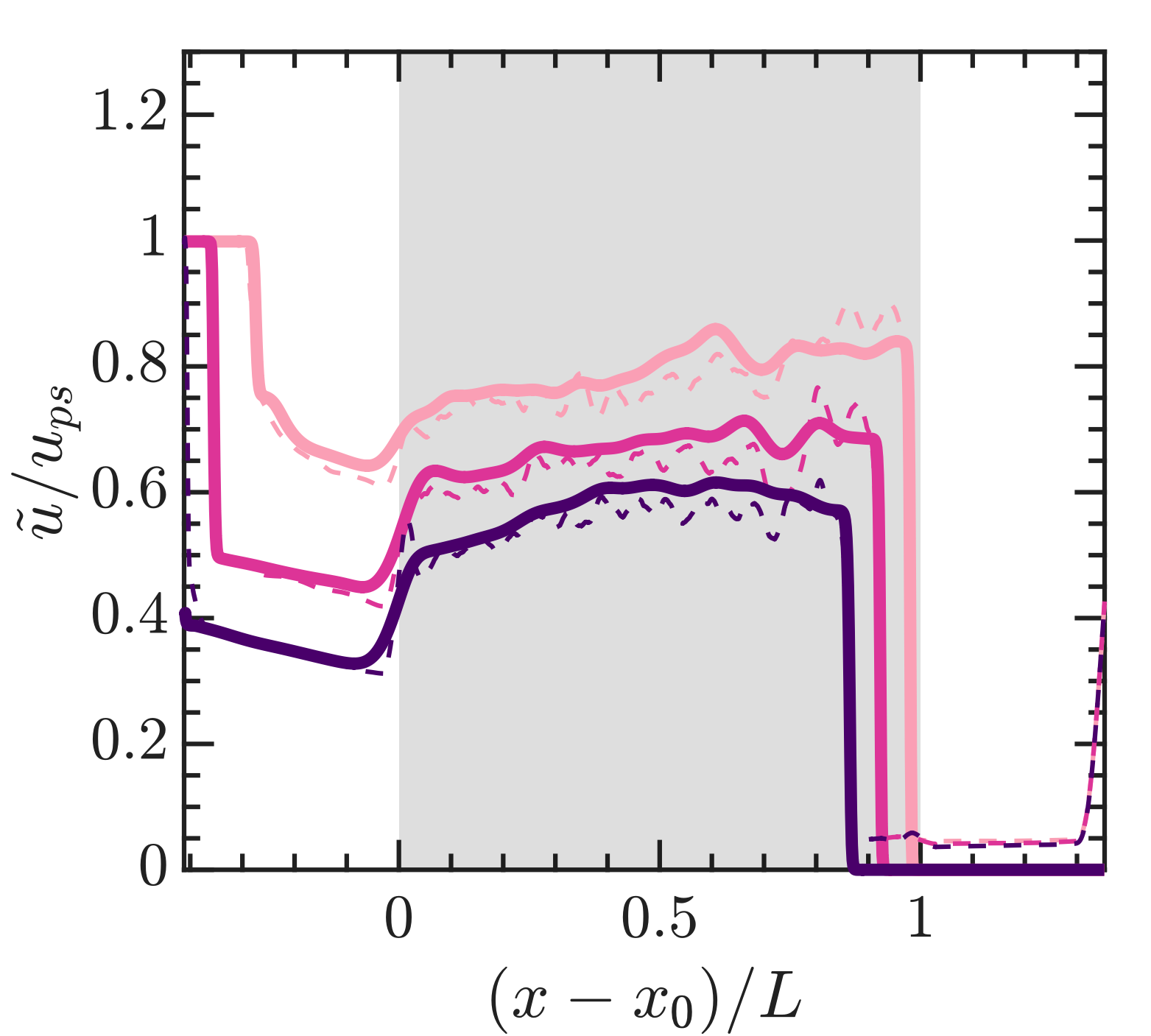}
      } 
      \subfloat{
          \includegraphics[width=0.31\textwidth]{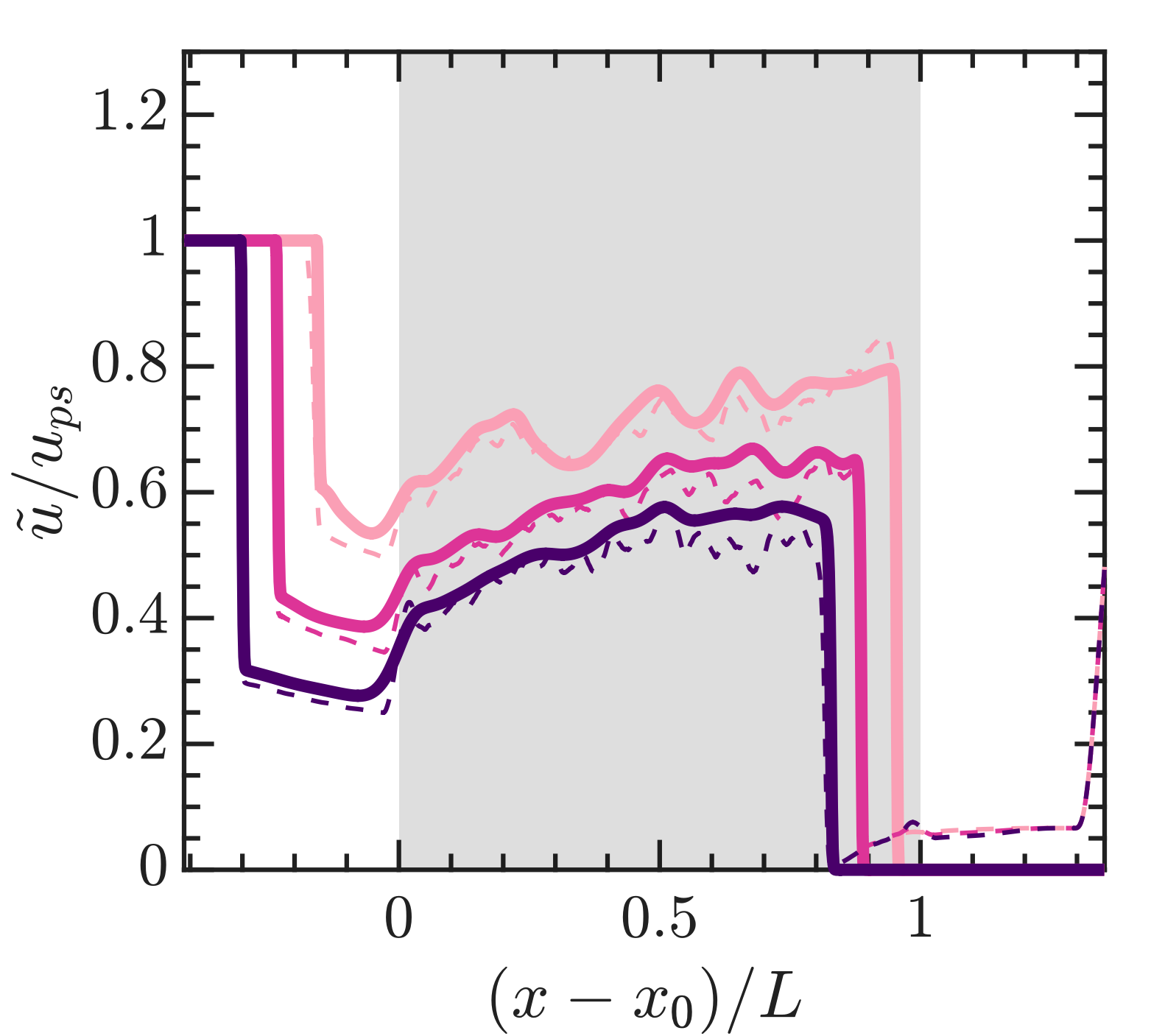}
      } \\
      \subfloat{
          \includegraphics[width=0.31\textwidth]{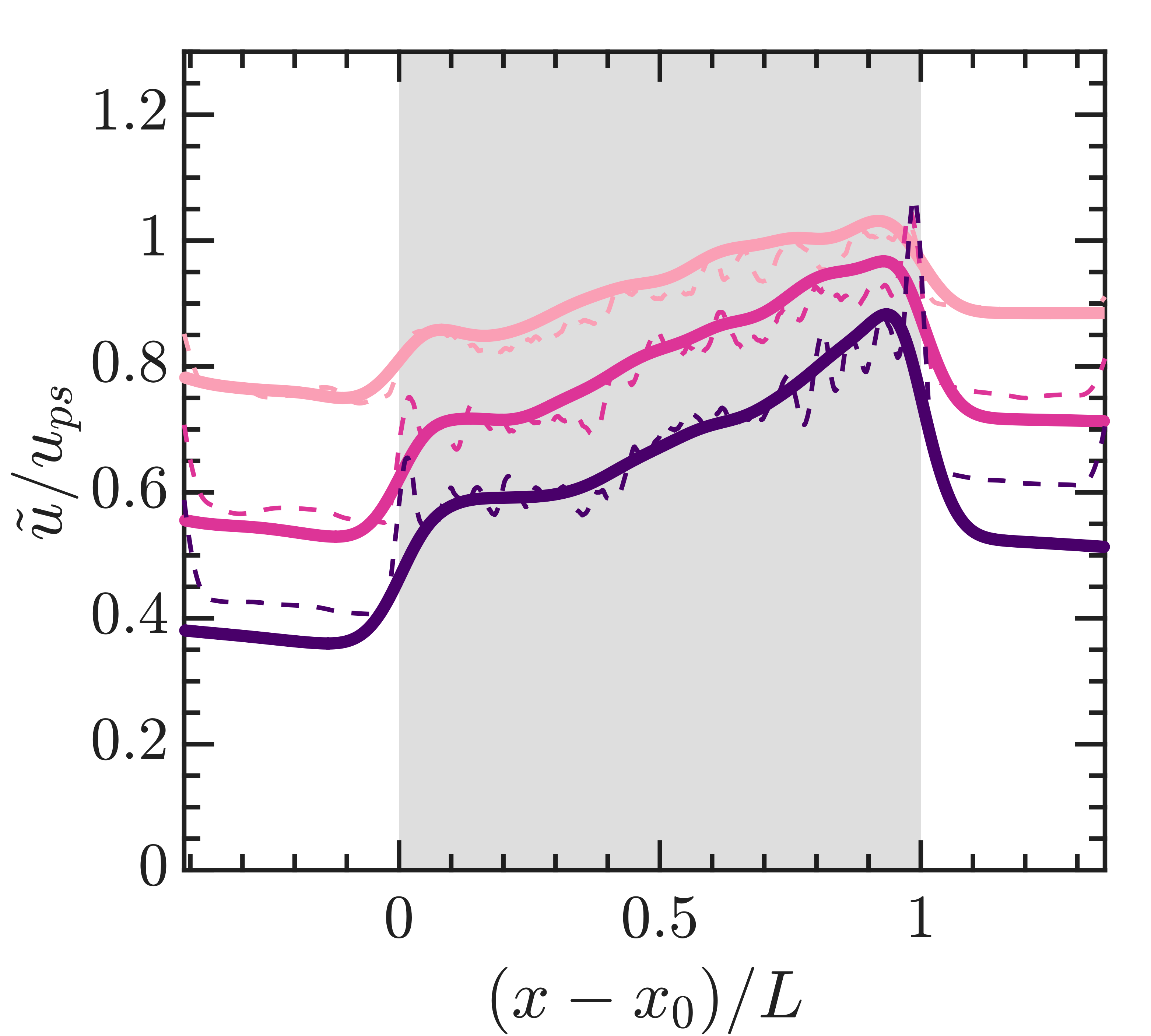}
      } 
      \subfloat{
          \includegraphics[width=0.31\textwidth]{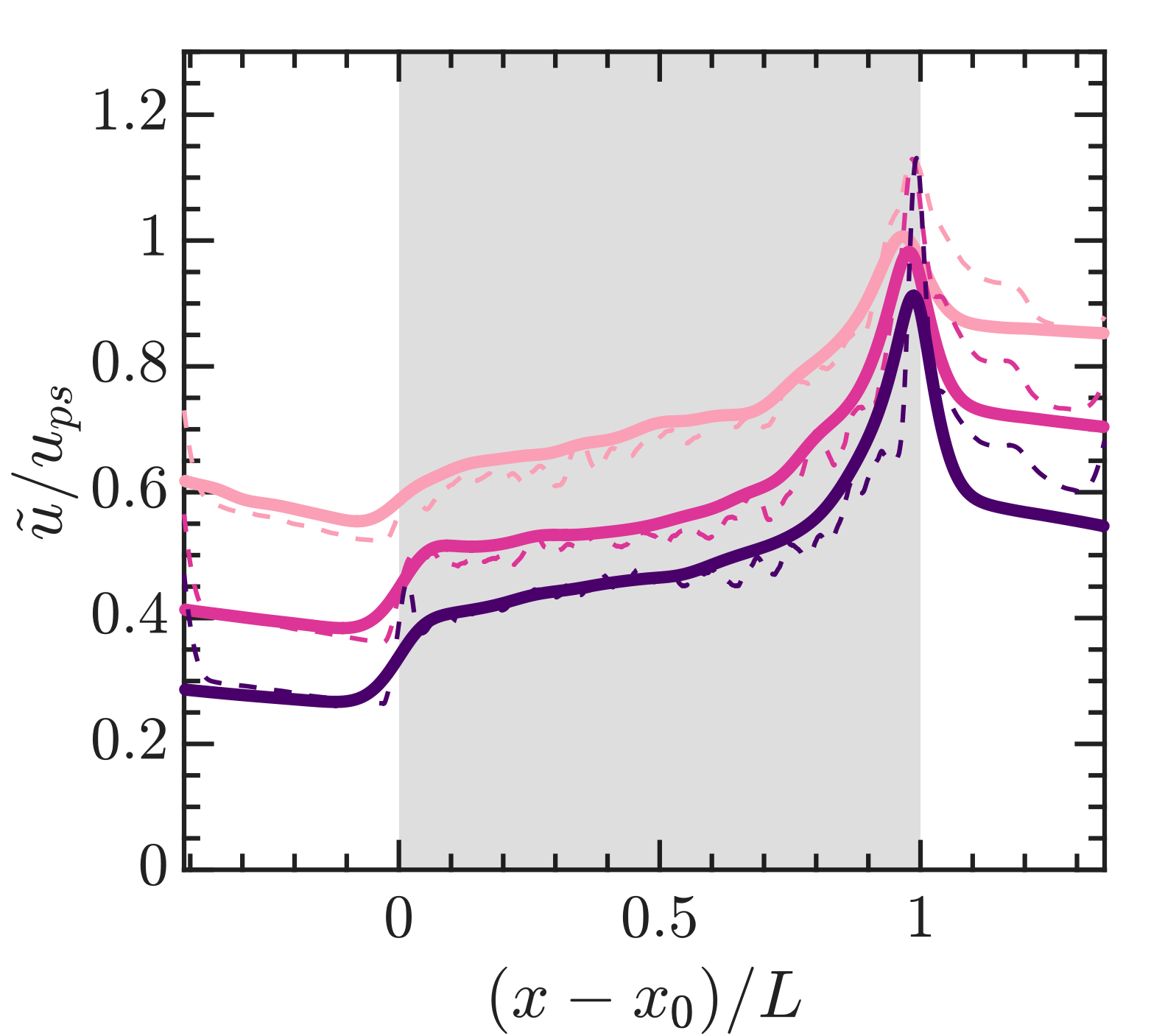}
      } 
      \subfloat{
          \includegraphics[width=0.31\textwidth]{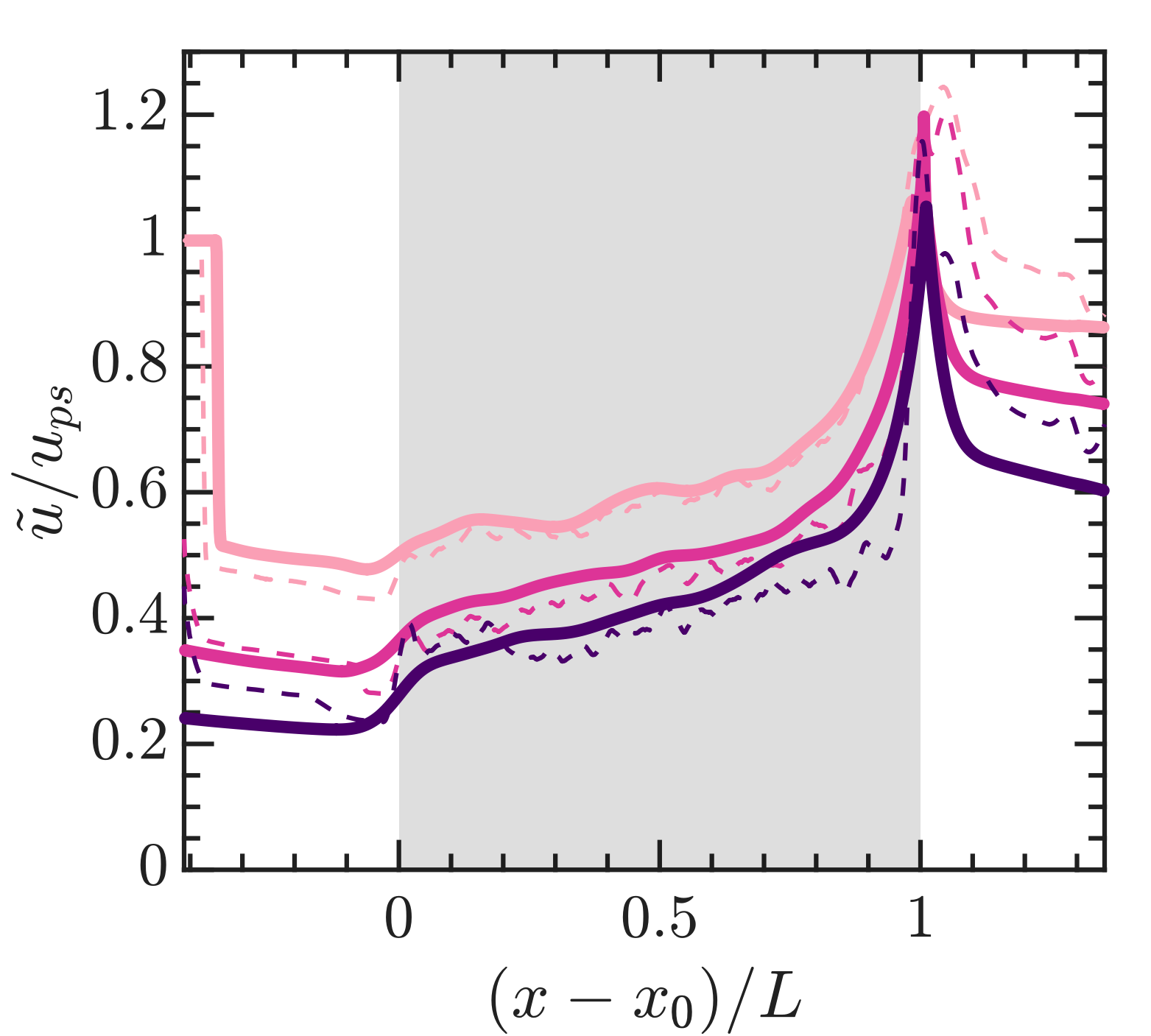}
      }
  \end{tabular}
  \begin{tikzpicture}[overlay, remember picture]
    \node at (-12,3.5) {$(a)$};
    \node at ([xshift=0.32\linewidth]-12,3.5) {$(b)$};
    \node at ([xshift=0.64\linewidth]-12,3.5) {$(c)$};
    \node at (-12,-0.4) {$(d)$};
    \node at ([xshift=0.32\linewidth]-12,-0.4) {$(e)$};
    \node at ([xshift=0.64\linewidth]-12,-0.4) {$(f)$};
    
\end{tikzpicture}
  \caption{Comparison of mean streamwise velocity from particle-resolved simulations (\dashed) with results from the two-equation model (\full). $(a,d)$ $M_s=1.2$, $(b,e)$ $M_s=1.66$, $(c,f)$ $M_s=2.1$. $t/\tau_L=1$ (top) and $t/\tau_L=2$ (bottom). The colour scheme for different volume fraction cases is the same as in figure~\ref{fig:volfrac}. }
  \label{fig:meanVelmodel}
\end{figure}
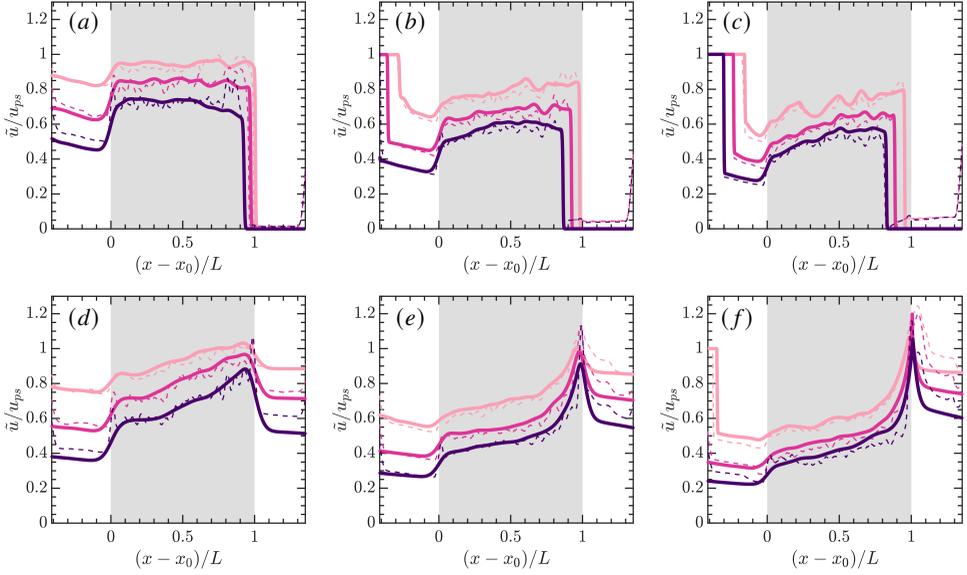

Figure~\ref{fig:meanVelmodel} shows comparisons of the mean streamwise velocity between the two-equation model and particle-resolved simulations. Overall excellent agreement is observed. The location of the transmitted and reflected shocks are predicted correctly. The model can be seen to predict choked flow at the downstream edge resulting in supersonic expansion, closely matching the particle-resolved simulations.

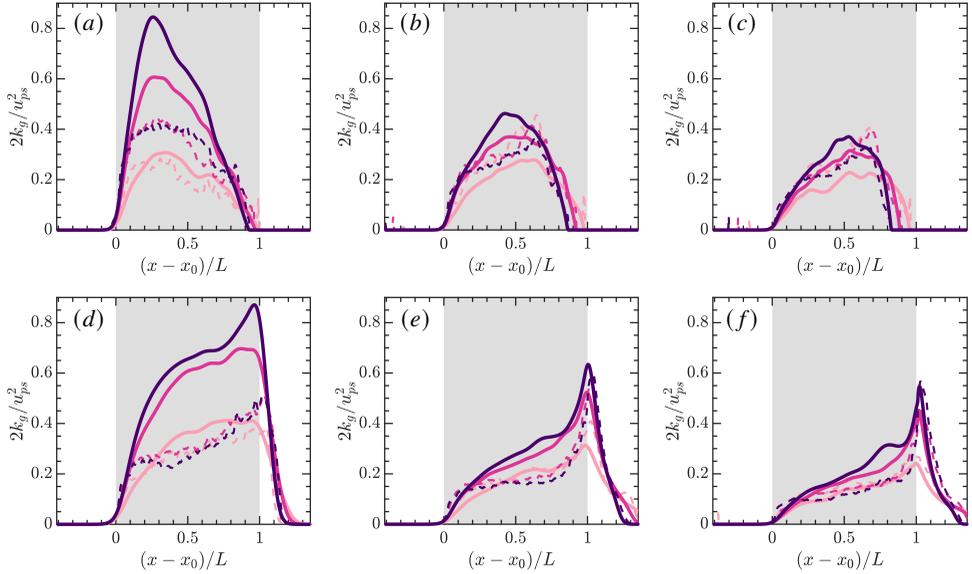
\begin{figure}

  \begin{tabular}{ccc}
      \subfloat{
          \includegraphics[width=0.31\textwidth]{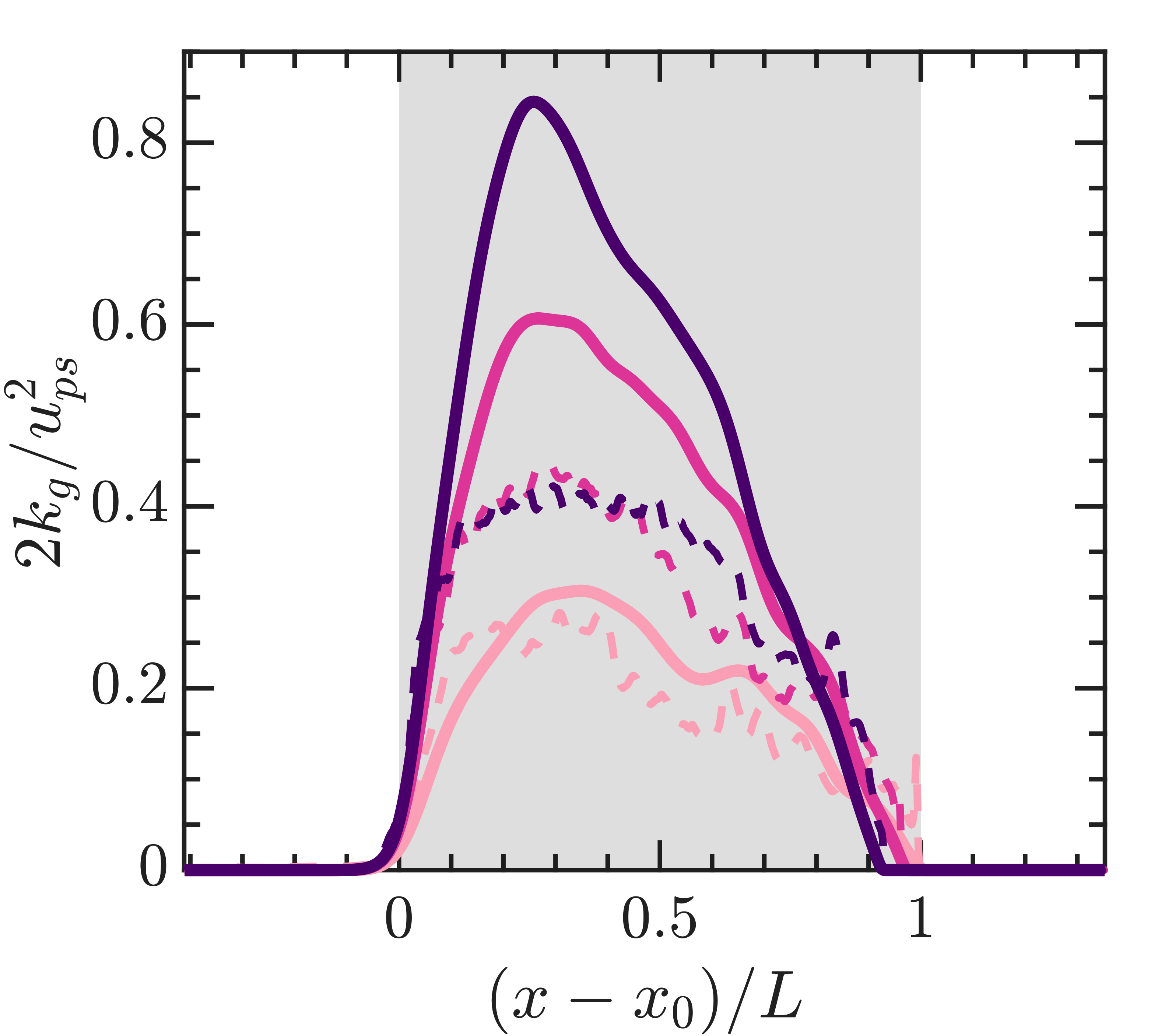}
      } 
      \subfloat{
          \includegraphics[width=0.31\textwidth]{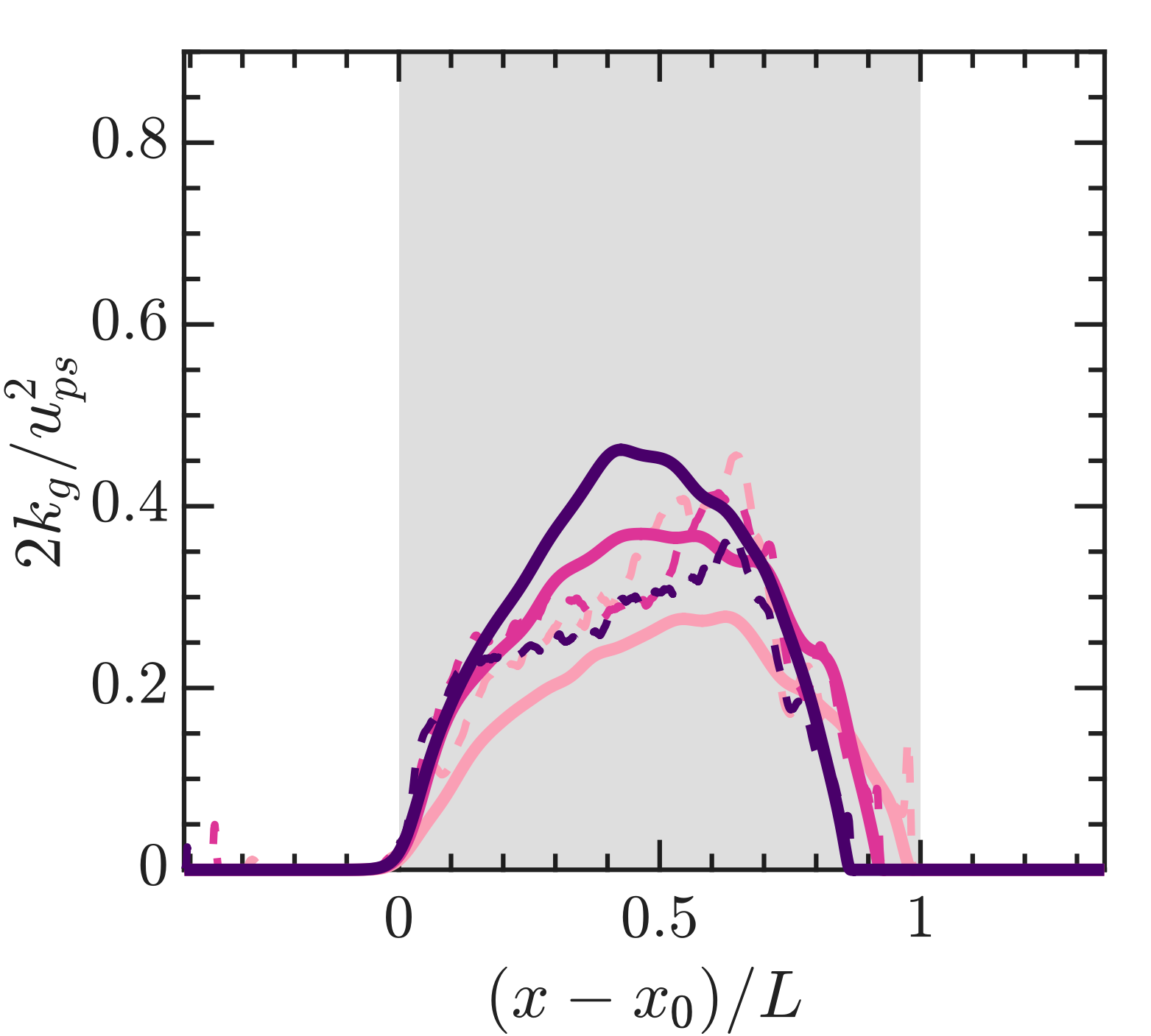}
      } 
      \subfloat{
          \includegraphics[width=0.31\textwidth]{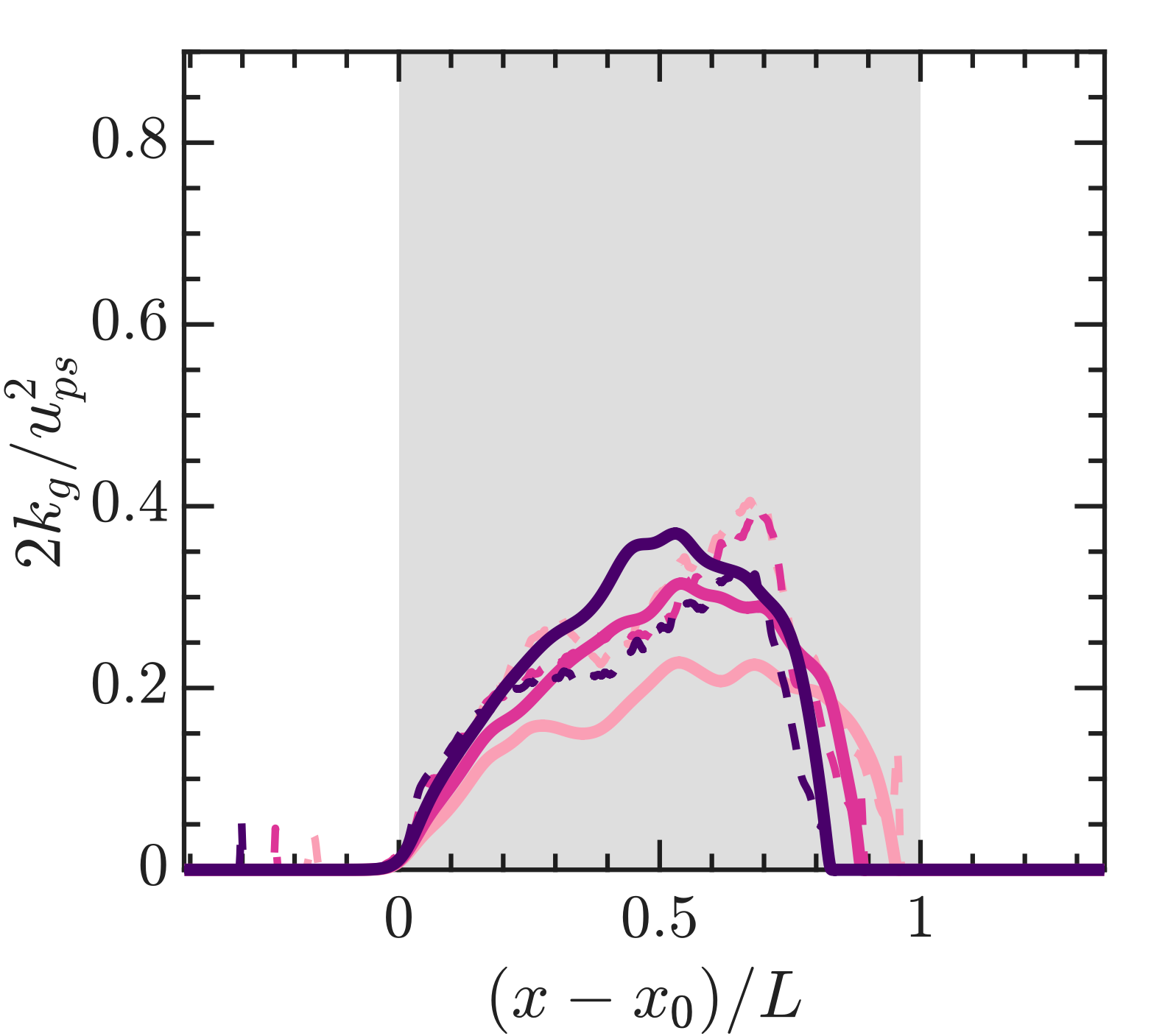}
      } \\
      \subfloat{
          \includegraphics[width=0.31\textwidth]{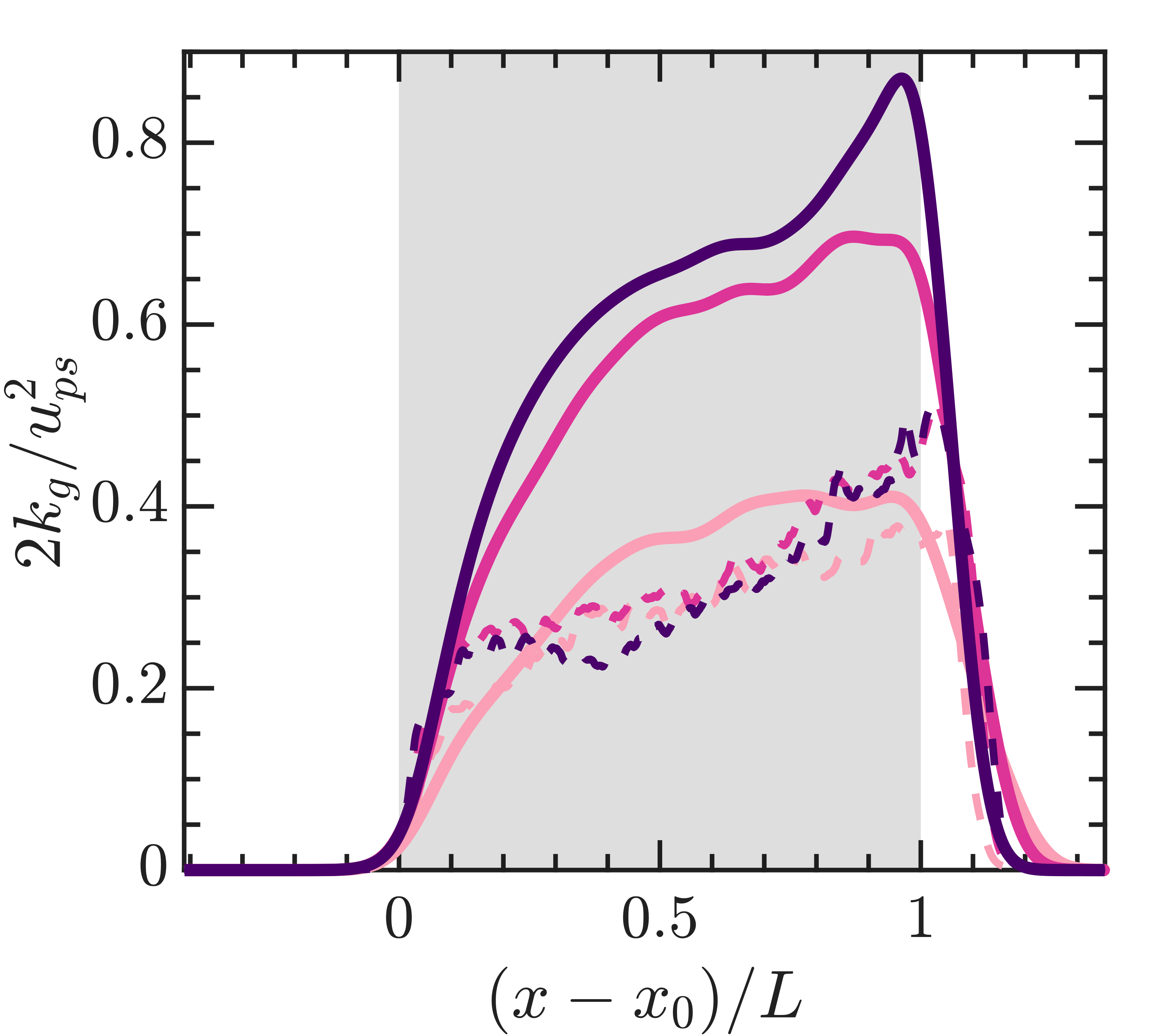}
      } 
      \subfloat{
          \includegraphics[width=0.31\textwidth]{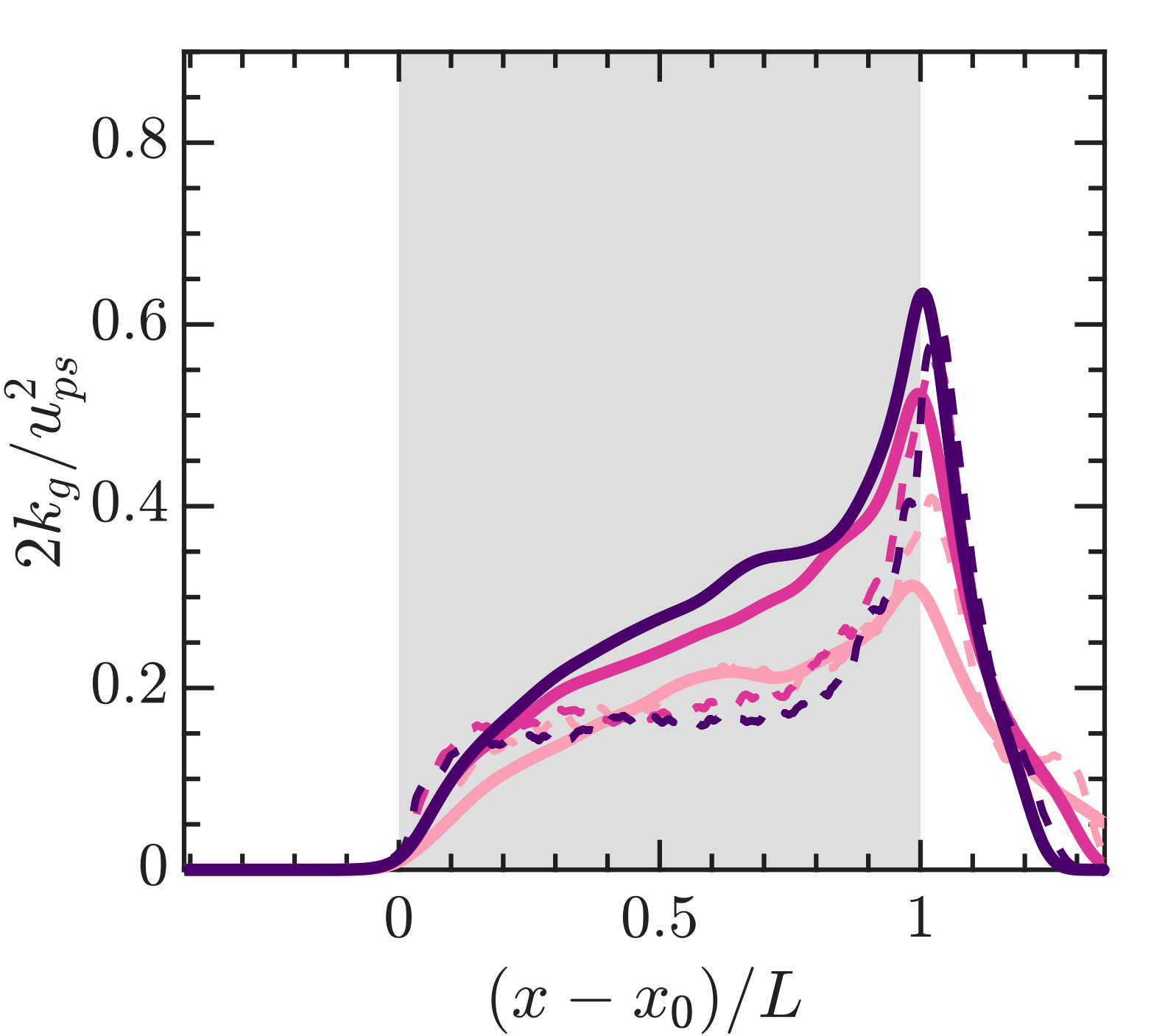}
      } 
      \subfloat{
          \includegraphics[width=0.31\textwidth]{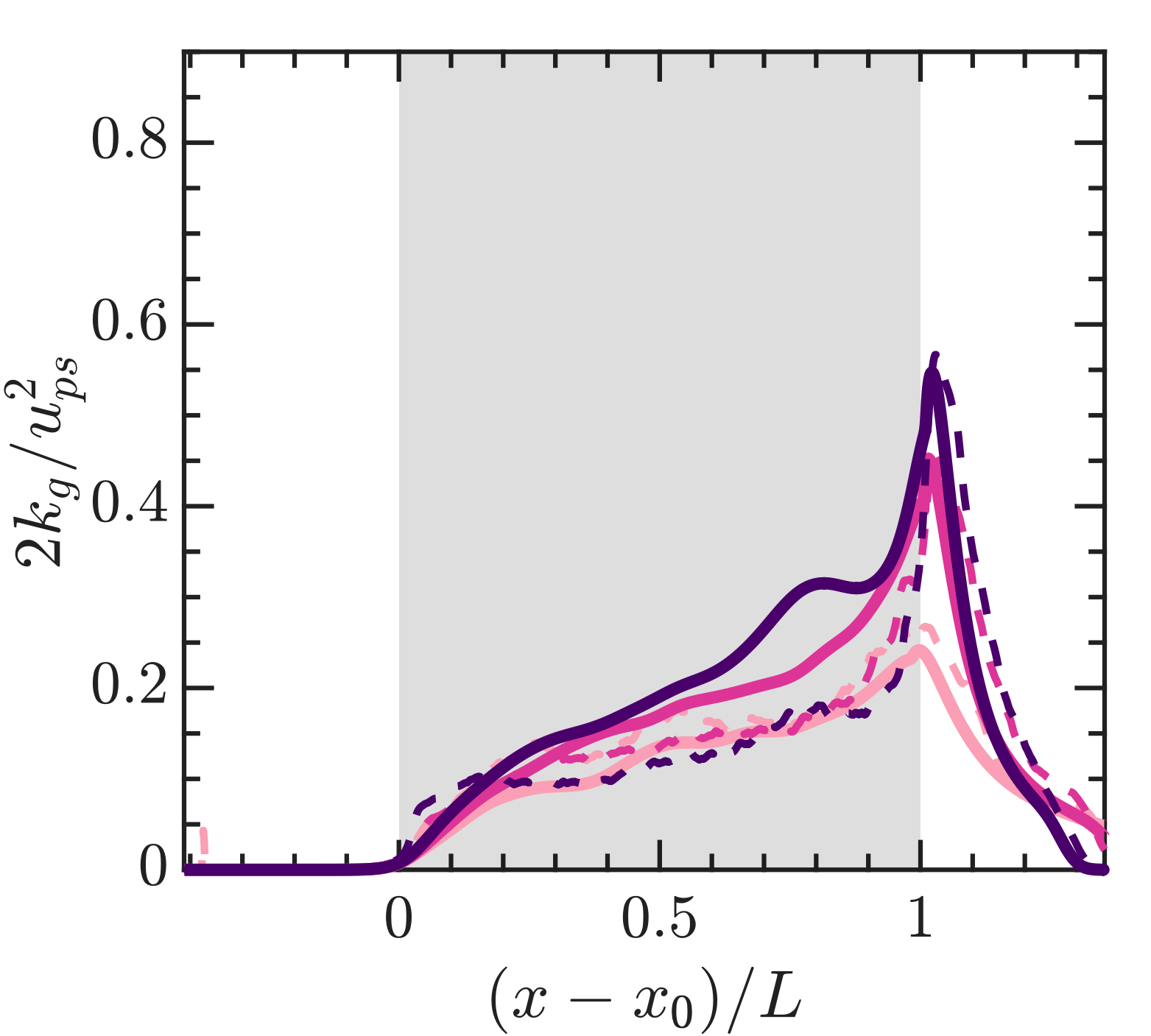}
      }
  \end{tabular}
  \begin{tikzpicture}[overlay, remember picture]
    \node at (-12,3.5) {$(a)$};
    \node at ([xshift=0.32\linewidth]-12,3.5) {$(b)$};
    \node at ([xshift=0.64\linewidth]-12,3.5) {$(c)$};
    \node at (-12,-0.4) {$(d)$};
    \node at ([xshift=0.32\linewidth]-12,-0.4) {$(e)$};
    \node at ([xshift=0.64\linewidth]-12,-0.4) {$(f)$};
    
\end{tikzpicture}
  \caption{Comparison of pseudo-turbulent kinetic energy between particle-resolved simulations (\dashed) with the two-equation model (\full). $(a,d)$ $M_s=1.2$, $(b,e)$ $M_s=1.66$, $(c,f)$ $M_s=2.1$. $t/\tau_L=1$ (top) and $t/\tau_L=2$ (bottom). The colour scheme for different volume fraction cases is the same as in figure~\ref{fig:volfrac}. }
  \label{fig:ptke_model}
\end{figure}

Figure~\ref{fig:ptke_model} shows comparisons of PTKE between the two-equation model and particle-resolved simulations at two time instances. Results show good agreement for all cases considered except for the cases with $M_s=1.2$ at higher volume fractions. The model predicts an increase in PTKE with $\Phi_p$, which is not observed in the particle-resolved simulations. Despite this, the model results show overall good agreement both within the curtain and downstream.

\begin{figure}
\centering
    \begin{tabular}{ccc}
    \subfloat{\includegraphics[width=0.31\textwidth]{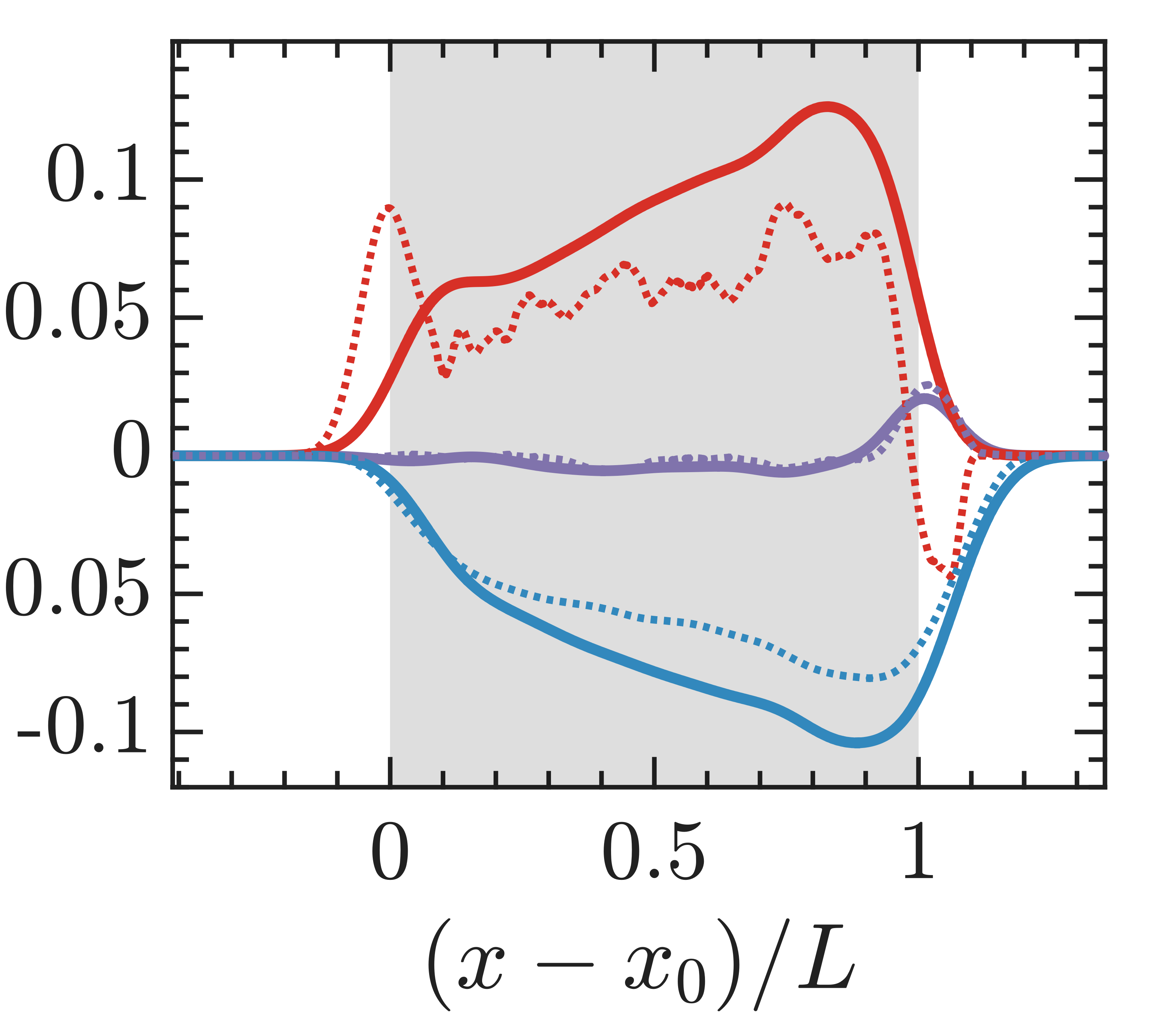}} & 
    \subfloat{\includegraphics[width=0.31\textwidth]{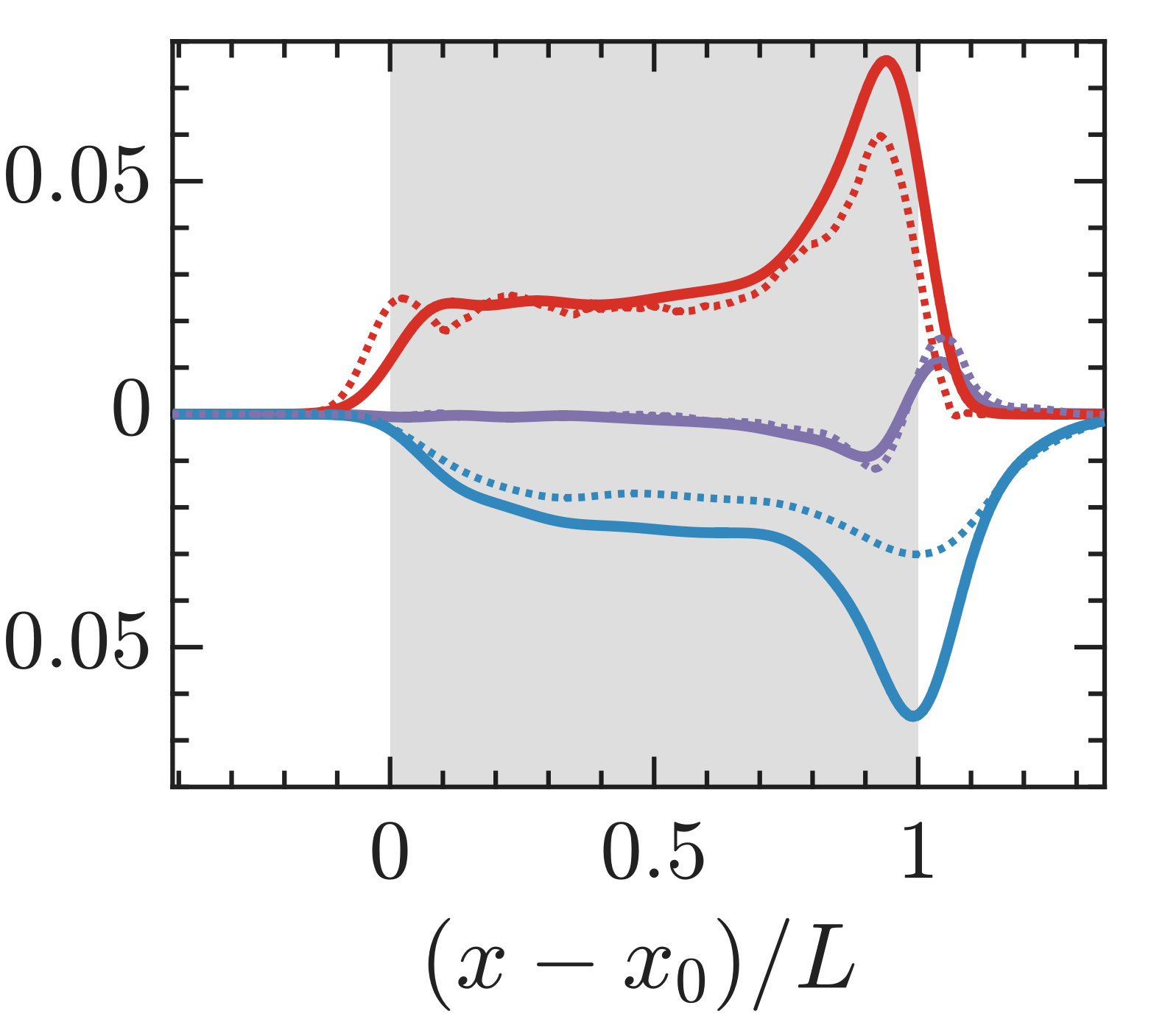}} & 
    \subfloat{\includegraphics[width=0.31\textwidth]{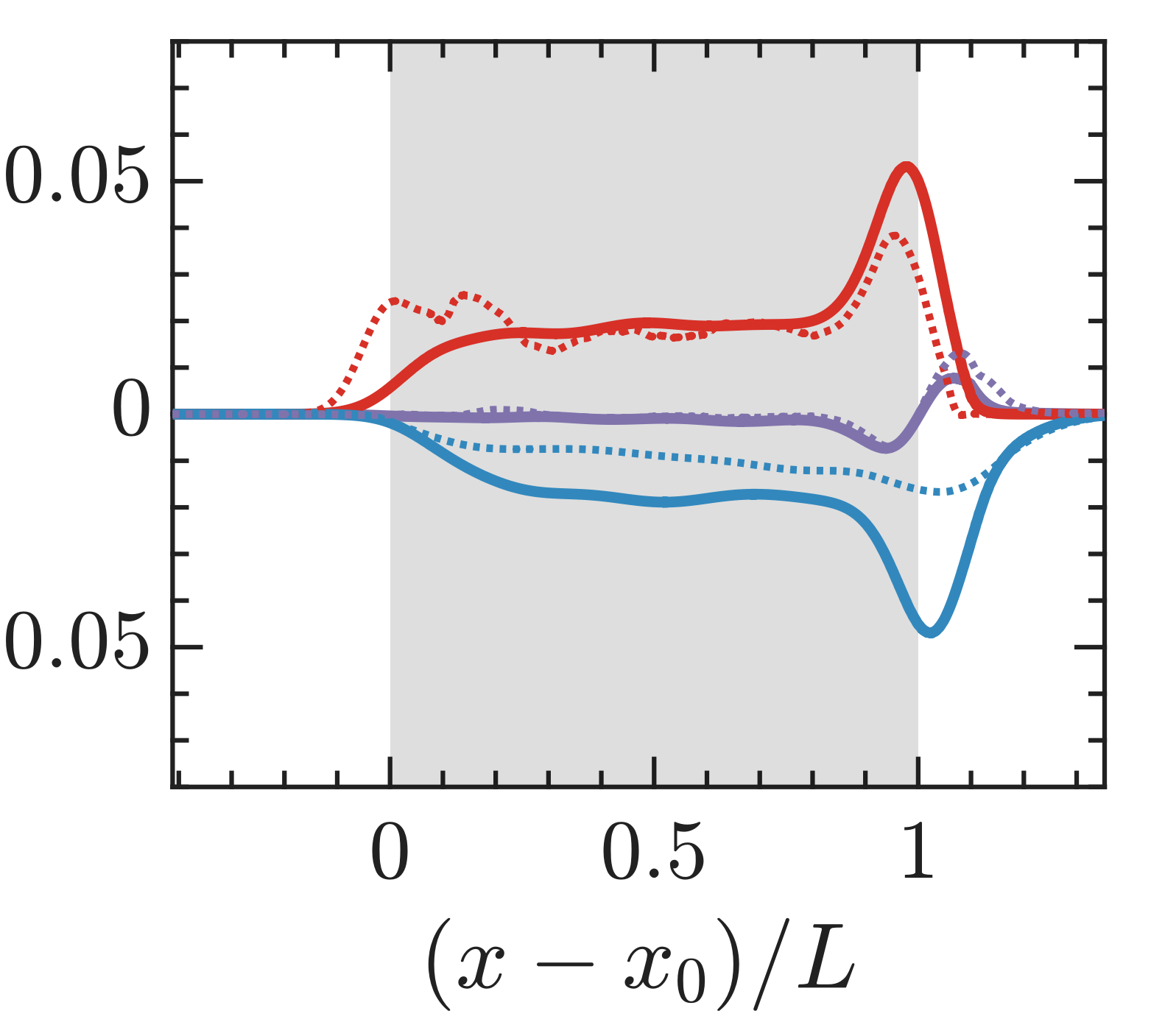}} \\ 
    \end{tabular}
    \caption{Comparison of terms in the PTKE budget between the two-equation model (\full) and particle-resolved simulations (\dotted) for $(a)$ $M_s=1.2$ and $\Phi_p=0.2$, $(b)$ $M_s=1.66$ and $\Phi_p=0.2$ and $(c)$ $M_s=2.1$ and $\Phi_p=0.3$ at $t/\tau_L=2$. $\mathcal{P}_{D}^P$ (red), $\mathcal{P}_s$ (purple) and $\alpha_g \rho_g \epsilon_g$ (blue).}
    \label{fig:budgetmodel}
    \begin{tikzpicture}[overlay, remember picture]
        \node at (-5.5,5.5) {$(a)$};
        \node at ([xshift=0.33\linewidth]-5.5,5.5) {$(b)$};
        \node at ([xshift=0.65\linewidth]-5.5,5.5) {$(c)$};
    \end{tikzpicture}
\end{figure}
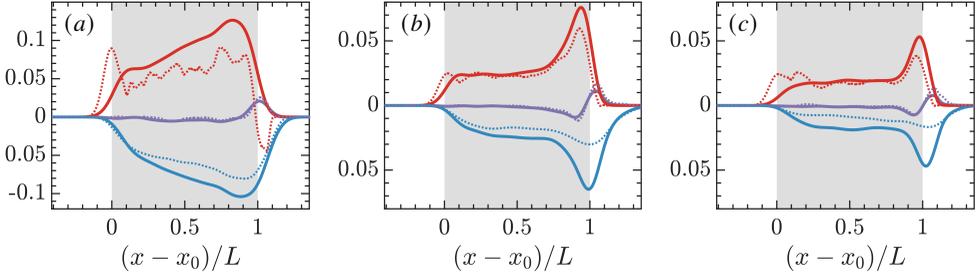

The terms in the PTKE budget computed from the two-equation model are compared with particle-resolved simulation data to identify and explain the observed discrepancies in PTKE. Specifically, the dominant terms--drag production $\mathcal{P}_D$, viscous dissipation $\alpha_g \rho \epsilon$, and mean-shear production $\mathcal{P}_s$--are examined. Figure~\ref{fig:budgetmodel} presents the comparison for one case, with similar results observed across all cases. Overall, excellent agreement is found over the three time instances shown. The largest discrepancies occur at the upstream and downstream edges of the curtain, where the particle-resolved simulations predict a sharper increase in drag production at the upstream edge and greater dissipation at the downstream edge. These differences may be attributed to numerical diffusion in the coarse-grained model.

\begin{figure}
  \centering
  \begin{tabular}{ccc}
      \subfloat{
          \includegraphics[width=0.45\textwidth]{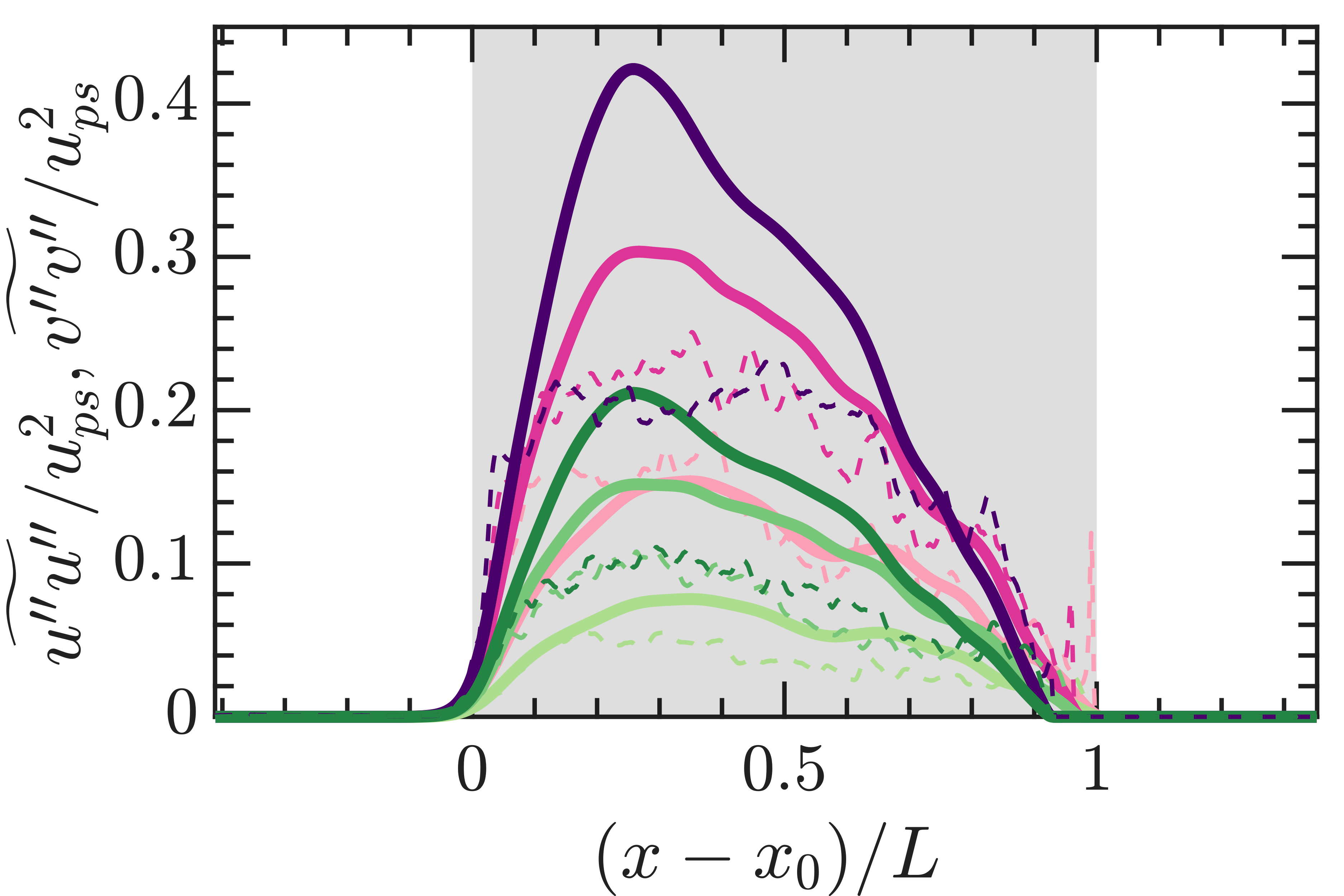}
      } &
      \subfloat{
          \includegraphics[width=0.45\textwidth]{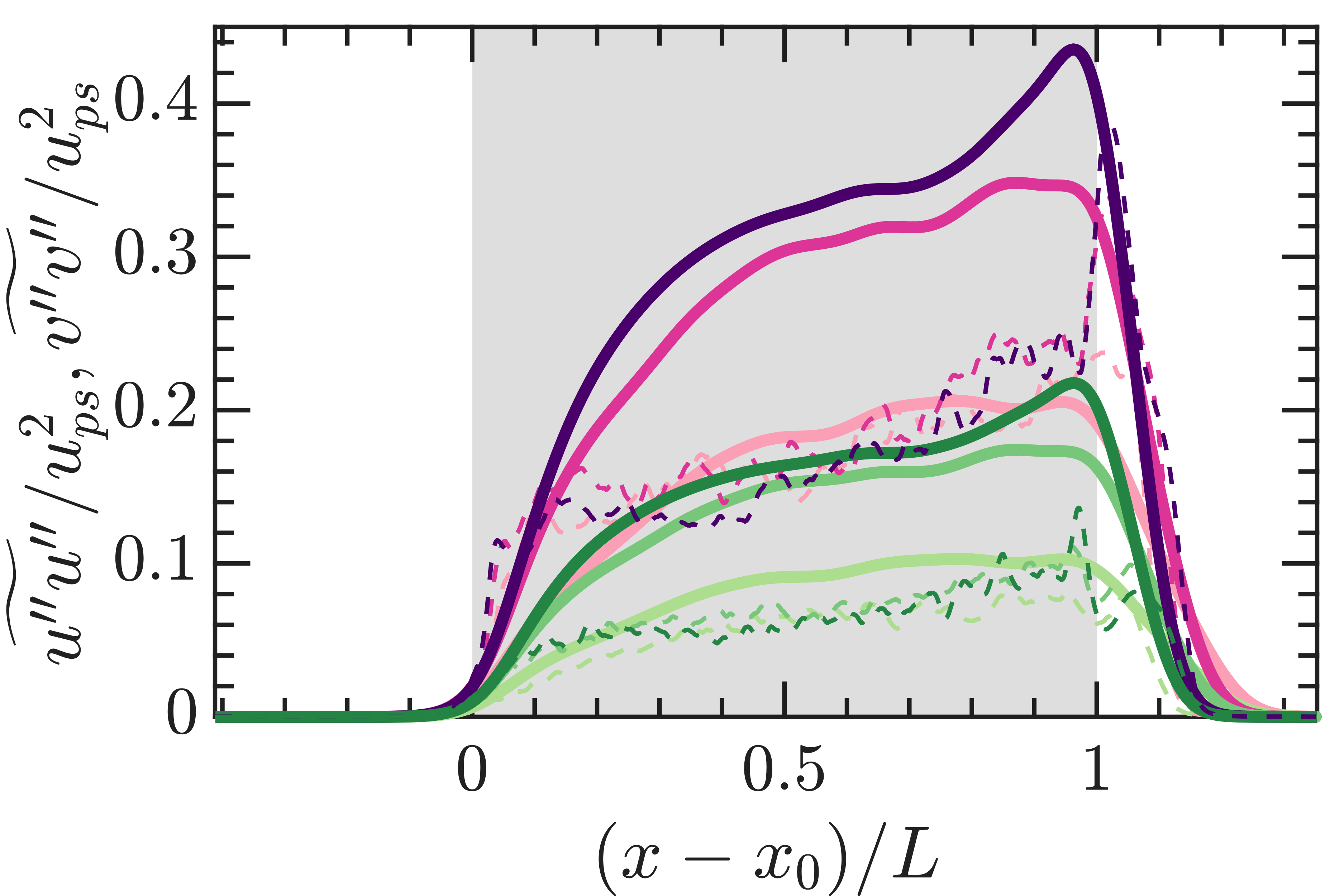}
      } \\
      \subfloat{
          \includegraphics[width=0.45\textwidth]{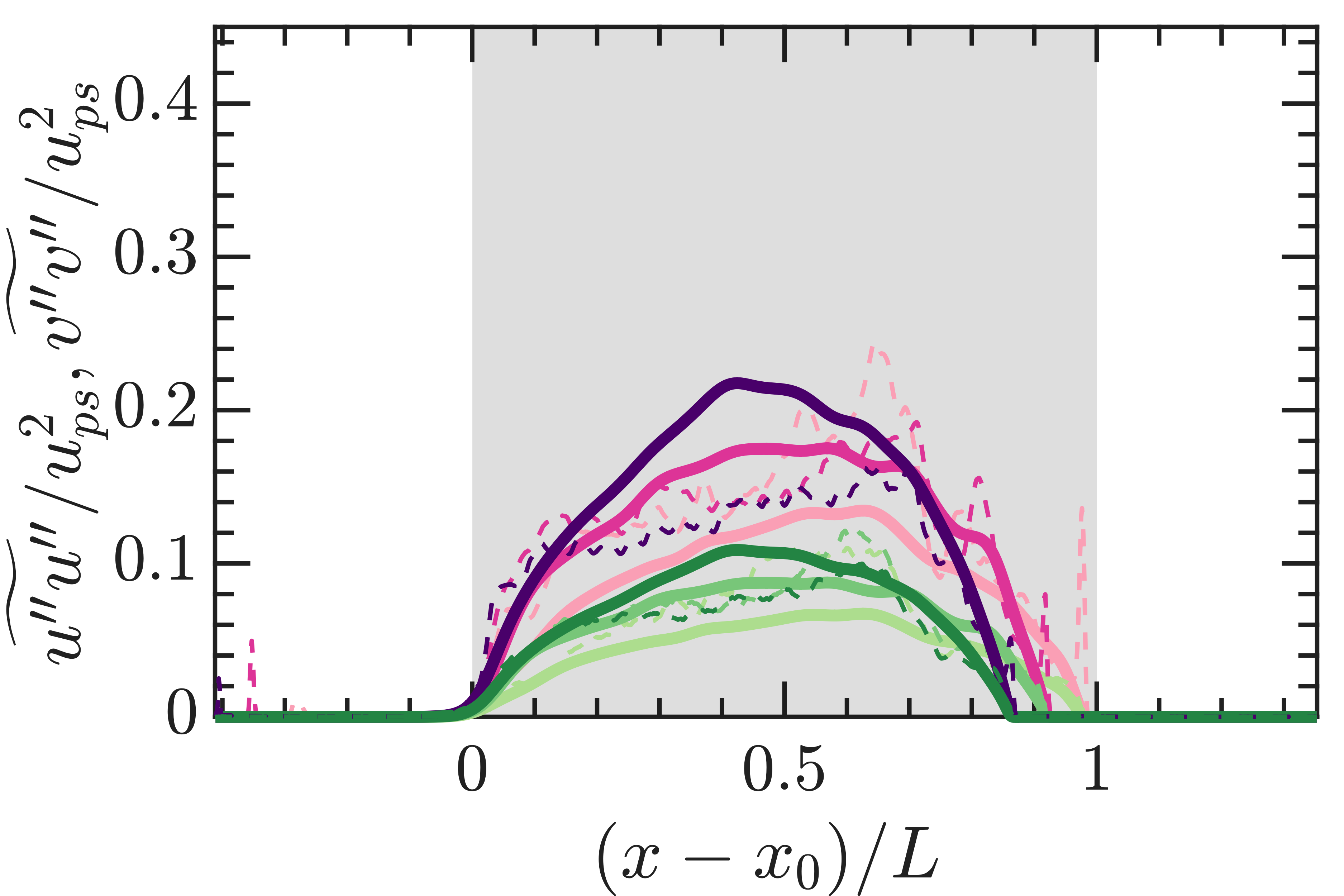}
      } &
      \subfloat{
          \includegraphics[width=0.45\textwidth]{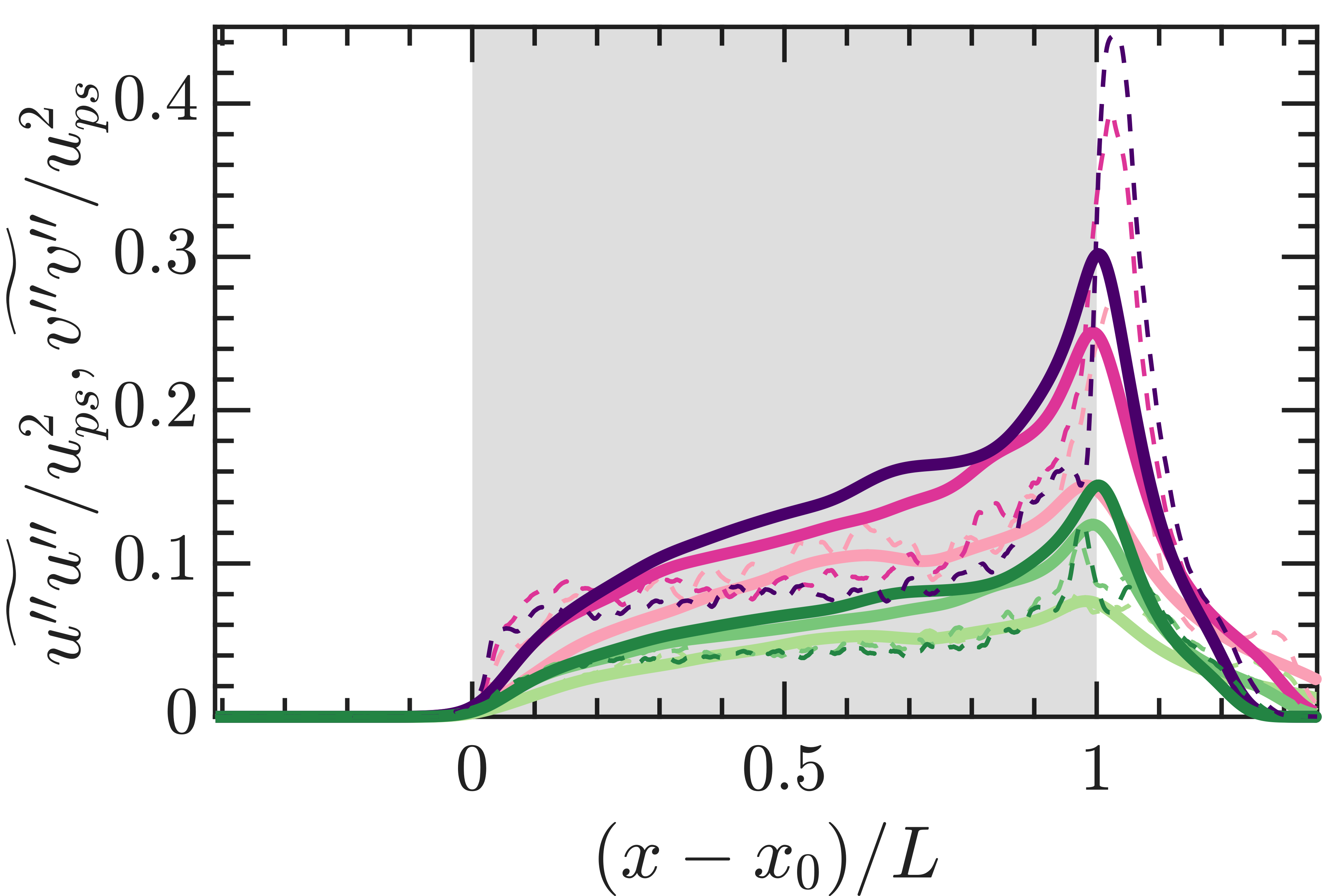}
      } \\
      \subfloat{
          \includegraphics[width=0.45\textwidth]{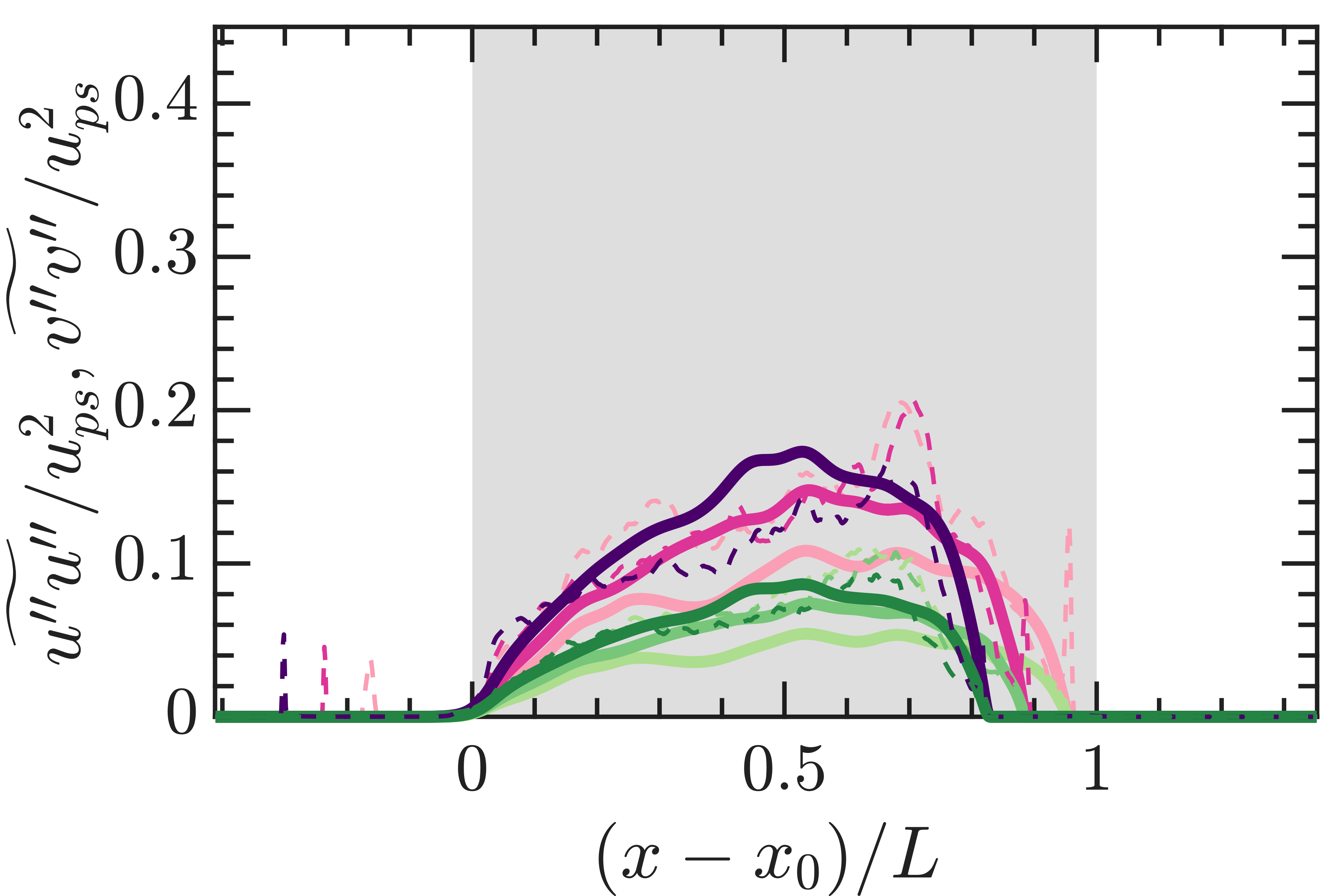}
      } &
      \subfloat{
          \includegraphics[width=0.45\textwidth]{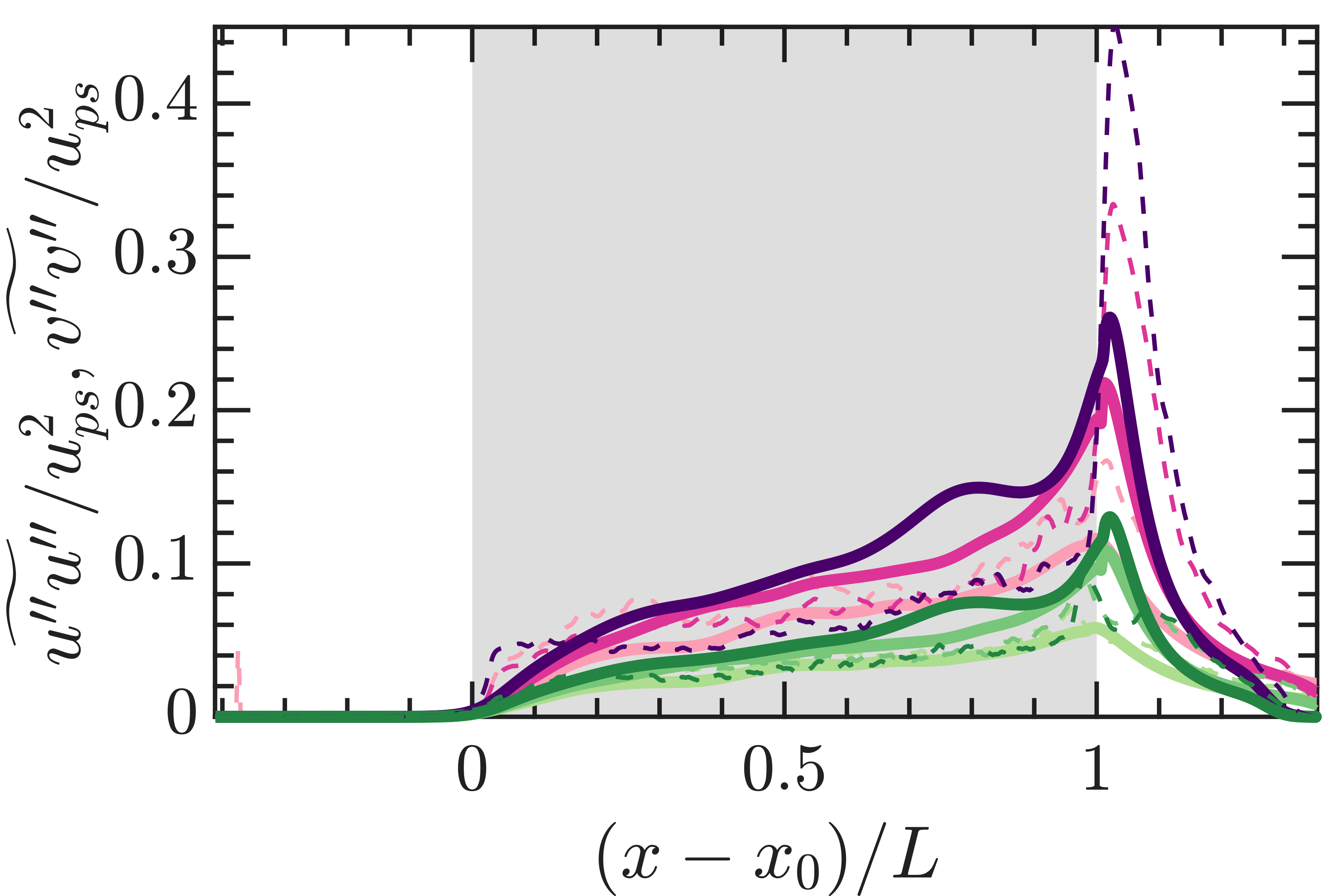}
      }\\
  \end{tabular}
    \begin{tikzpicture}[overlay, remember picture]
          \node at (-11.3,6.) {$(a)$};
          \node at ([xshift=0.48\linewidth]-11.3,6) {$(b)$};
          \node at (-11.3,1.8) {$(c)$};
          \node at ([xshift=0.48\linewidth]-11.3,1.8) {$(d)$};
          \node at (-11.3,-2.4) {$(e)$};
          \node at ([xshift=0.48\linewidth]-11.3,-2.4) {$(f)$};
    \end{tikzpicture}
  \caption{Comparison of pseudo-turbulent Reynolds stresses between the particle-resolved simulations (\dashed) and the model (\full). $(a,d)$ $M_s=1.2$, $(b,e)$ $M_s=1.66$, $(c,f)$ $M_s=2.1$. Colour scheme defined in figure~\ref{fig:uuvv}.}
  \label{fig:uuvvmodel}
\end{figure}
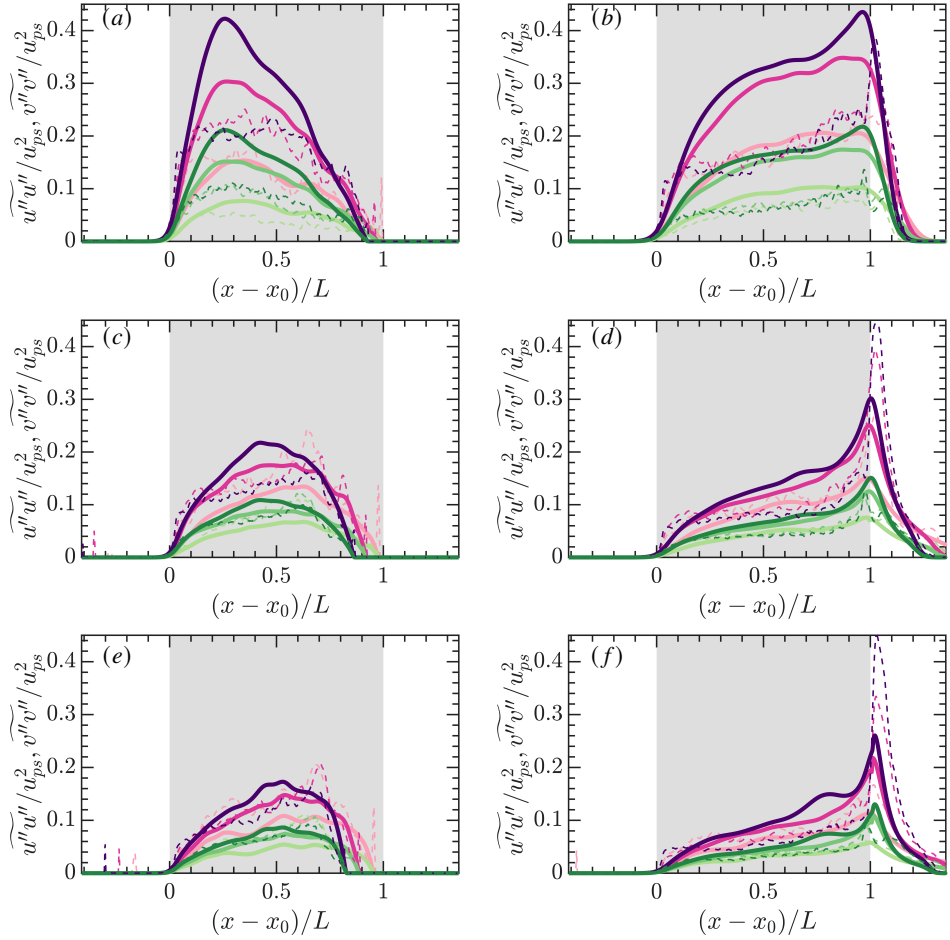

The streamwise and spanwise fluctuations are reconstructed using \eqref{eq:Rij} and compared with particle-resolved simulations in figure~\ref{fig:uuvvmodel}. Overall, the results show good agreement, with cases $2$ and $3$ exhibiting the most discrepancies. These discrepancies may arise from the drag model, the choice of $C_{\epsilon,D}$, or the omission of the viscous term in the two-fluid model. At the downstream edge, streamwise fluctuations are slightly underpredicted, likely due to an overprediction of viscous dissipation, as observed in the previous figure. This overprediction is ultimately linked to the choice of $C_{\epsilon,D}$ or the drag model. Despite these issues, the two-equation model predicts the overall behaviour well, including the PTKE downstream of the curtain, in the pure gas.

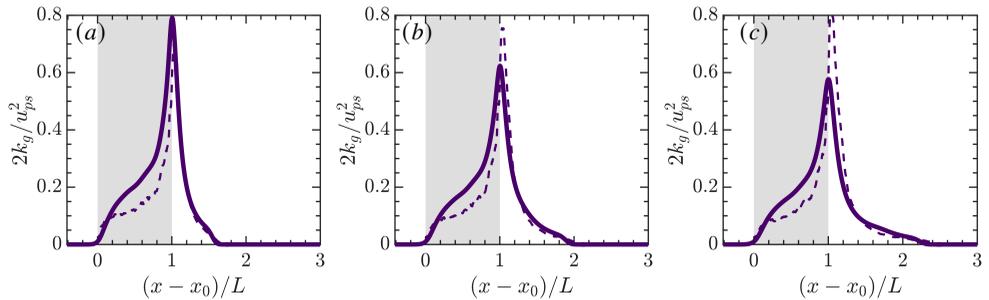
\begin{figure}
  
  \begin{tabular}{ccc}
      \subfloat{
          \includegraphics[width=0.31\textwidth]{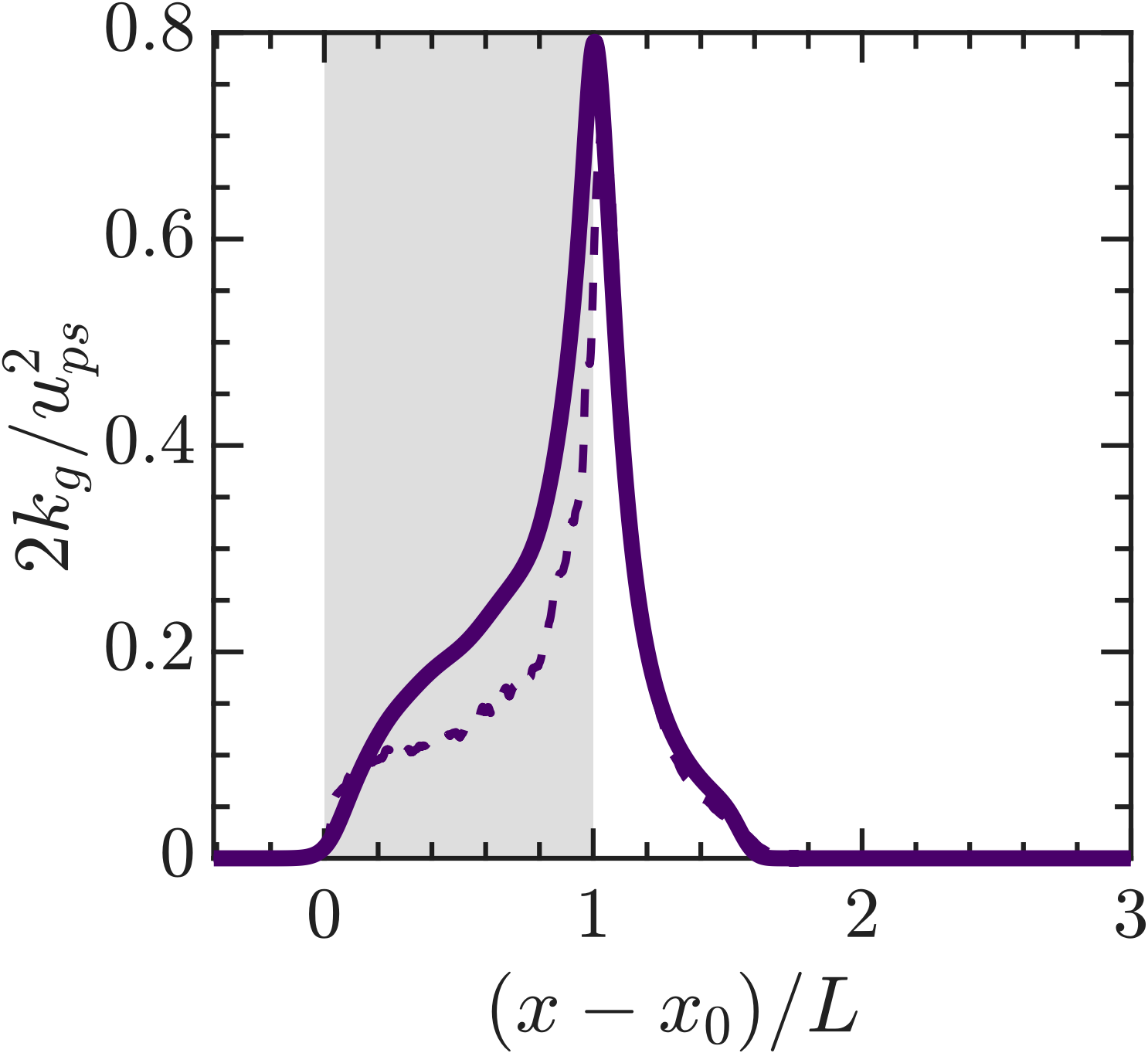}
      } 
      \subfloat{
          \includegraphics[width=0.31\textwidth]{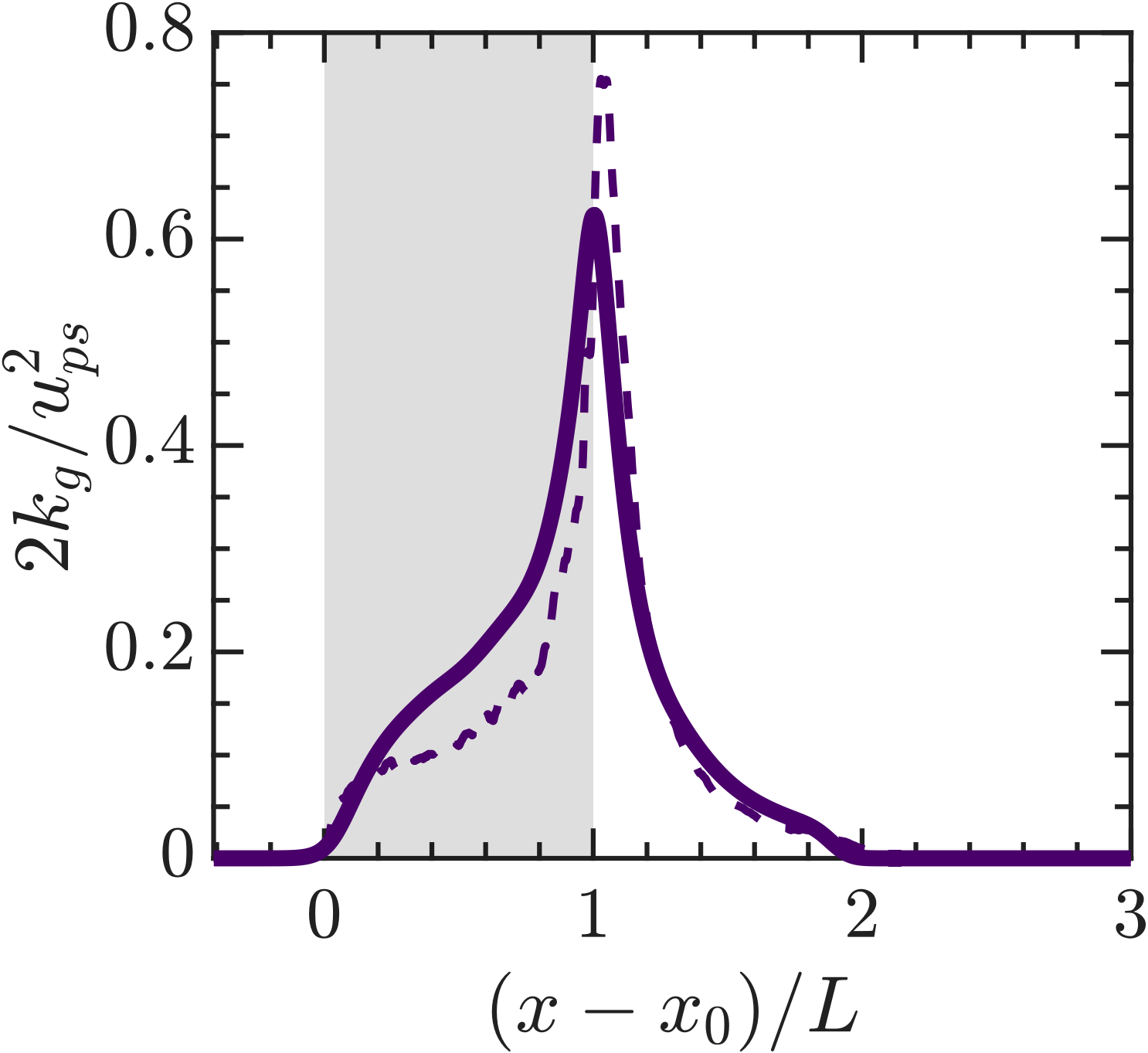}
      } 
      \subfloat{
          \includegraphics[width=0.31\textwidth]{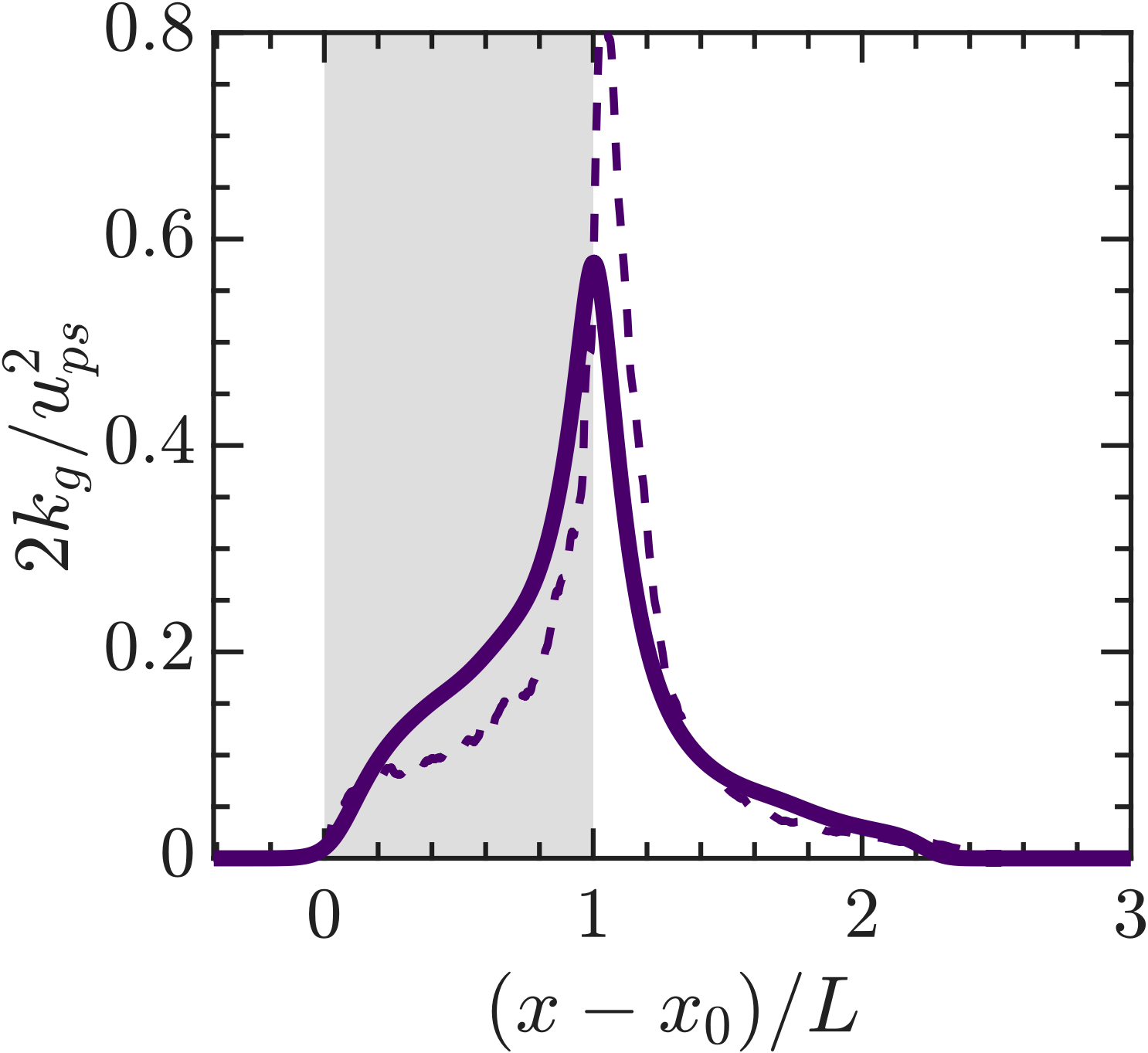}
      }\\      
  \end{tabular}
\begin{tikzpicture}[overlay, remember picture]
        \node at (-12,1.7) {$(a)$};
        \node at ([xshift=0.32\linewidth]-12,1.7) {$(b)$};
        \node at ([xshift=0.65\linewidth]-12,1.7) {$(c)$};
    \end{tikzpicture}
  \caption{Comparison of pseudo-turbulent kinetic energy of the longer domain (case 10) between particle-resolved simulations (\dashed) and the two-equation model (\full) at $(a)$ $t/\tau_L = 3$, $(b)$ $t/\tau_L = 4$ and $(c)$ $t/\tau_L = 5$.}
  \label{fig:ptkelong}
\end{figure}

The gas-phase turbulence downstream of the curtain lacks any production terms and, according to the budget, should only advect and diffuse. The cases considered so far extend only $6D$ from the downstream curtain edge to the right domain boundary. Here, we examine Case 10 from table~\ref{tab:kd}, with $M_s=1.66$, $\Phi_p=0.3$, and a domain extending $34D$ ($2L$) downstream. Figure~\ref{fig:ptkelong} shows PTKE comparisons after the flow reaches a statistically stationary state. The PTKE decay resembles grid-generated turbulence, and the model captures the turbulence transport and decay well.


\section{Conclusions}\label{sec:conclusions}
When a shock wave interacts with a suspension of solid particles, momentum and energy exchanges between the phases give rise to complex flow. Particle wakes induced by the transmitted shock generate velocity fluctuations referred to as `pseudo-turbulence.' Phase-averaging the viscous compressible Navier--Stokes equations reveals a route for turbulence generation through drag production within the particle curtain and localized mean-shear production at the edge of the curtain. This turbulence generation is balanced by viscous and dilatational dissipation.

Three-dimensional particle-resolved simulations of planar shocks interacting with stationary spherical particles were used to analyse the characteristics of pseudo-turbulence for a range of shock Mach numbers and particle volume fractions. In each case, pseudo-turbulent kinetic energy (PTKE) is generated through interphase drag coupling, contributing to $20-50\%$ of the post-shock kinetic energy. The abrupt change in volume fraction at the downstream edge of the curtain chokes the flow, resulting in supersonic expansion where PTKE is maximum. The pseudo-turbulent Reynolds stress is highly anisotropic but approximately constant throughout for the range of volume fractions and Mach numbers considered. The energy spectra of the streamwise gas-phase velocity fluctuations reveal an inertial subrange that begins at the mean interparticle spacing and decays with a $-5/3$ power law then steepens to $-3$ at smaller scales. This $-3$ scaling only exists in the solenoidal component of the velocity field and is attributed to particle wakes.

A one-dimensional two-equation turbulence model was formulated for PTKE and its dissipation and implemented within a hyperbolic two-fluid framework. Drag production is closed using a drag coefficient that takes into account local volume fraction, Reynolds number and Mach number. A new closure is proposed for drag dissipation that ensures the proper amount of PTKE is obtained in the limit of statistically stationary and homogeneous flow. An a-posteriori analysis demonstrated the ability of the model to predict PTKE accurately during shock-particle interactions and capture flow-choking behaviour. Such a turbulence model can be adopted into Eulerian two-fluid models or Eulerian--Lagrangian frameworks.

\appendix

\section{Convergence studies}\label{appA}
This section quantifies the effects of varying domain size and particle configurations within the curtain in particle-resolved simulations. A grid refinement study of the numerical solver for periodic compressible flow over a homogeneous suspension is detailed in our previous work \citep{khalloufi2023drag}.

\subsection{Effect of domain size}
In this section, we examine the effects of varying the domain size in the periodic spanwise ($y$ and $z$) directions. A series of three-dimensional simulations were performed with $M_s=1.66$ and $\Phi_p=0.2$ to evaluate the impact of domain size on the individual terms in the PTKE budget. Table~\ref{tab:dom} summarizes the cases considered. The streamwise domain length $L_x$ is kept constant, while the spanwise dimensions $L_y$ and $L_z$ are varied. Uniform grid spacing is maintained at $\Delta=D/40$. 

\begin{figure}
    \centering
    \includegraphics[width=0.55\linewidth]{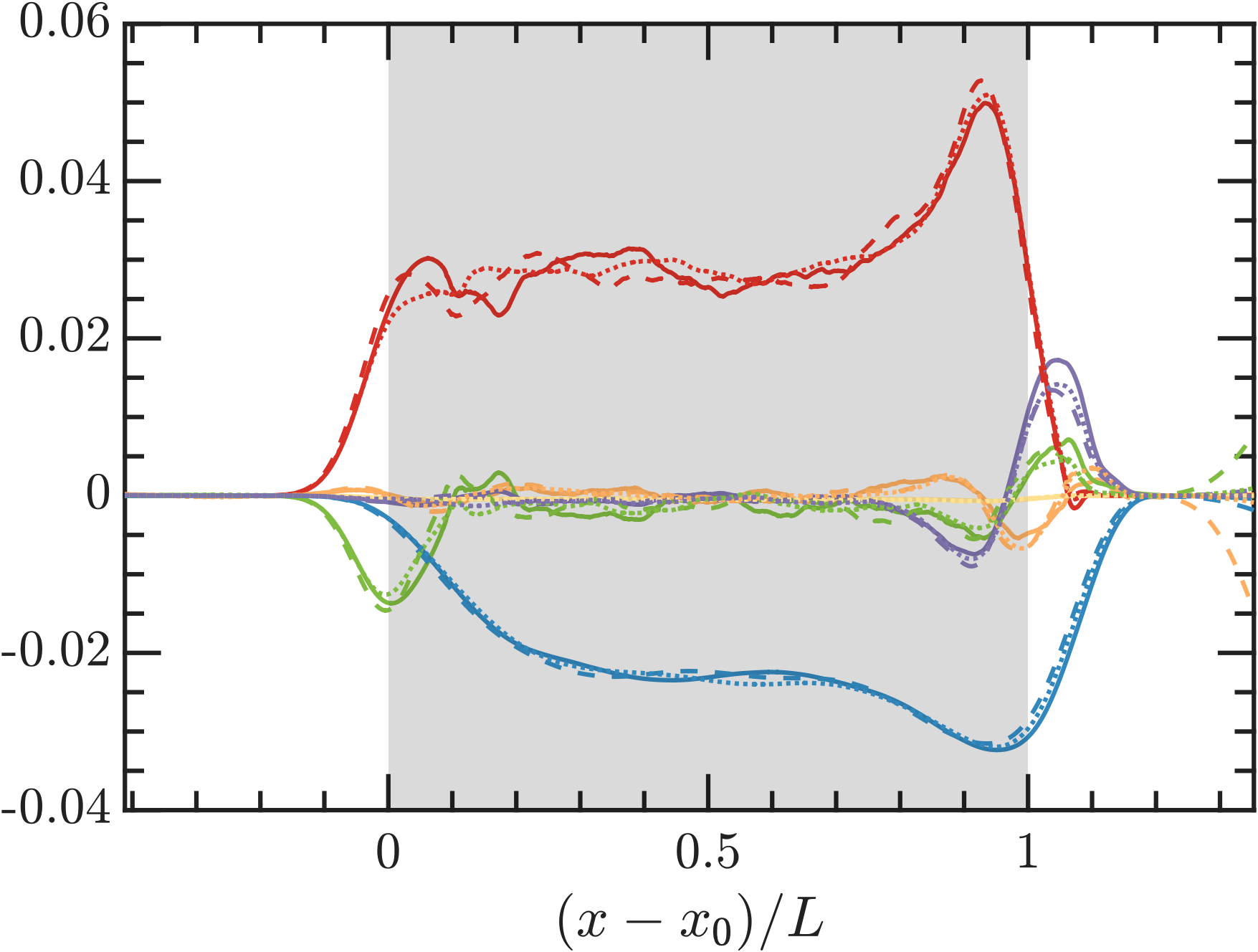}
    \caption{Effect of domain size on the PTKE budget for $M_s=1.66$ and $\Phi_p=0.2$ at $t/\tau_L=1.5$. Case A (\full), Case B (\dashed), Case C (\dotted). Colours correspond to figure~\ref{fig:budget}.}
    \label{fig:dom-conv}
\end{figure}

Figure~\ref{fig:dom-conv} presents comparisons of the individual PTKE budget terms. The results indicate that variations in the periodic domain lengths have minimal influence on the budget terms, suggesting that volume-averaging over two-dimensional $y-z$ slices can be performed without significantly affecting the one-dimensional statistics. Consequently, for the case studies presented in the main paper, we adopt $L_y=L_z=12D$.

\begin{table}
    \begin{center}
  \def~{\hphantom{0}}
    \begin{tabular}{cccccc}
        Case  & ~$L_x/D$~ & ~$L_y/D$~ & ~$L_z/D$~ &~$N_x \times N_y \times N_z$~ & ~$N_p$~\\[3pt]
        A  & $30$  & $8.5$ & $8.5$ & $1200 \times 340 \times 340$ & $470$ \\ 
        B  & $30$ & $12$ & $12$ & $1200 \times 480 \times 480$ & $936$\\ 
        C  & $30$ & $24$ & $24$ & $1200 \times 960 \times 960$ & $3740$\\
    \end{tabular}
    \caption{Parameters used for the domain size study. For each case, ${M}_s=1.66$ and $\Phi_p=0.2$.}
    \label{tab:dom}
    \end{center}
  \end{table}

\subsection{Effect of varying random particle distributions}
Particles are randomly distributed within the curtain while avoiding overlap. The drag force on individual particles is known to depend on the arrangement of their neighbours~\citep{akiki2017pairwise,lattanzi2022stochastic,osnes2023comprehensive}. This section investigates the impact of different random particle configurations within the curtain on PTKE for $M_s=1.2$ and $\Phi_p=0.3$. Three distinct realizations are considered, keeping all parameters constant except for the random arrangement of particles. 

\begin{figure}
    \centering
    \includegraphics[width=0.55\linewidth]{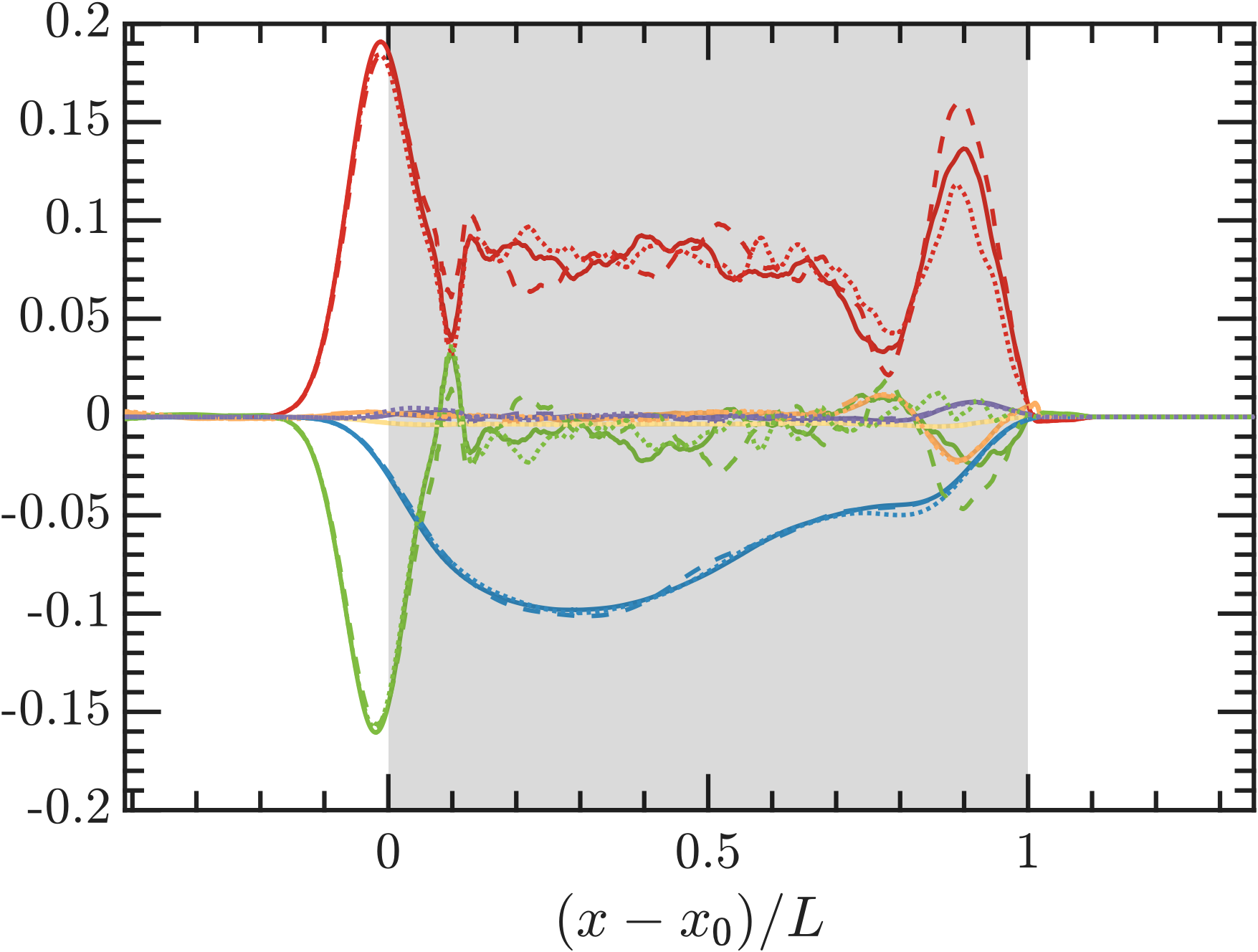}
    \caption{Effect of random particle placement on the PTKE budget for $M_s=1.2$ and $\Phi_p=0.3$ at $t/\tau_L=1$. Realization 1 (\full), realization 2 (\dashed), realization 3 (\dotted). Colours correspond to figure~\ref{fig:budget}.}
    \label{fig:real-conv}
\end{figure}

Figure~\ref{fig:real-conv} shows the PTKE budget terms for each realization at $t/\tau_L=1$, when the shock has just passed the downstream edge of the curtain. All realizations exhibit similar trends with negligible discrepancies. Therefore, we conclude that the random distribution of particles does not significantly affect the statistics.

\section{One-dimensional two-fluid model parameters}\label{appB}

Starting from the conserved variables $X_1 = \alpha_a \rho_a$ and $X_2 = \alpha_g^\star \rho$ with known $\alpha_p$, the primitive variables are found using the following formulae:
 \begin{equation}
    \hat{\kappa} = \frac{X_1}{X_2}; \ \kappa = \frac{T}{T_a}; \ T = \frac{\gamma e}{C_p}; \ T_a = \frac{\gamma e_a}{C_p}; \ e = E - \frac{1}{2}u^2 - k_g ; \
\end{equation}
 \begin{equation}
    \alpha_g = 1- \alpha_p; \ \alpha_a = \frac{\hat{\kappa}}{\hat{\kappa}+\kappa} \alpha_g; \ \alpha_p^\star = \alpha_p+ \alpha_a; \ \alpha_g^\star = \alpha_g - \alpha_a; \ \rho = \frac{X_2}{\alpha_g^\star}; \
\end{equation}
\begin{equation}
    p = (\gamma -1)\rho e ; \ \hat{p} = p + 2/3 \rho k_g.
\end{equation}
The remaining model parameters are defined as follows:
 \begin{equation}
    P_{pfp} = \rho ( \alpha_p^\star u )^2 ; \ F_{pg} = u^2 \partial_x \rho + 2/3 \rho (\partial \alpha_g^\star u / \partial x)u 
\end{equation}
\begin{equation}
    S_a = \frac{\rho}{\tau_a}(c_m^\star\alpha_p \alpha_g - \alpha_a); \ S_{gp}=\max(S_a,0)u ; \ S_E = \max(S_a,0)E + \min(S_a,0)e_a
\end{equation}
 \begin{equation}
    \Rey_p = \frac{\rho D u}{\mu}; \ \Pran = \frac{C_p \mu}{\lambda}; 
\end{equation}
 \begin{equation}
    c_m^\star = \frac{1}{2}\min(1+2\alpha_p,2); \ \tau_a = 0.001\tau_p ; \ \tau_p = \frac{4\rho D^2}{3 \mu C_DRe_p}. 
\end{equation}
The drag coefficient $C_D$ is given by \citet{osnes2023comprehensive}.

\bibliographystyle{jfm}

\bibliography{bibli}

\end{document}

%% file: 0Caseshockphysics.tex
\begin{tikzpicture}
    \node (mainpic) at (0,0) {\includegraphics{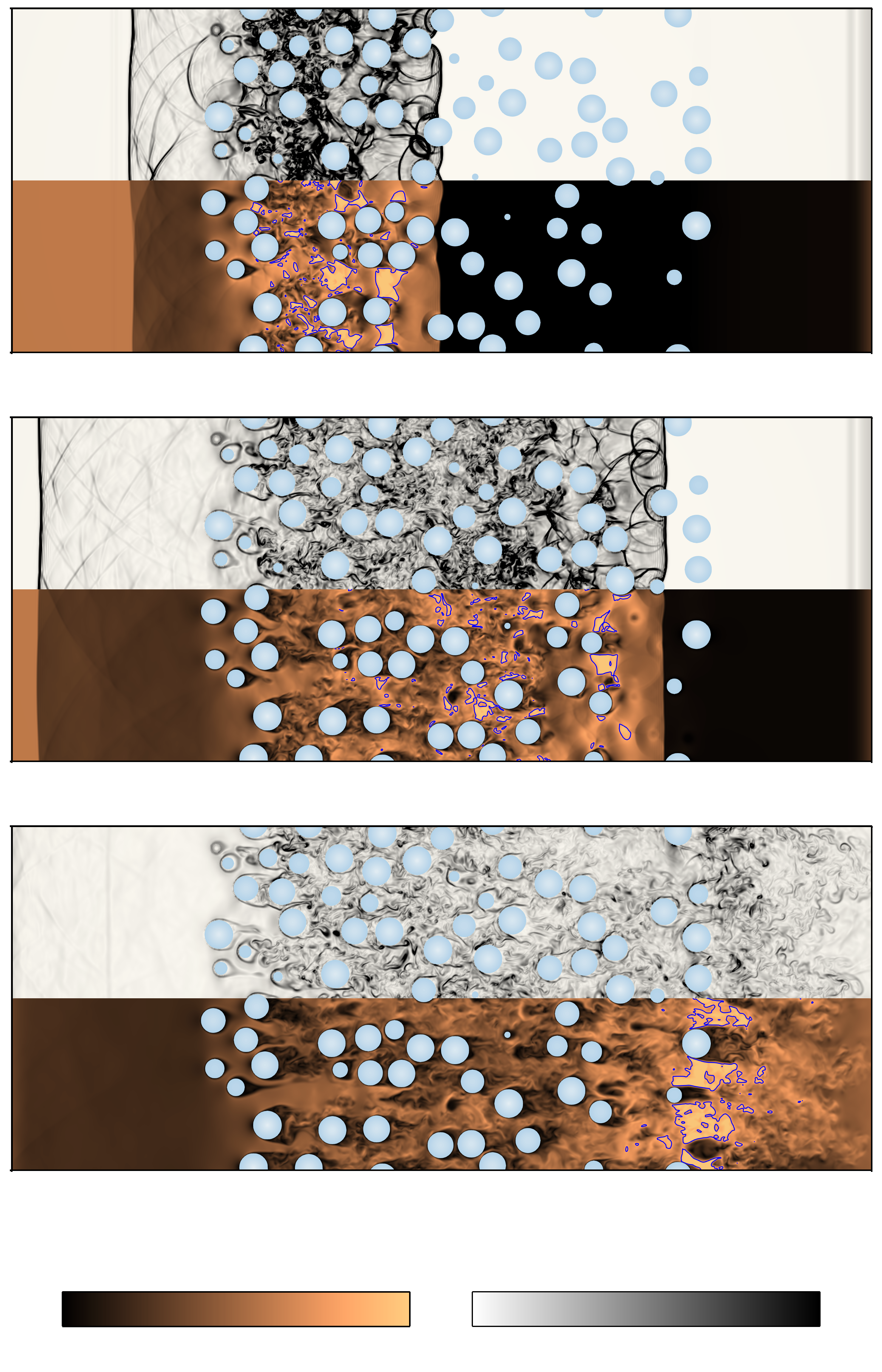}};
    \node[anchor=center, font=\normalsize,scale=10,rotate=0] (ttau1) at (-26,-84) {{$M$}};  
    \node[anchor=center,font=\normalsize, scale=10,rotate=0] (ttau1) at (24, -84) {$\Vert \nabla \rho \Vert \frac{D}{\rho_\infty}$}; 

    \node[anchor=west, scale=10] (ttau1) at (-47, -78) {0};
    \node[anchor=west, scale=10] (ttau1) at (-28,-78) {0.6};
    \node[anchor=west, scale=10] (ttau1) at (-5,-78) {1.2};  
    \node[anchor=west, scale=10] (ttau1) at (2, -78) {0};
    \node[anchor=west, scale=10] (ttau1) at (22, -78) {2.5};
    \node[anchor=west, scale=10] (ttau1) at (42, -78) {5.0};  

    \node[anchor=west, scale=10] (ttau1) at (-50,83) {$(a)$}; 
    \node[anchor=west, scale=10] (ttau1) at (-50,34.5) {$(b)$}; 
    \node[anchor=west, scale=10] (ttau1) at (-50, -14) {$(c)$}; 
    
    \node[anchor=west, scale=10pt] (ttau1) at (-52.3,-58.5) {$|$};     
    \node[anchor=west, scale=10pt] (ttau1) at (-26.8,-58.5) {$|$};      
    \node[anchor=west, scale=10] (ttau1) at (-1.25,-58.5) {$|$};
    \node[anchor=west, scale=10] (ttau1) at (24.2,-58.5) {$|$};
    \node[anchor=west, scale=10] (ttau1) at (49.4,-58.5) {$|$};

    \node[anchor=west, scale=10pt] (ttau1) at (-53,-63) {$0$};     
    \node[anchor=west, scale=10pt] (ttau1) at (-28,-63) {$7.5$};      
    \node[anchor=west, scale=10] (ttau1) at (-2,-63) {$15$};
    \node[anchor=west, scale=10] (ttau1) at (20.125,-63) {$22.5$};
    \node[anchor=west, scale=10] (ttau1) at (46,-63) {$30$};

    \node[anchor=center, scale=10] (ttau1) at (-0.0,-67.5) {$x/D$}; 
    \node[anchor=center, scale=10,rotate=90] (ttau1) at (-60,-36.5) {$y/D$}; 
    \node[anchor=center, scale=10,rotate=90] (ttau1) at (-60,11.5) {$y/D$}; 
    \node[anchor=center, scale=10,rotate=90] (ttau1) at (-60,60) {$y/D$}; 
    \node[anchor=east, scale=10] (ttau1) at (-50, -57.0) {-6  --};  
    \node[anchor=east, scale=10] (ttau1) at (-50, -16.2) {6  --};
    \node[anchor=east, scale=10] (ttau1) at (-50, -36.6) {0  --};


    \node[anchor=east, scale=10] (ttau1) at (-50, -8.5) {-6  --};
    \node[anchor=east, scale=10] (ttau1) at (-50, 11.5) {0  --};
    \node[anchor=east, scale=10] (ttau1) at (-50, 32.0) {6  --}; 

    \node[anchor=east, scale=10] (ttau1) at (-50, 39.8) {-6  --};
    \node[anchor=east, scale=10] (ttau1) at (-50, 60) {0  --};
    \node[anchor=east, scale=10] (ttau1) at (-50, 80.5) {6  --}; 
\end{tikzpicture}   